\documentclass{emulateapj}
\usepackage{graphicx}
\usepackage{epstopdf}
\usepackage{subfigure}
\usepackage{amsmath}
\usepackage{hyperref}
\usepackage{natbib}
\usepackage{appendix}
\makeatletter

\newcommand{\Rmnum}[1]{\expandafter\@slowromancap\romannumeral #1@}
\makeatother

\newcommand{\mkpc}{$\mathrm{M_\odot\,kpc^{-2}}$}

\newcommand {\OIIIHB}{$\mbox{[O\,\small\Rmnum{3}]/H$\beta$}$}

\newcommand {\NIIHA}{$\mbox{[N\,\small\Rmnum{2}]/H$\alpha$}$}
\newcommand {\HII}{$\mbox{H\,\small\Rmnum{2}}$}

\newcommand {\OII}{$\mbox{[O\,\small\Rmnum{2}]}$}

\newcommand {\NIIOII}{$\mbox{[N\,\small\Rmnum{2}]/[O\,\small\Rmnum{2}]}$}

\shortauthors{Hassen Yesuf, et al.}
\shorttitle{From Starburst to Quiescence}

\begin{document}
 
\author{Hassen M. Yesuf\altaffilmark{1}, S. M. Faber\altaffilmark{1}, Jonathan R. Trump\altaffilmark{1,2}, David C. Koo\altaffilmark{1}, Jerome J. Fang\altaffilmark{1}, F. S. Liu\altaffilmark{1,3}, Vivienne Wild\altaffilmark{4}, Christopher C. Hayward\altaffilmark{5}}

\altaffiltext{1}{Department of Astronomy \& Astrophysics, University of California, Santa Cruz, CA 95064 USA}
\altaffiltext{2}{Department of Astronomy \& Astrophysics, Penn State, 515A Davey Lab, University Park, PA 16802  CA 95064 USA}
\altaffiltext{3}{College of Physical Science and Technology, Shenyang Normal University, Shenyang 110034, China}
\altaffiltext{4}{School of Physics and Astronomy, University of St Andrews, North Haugh, St Andrews, KY16 9SS U.K.} 
\altaffiltext{5}{Heidelberger Institut f\"ur Theoretische Studien, Schloss--Wolfsbrunnenweg 35, 69118 Heidelberg, Germany}
\title{From Starburst to Quiescence: Testing AGN feedback in Rapidly Quenching Post-Starburst Galaxies.}
\slugcomment{{\sc Accepted to ApJ:} July 7, 2014}

\begin{abstract}

Post-starbursts are galaxies in transition from the blue cloud to the red sequence. Although they are rare today, integrated over time they may be an important pathway to the red sequence. This work uses SDSS, GALEX, and WISE observations to identify the evolutionary sequence from starbursts to fully quenched post-starbursts in the narrow mass range $\log M(M_\odot) = 10.3-10.7$, and identifies ``transiting'' post-starbursts which are intermediate between these two populations. In this mass range, $\sim 0.3\%$ of galaxies are starbursts, $\sim 0.1\%$ are quenched post-starbursts, and $\sim 0.5\%$ are the transiting types in between. The transiting post-starbursts have stellar properties that are predicted for fast-quenching starbursts and morphological characteristics that are already typical of early-type galaxies. The AGN fraction, as estimated from optical line ratios, of these post-starbursts is about $3$ times higher ($\gtrsim 36 \pm 8 \%$) than that of normal star-forming galaxies of the same mass, but there is a significant delay between the starburst phase and the peak of nuclear optical AGN activity (median age difference of $\gtrsim 200 \pm 100$\,Myr), in agreement with previous studies. The time delay is inferred by comparing the broad-band near NUV-to-optical photometry with stellar population synthesis models. We also find that starbursts and post-starbursts are significantly more dust-obscured than normal star-forming galaxies in the same mass range. About 20\% of the starbursts and 15\% of the transiting post-starbursts can be classified as the ``Dust-Obscured Galaxies'' (DOGs), with near-UV to mid-IR flux ratio of $\gtrsim 900$, while only 0.8\% of normal galaxies are DOGs. The time delay between the starburst phase and AGN activity suggests that AGN do not play a \emph{primary} role in the original quenching of starbursts but may be responsible for quenching later low-level star formation by removing gas and dust during the post-starburst phase.
 
\end{abstract}

\maketitle

\section{Introduction}

Galaxies show bimodality in their colors, morphologies, and star formation rates both locally and at high redshift \citep[e.g.,][]{strateva01,baldry04,bell04,brammer09}. It is thought that star formation quenching causes ``blue cloud'' galaxies to migrate to the ``red sequence'' \citep[][]{bell04,faber07}. A wide variety of quenching mechanisms have been proposed to explain the observed bi-modal galaxy properties \citep[e.g.,][]{dimatteo05,keres05,croton06,dekel06,hopkins06,somerville08,martig09}. These mechanisms quench star formation by heating up gas in the galaxy (halo), stabilizing it against collapse, or (rapidly) using it up or expelling it from the galaxy. These quenching mechanisms can be classified broadly into two modes: fast and slow. The slow mode occurs when star formation gradually fades, probably due to simple gas exhaustion over timescales longer than $\gtrsim 1$\,Gyr, and it does not require any special triggering event such as mergers \citep[e.g.,][]{noeske07,fang13}. On the other hand, rapid quenching is often identified with a triggering event associated with a merger-induced starburst and the resulting feedback (from either the starburst or from an associated AGN) that rapidly removes or exhausts the gas \citep[e.g.,][]{sanders88,hopkins06}. This work focuses on rapidly quenching or recently quenched galaxies.

(Quenched) post-starburst galaxies, also known as K+A or E+A \footnote{The term ``K+A'' refers to a galaxy with significant populations of both (old) K stars and (young) A stars, indicative of rapidly quenched recent star formation.  The term ``post-starburst'' traditionally refers to a K+A galaxy that was necessarily preceded by a starburst.  We use the terms K+A and post-starburst interchangeably. We often use the term K+A in a general sense when we refer to related past studies. We avoid the term ``E+A'' which refers to a quenched galaxy with early-type morphology and a young stellar population, because we show that many starburst galaxies already have compact and early-type morphologies before quenching into post-starbursts.}galaxies \citep[e.g.,][]{dressler83,zabludoff96,quintero04}, offer a unique view into galaxy evolution because they are believed to be recently quenched starbursts rapidly transitioning from the blue cloud to the red sequence. They may contain lingering signatures of a quenching process imprinted on their spectral and structural properties. The fact that these galaxies have unusually large A-star populations but lack younger stars has been interpreted as evidence for recently quenched starbursts.

Theoretically, post-starburst galaxies might be the end-product of galaxy mergers \citep[][]{hopkins06,hopkins08a,bekki01,bekki05,snyder11}. In gas-rich model mergers, tidal torques channel gas to galaxy centers and power intense nuclear starbursts \citep{barnes91,barnes96}. The gas channeled to the centers may also lead to the onset of obscured nuclear AGN activity \citep{dimatteo05,hopkins06}. At the end of the starburst, after gas has been exhausted by the starburst itself and/or expelled by stellar feedback, the leftover gas and dust obscuring the active galactic nucleus (AGN) are cleared out due to feedback from the AGN \citep{springel05a,springel05b,hopkins06,kaviraj07,hopkins08a,snyder11,cen12}. Consequently, star formation and further black hole growth are halted. Then the galaxies pass through the quenched post-starburst phase before they passively age and become ``red and dead''. This work aims to test this hypothesis in detail.

Using galaxy merger simulations, \citet{snyder11} have constrained the typical K+A life-time of merger-induced post-starbursts to be $\lesssim 0.1-0.3$\,Gyr \citep[cf.][]{falkenberg09a,wild09}. They find that the presence of AGN makes almost no difference to the evolution of the post-merger spectrum in their simulation without diffuse dust. However, with diffuse dust, they find that including AGN accretion results in a longer K+A phase with stronger Balmer absorption lines than without this accretion. Their interpretation of this result is that AGN feedback does not itself directly shut off a starburst but rather serves to remove the leftover obscuring dust around the post-starburst population in the nucleus, thereby enhancing the Balmer absorption features. 

Consistent with the merger origin, past observational studies of morphology and kinematics of K+A galaxies in the field hint that they are merger remnants on the way to becoming early-type galaxies \citep{norton01,quintero04,goto07}. As such, they display both early-type morphologies and signs of interactions \citep{schweizer82,zabludoff96,quintero04, balogh05,yang08,pracy09}. Similarly, many post-starburst galaxies are observed to be blue-centered with positive color gradients and have central (within $\sim 4$\,kpc) A star populations \citep[e.g.,][]{yamauchi05,goto08,yang08,swinbank12}.

Identifying the mechanisms responsible for quenching star formation in post-starbursts is still an outstanding problem after thirty years since their discovery. The rest of this section highlights special problems that plague post-starburst studies and have consequently hampered progress in understanding feedback in these systems. A brief discussion of how these problems are addressed in this paper is also presented. The key novel feature of this work is that it broadens the definition of post-starburst using multi-wavelength galaxy colors and spectral indices and thereby identifies AGN in post-starbursts more completely and consistently.

\subsection{Finding a more complete sample of post-starbursts} 

Post-starbursts are rare galaxies, especially at low red-shift \citep{wild09}. They comprise $\lesssim 1$\,\% of all galaxies at $z \sim 0.1$ \citep[]{wong12}. Furthermore, these galaxies 
tend to evolve through regions of parameter space populated by normal galaxies. For these two reasons, post-starbursts are hard to distinguish from the underlying, slowly quenching normal galaxies. Because of this difficulty, the conventional definition of post-starburst is restricted to quenched post-starbursts with weak or no emission lines. The problem is that this definition excludes any transiting (quenching) post-starbursts with on-going star formation or strong AGN activity, which may be a key link between starbursts and the quenched post-starbursts.

In this work, we strive to directly link starbursts and post-starbursts by using a variety of novel yet plausible criteria to identify a more complete sample of objects that are in transit between the starburst phase and the fully quenched post-starburst phase. \citet{wild10} traced the evolution of local bulge-dominated galaxies during the first 600 Myr after a starburst using principle component analysis of stellar continuum indices around the 4000\,\AA\ break.

Our work is complementary to \citet{wild10} in some of its main results but it is quite different in its overall methodology and analysis. For instance, we take a multi-wavelength approach: while we use stellar continuum indices to define some post-starbursts, we also apply additional constraints, such as the dust-corrected global near-UV to optical colors or near-UV to mid-IR colors. The continuum indices are SDSS values measured in a 3\arcsec\ aperture and may not be representative of a galaxy as a whole. They are also not sufficient to identify heavily dust-obscured post-starbursts, which happen to overlap with normal galaxies in their spectral indices. To identify such objects we find that mid-IR colors are useful augmentation.

Finally, one of the methods used in this work is to identify common members of the starburst sequence in a narrow mass slice around $\log M(M_\odot) \sim 10.5$. This value corresponds to the transition mass in the color-mass diagram where both star-forming and quiescent galaxies are presently observed \citep{kauffmann03b}. A narrow mass window approximately captures galaxies on the starburst evolutionary sequence because the quenching timescale \citep[$\lesssim$ 1 Gyr;][]{martin07,snyder11} is much faster than the merging timescale ( a typical local galaxy has a major merger rate of $\lesssim 0.05$\,Gyr$^{-1}$ \citep{hopkins10b}). Therefore, the precursors of post-starbursts are unlikely to grow in mass via multiple mergers in the time it takes them to quench. The burst fraction (amount of new stars formed) in a merger is typically $<20$\% of the total mass \citep[][]{norton01,balogh05,kaviraj07,wild09,swinbank12}. Hence, it also does not significantly contribute to the mass increase on the starburst sequence.

Even though a galaxy can be linked to its immediate progenitors by its mass, mass by itself is not a sufficient predictor of galaxy properties. Recent studies show that structural parameters that combine mass and radius (i.e., stellar mass surface density, $\mu_\ast$ or velocity dispersion, $\sigma$) are better tracers of galaxy quenching \citep[][]{kauffmann06,franx08,cheung12,wake12,fang13}. This work will examine whether $\mu_\ast$ and $\sigma$ of starbursts and post-starbursts are indicative of quenching in these galaxies.
    
\subsection{Finding all AGN, including a population of highly
    obscured AGN.}

Most previous studies on post-starburst galaxies excluded strong AGN because they adopted a restrictive definition of post-starbursts as galaxies with weak or no emission lines. \citet{wild10} attempted to improve this by defining post-starbursts using spectral indices only, bypassing the need for (weak) emission line requirements. However, heavily dust-obscured post-starbursts and broad-line AGN are still excluded or missing from their sample. This work attempts to include dust-obscured post-starbursts as part of the starburst sequence. It also constrains the star formation rates of broad-line AGN and investigates whether they are preferential to a specific stage in the merger sequence \citep[e.g.,][]{hopkins06}. We use GALEX and WISE photometery in our selection criteria of obscured post-starbursts, and  we use the WISE $12\,\mu \mathrm{m}$ luminosity as a proxy for star formation rates (upper limits) of broad-line AGN. In a follow up paper, we plan to do further study on the star formation rates of broad-line AGN using far-infrared data.   

Past studies of quenched post-starburst galaxies hint that AGN are more common in these galaxies than in normal galaxies \citep{yan06,georgakakis08,brown09}. However, these past studies were explicitly biased against strong AGN (Seyferts) since they excluded emission-line galaxies from their post-starburst samples. These studies also cannot exclude the possibility that the weak AGN signatures in their post-starbursts are from ``LINER-like'' emission unrelated to AGN activity \citep{cid11,yan12,singh13}. Regardless, \citet{yan06} have found that 95\%  of their K+A galaxies have LINER-like line ratios. Using a sample of 44 K+A galaxies at $z \sim 0.8$, \citet{georgakakis08} have found  a higher fraction of X-ray sources in post-starbursts ($\sim 15\%$) than in normal red-sequence galaxies $(\sim 5\%$). These sources are mostly low luminosity AGN at best and have a hard mean stacked X-ray spectrum suggesting moderate levels of obscured AGN activity in the bulk of this population. Similarly, \citet{brown09} have found that a third (8/24) of their K+A galaxies at $z \sim 0.2$ are X-ray sources with luminosities of $\sim 10^{42}$ erg\,s$^{-1}$. 

To improve on these previously incomplete estimates of the AGN fraction in post-starbursts, we assemble a large and less biased sample of post-starbursts which includes emission-line galaxies to robustly identify AGN. This enables us to estimate the AGN fraction in transiting post-starbursts for the first time. We will infer the relationship between AGN and recent quenching in post-starbursts from the AGN fraction, and the time interval between the peak of starburst to the peak of AGN activity. If the AGN fraction is low, it indicates that AGN and quenching of starbursts are likely not related. A significant AGN delay might indicate a non-causal or secondary relationship (e.g., a common fueling mechanism or later additional quenching) between starbursts and AGN even if AGN are more common in post-starbursts than in normal galaxies. 

The rest of the paper is structured as follows. Section 2 describes the multi-wavelength data. Section 3 presents the sample selection. Starbursts and the different classes of post-starbursts are defined in this section. Section 4 investigates the AGN properties of post-starbursts. Section 5 presents the bulge properties of post-starbursts as an independent check on our sample selection. Section 6 presents a discussion on the importance of post-starbursts in the build-up of the red sequence. This section also summarizes the main results of the paper. Throughout the paper, an $(\Omega_m,\Omega_\Lambda, h) = (0.27, 0.73, 0.7)$ cosmology is used. All magnitudes and colors are on the AB system unless indicated otherwise.

\section{DATA AND MEASUREMENTS}

In the first three subsections, we will briefly describe the SDSS, GALEX and WISE data used. In the later subsections, we will describe the dust correction, galaxy structural parameters and stellar population modeling employed in the following sections.

\subsection{SDSS}

The Sloan Digital Sky Survey \citep[][SDSS]{york00} is a large photometric and spectroscopic survey. It has mapped out about a third of the celestial sphere with its five filter band-passes, $ugriz$ \citep{fukugita96}. The parent sample (\S3.1) used in this paper comes from SDSS Data Release 8 \citep[DR8,][]{aihara11}. The SDSS DR8 has more value-added quantities essential for the paper. 

As described in \citet{aihara11}, DR8 includes various galaxy physical parameters such as stellar masses. Briefly, the stellar masses are estimated from $ugriz$ photometery using the Bayesian methodology to calculate the likelihood of each model star formation history (SFH) given the data \citep{kauffmann03a}. The mass estimate assumes that the SFH is approximated by a sum of discrete bursts and uses templates over a wide range in age and metallicity. Thus, there should be no concern over systematic differences between the stellar mass estimates of starbursts, post-starbursts and normal galaxies. In addition, the masses of these galaxies are dominated by their old pre-burst stellar populations as the contribution to the total mass from newly formed stars in a burst is only $3-20$\% of the total mass \citep[][]{norton01,balogh05,kaviraj07,wild09,swinbank12}. 

DR8 also provides spectral indices and emission line measurements \citep{tremonti04,aihara11}. To measure the nebular emission lines of a galaxy, the continuum is modeled as a non-negative linear combination of single stellar population (SSP) template spectra generated using the \citet[][]{bc03} (hereafter BC03) population synthesis code, and the best fitting model is subtracted from the galaxy spectrum.

\subsection{GALEX}

We use UV data from the Galaxy Evolution Explorer \citep[GALEX,][]{martin05} to to exploit the greater sensitivity of its near-UV ($m_{NUV}<20.8$) band to recent star formation. The near-UV (1771 - 2831 \AA) imaging data have a spatial resolution of 6-8\arcsec\, and 1\arcsec\ astrometry. The data come from the cross-matched catalog between GALEX GR6 against SDSS DR7. This catalog is available through the GALEX CASJobs interface\footnote{http://galex.stsci.edu/casjobs/}. At fainter UV magnitudes, GALEX loses red galaxies because they drop below the GALEX detection threshold. About 82\% ($\sim$ 220,000) of galaxies in SDSS spectroscopic sample ($m_r<17.77$), in the redshift of interest for this work ($0.03<z<0.1$), have a GALEX counterpart within $5\arcsec$. Adopting brighter r-band limit gives higher completeness ($\gtrsim 90\%$) but excludes significantly more post-starbursts (see \citet{wyder07} for a discussion of GALEX completeness relative to SDSS). Since post-starbursts (including dust-obscured ones) are mainly in the blue cloud and green valley, the GALEX incompletenss is less likely to affect our results significantly. Furthermore, about 10\% of the post-starburst galaxies have multiple GALEX matches within $5\arcsec$. Although the GALEX photometery for post-starbursts with multiple matches may not be accurate, we do not exclude them lest we systematically exclude merging systems. About 90\% of these post-starbursts are significantly dust obscured compared to normal galaxies. The exclusion of these post-starbursts does not significantly alter any of our main results.
 
\subsection{WISE}

The Wide-field Infrared Survey Explorer \citep[WISE,][]{wright10} performed an all-sky survey with photometery in the $3.4\,\mu$m, $4.6\,\mu$m, $12\,\mu$m, and $22\,\mu$m bands. We used the Infrared Science Archive (IRSA)\footnote{http://irsa.ipac.caltech.edu/Missions/wise.html} to match SDSS galaxies with the closest WISE sources within a 5\arcsec\ radius. About 99 (92)\% of SDSS galaxies with 5\arcsec (2\arcsec) GALEX matches have corresponding matches in WISE. We use WISE data to study obscured star formation and AGN properties of post-starburst galaxies. 
 
\subsection{Dust correction}
\label{sec:dustcorr}

The main purpose of the dust correction is to reduce the number of dusty obscured emission-line galaxies, which otherwise masquerade as post-starbursts. We use the Balmer decrements, $\mathrm{H}\alpha/\mathrm{H}\beta$, with the physically motivated two-component dust attenuation model of \cite{charlot00} to correct for attenuation of the nebular emission lines by dust. In the two-component model, the diffuse dust accounts for 40\% of the optical depth at 5500\,\AA\ while the denser birth-cloud dust accounts for the other 60\% \citep{wild11a}. The optical depth of the dust is assumed to be a power-law of the form $\tau_\lambda \propto \lambda ^{-0.7}$ for the diffuse dust and $\tau_\lambda \propto \lambda ^{-1.3}$ for the birth-cloud dust. We adopt this model because it has a physical basis and is broadly consistent with observations \citep{wild11a}.     

In addition, we correct the continuum fluxes (i.e., integrated magnitudes) using the empirical relationship between the emission line and continuum optical depths found in \cite{wild11b} and their empirical stellar attenuation curve. They found that Balmer emission lines experience two to four times more attenuation than the continuum at 5500\,\AA. We apply the dust correction only on galaxies whose $\mathrm{H\alpha}$ and $\mathrm{H\beta}$ lines are well measured (with signal-to-noise ratio (SNR) $> 1$). Galaxies with undetected or low signal-to-noise Balmer emission lines are not dust-corrected and their observed quantities are used as the intrinsic ones. We assume the dust-free Case B recombination ratio of $\mathrm{H\alpha/H\beta = 2.86}$  for \HII\; regions \citep[][]{osterbrock89} and $\mathrm{H\alpha/H\beta = 3.1}$ for type \Rmnum{2} AGN \citep{veilleux87}.

The Balmer decrements are measured within the 3\arcsec\ fiber and do not reflect the galaxy-wide values, as there are dust gradients across galaxies \citep{munoz09,wild11b}. We make an approximate correction for this effect following \cite{wild11b}.

We also correct for Galactic extinction of optical fluxes using the catalog values provided in SDSS DR8 and of the NUV fluxes assuming a ratio $A_\mathrm{NUV}/E(B-V)= 8.2$ \citep{wyder07}, where $A_\mathrm{NUV}$ is the NUV Galactic extinction and $E(B-V)$ is the $B-V$ color excess.

More details on the dust correction can be found in Appendix~\ref{sec:appA}, where it is shown that our post-starburst selection does not significantly depend on the detailed assumptions of the dust correction described above. For instance, using single foreground screen model for dust distribution \citet{calzetti00}, we recover 85\% of PSBs selected using the two-component dust attenuation model. However, the single-component model identifies $\sim 15-25\%$ more PSB candidates, which may also be dusty contaminants. Throughout the paper, the subscript `$\mathrm{dc}$' on a given quantity denotes dust-correction. For example, $\mathrm{W_{H\alpha,dc}}$ denotes a dust-corrected $\mathrm{H\alpha}$ equivalent width ($\mathrm{W_{H\alpha}}$).

\subsection{K-correction}
In addition to the dust correction, all galaxy magnitudes and colors used in this work are k-corrected to $z=0$ using the public $kcorrect$ IDL code \citep{blanton07}. The GALEX NUV magnitude and the five SDSS $ugriz$ magnitudes are used in estimating the k-correction.

\subsection{Structural parameters}

This section describes three structural parameters used to study the relationship between star formation quenching and bulge growth. 

The stellar surface mass density is defined as the ratio of half the total stellar mass to the half-light Petrosian $z$ band area, $\mu_\ast = M_\ast/2 \pi R_\mathrm{50,z}^2$. \citet{kauffmann06} found that $\mu_\ast$ is inversely proportional to the consumption time of the accreted gas from a galaxy halo (i.e, the burst decline time). They suggested that a high stellar surface mass density may be connected to bulge formation through a nuclear starburst and quenching of star formation. However, \citet{fang13} showed that the mass surface densities as defined above may exaggerate structural differences between blue and red galaxies because they use a light-profile based radius as opposed to mass-profile based radius. We use $\mu_\ast$ as defined above only to show that starbursts and post-starbursts are both bulge-dominated galaxies, unlike most normal star-forming galaxies.

The velocity dispersion, $\sigma$, corrected to 1/8 of the effective radius, $r_\mathrm{e}$, is estimated from the velocity dispersion measured within the 1.5\,\arcsec\ radius fiber, $\sigma_{1.5}$, using the relation: $\sigma = \sigma_{1.5}(8\times 1.5\arcsec/r_\mathrm{e})^{0.066}$ \citep{cappellari06}. $\sigma_{1.5}$ is measured by the SDSS \textit{idlspec2d} pipeline using broadened stellar PCA templates \citep{aihara11}. For $r_\mathrm{e}$, we use the weighted average of the circularized $r$-band radi of the de Vaucouleurs profile $(r_\mathrm{e,dev})$ and exponential profile $(r_\mathrm{e,exp})$ : $r_\mathrm{e} = f_\mathrm{dev} \times r_\mathrm{e,dev} \sqrt{b/a} + (1-f_\mathrm{dev}) \times r_\mathrm{e,exp}\sqrt{b/a}$, where $f_\mathrm{dev}$ is a coefficient that characterizes a galaxy image as a linear combination of a de Vaucouleurs profile and an exponential profile (available in the SDSS catalog). 

The color gradient, $\nabla_\mathrm{color}$, is defined as the difference between the $g-r$ galaxy-wide color and the $2\arcsec$ $g-r$ aperture color. The $2\arcsec$ aperture magnitudes are available in SDSS DR8. The global galaxy colors are derived from model magnitudes by fitting the galaxy light with either de Vaucouleurs or exponential profile. 

Previous studies have used color gradients defined based on $3\arcsec$ apertures \citep[][]{roche09,bernardi11}. We define $\nabla_\mathrm{color}$ using the $2\arcsec$ aperture instead to better probe galaxy centers. For instance, about 90\% of galaxies in the parent sample have half-light $r$-band areas that are twice the $2\arcsec$ aperture areas at the corresponding redshifts. In comparison, only $\sim 60\%$ of the galaxies have half-light areas that are twice the $3\arcsec$ aperture areas. We note that this is the only time we use a quantity measured within a $2\arcsec$ aperture.

Positive $\nabla_\mathrm{color}$ means blue-centered (young bulge), negative $\nabla_\mathrm{color}$ means red-centered (old bulge) and $\nabla_\mathrm{color} \sim 0$ means a uniform color throughout a galaxy.

\subsection{Stellar population modeling}

To illustrate how a starburst evolves in some of our diagrams, we overplot \citet{bc03} model tracks on these diagrams. To do so, we model SFHs of a post-starburst as a superposition of an old stellar population initially starting to form at time $t=0$ and following a delayed exponential SFH of the form $\psi \propto t\exp(-t/\tau_1)$ with e-folding time $\tau_1=1$ Gyr \citep[cf.][]{kriek11} plus a young stellar population formed in a recent burst at $t=12.5$\,Gyr ($z \sim 0.1$) with exponentially declining SFH, $\psi \propto \exp(-t/\tau_2)$ and $\tau_2 = 0.1$\,Gyr \citep[cf.][]{kaviraj07,falkenberg09a}. The SSP models assume \cite{chabrier03} IMF, a solar metallicity for SFH before the recent burst, and 2.5 solar metallicity for the recent burst. A superposition of the two SFHs with varying burst mass fraction ($bf \sim 3\%-20\%$) generally describes the starburst to post-starburst evolution. Because of the well-known burst mass-age degeneracy, the ages of the post-starbursts depend on the assumed decay timescale. In Appendix~\ref{sec:appB}, we quantify the effect of using different decay timescales  ($\tau_2 = 0.05$\,Gyr or $\tau_2 = 0.2$\,Gyr) instead of our adopted one. The model tracks overplotted on the data in some of our figures mainly serve to facilitate the interpretation of the data, and our post-starburst selection is purely empirical: it does not explicitly use the models. 

\subsection{Galaxy merger simulation}

To further justify our selection of dust-obscured post-starburst galaxies, we use results from the M2M2 simulation presented in \citet{lanz14} and \citet{hayward14}, which is an equal-mass merger of two disk galaxies. Each disk galaxy is composed of a dark matter halo, gaseous and stellar exponential disks, and a bulge. The progenitor galaxies each have a stellar mass of $1.1 \times 10^{10} M_{\odot}$ and a gas mass of $3.3 \times 10^{10} M_{\odot}$. See \citet{lanz14} for full details of the specific simulation used.

The merger was simulated using the smoothed-particle hydrodynamics code {\sc Gadget-3} \citep{springel05a}. The simulation includes models for star formation and stellar feedback \citep{springel03} and black hole accretion and AGN feedback \citep{springel05b}. In post-processing, the three-dimensional dust radiative transfer code {\sc Sunrise} \citep{jonsson06,jonsson10} was used to calculate synthetic UV--mm SEDs of the simulated merger at various times throughout the merger. {\sc Sunrise} uses the stellar and AGN particles from the {\sc Gadget-3} simulation as sources of radiation and calculates the effects of dust absorption, scattering, and re-emission as the radiation propagates through the dusty ISM of the simulated galaxies. {\sc Sunrise}
calculates SEDs and images from arbitrary viewing angles. For clarity, we show only results from a single viewing angle in this work. See \citet{jonsson10} and \citet{hayward11} for further details of the {\sc Sunrise} calculations.

\section{Sample Selection}

This section presents the parent sample, and details on how starbursts and post-starbursts are selected from this sample. 

\subsection{The Parent Sample}

\begin{figure*}
\mbox{\subfigure{\includegraphics[width=3.5in]{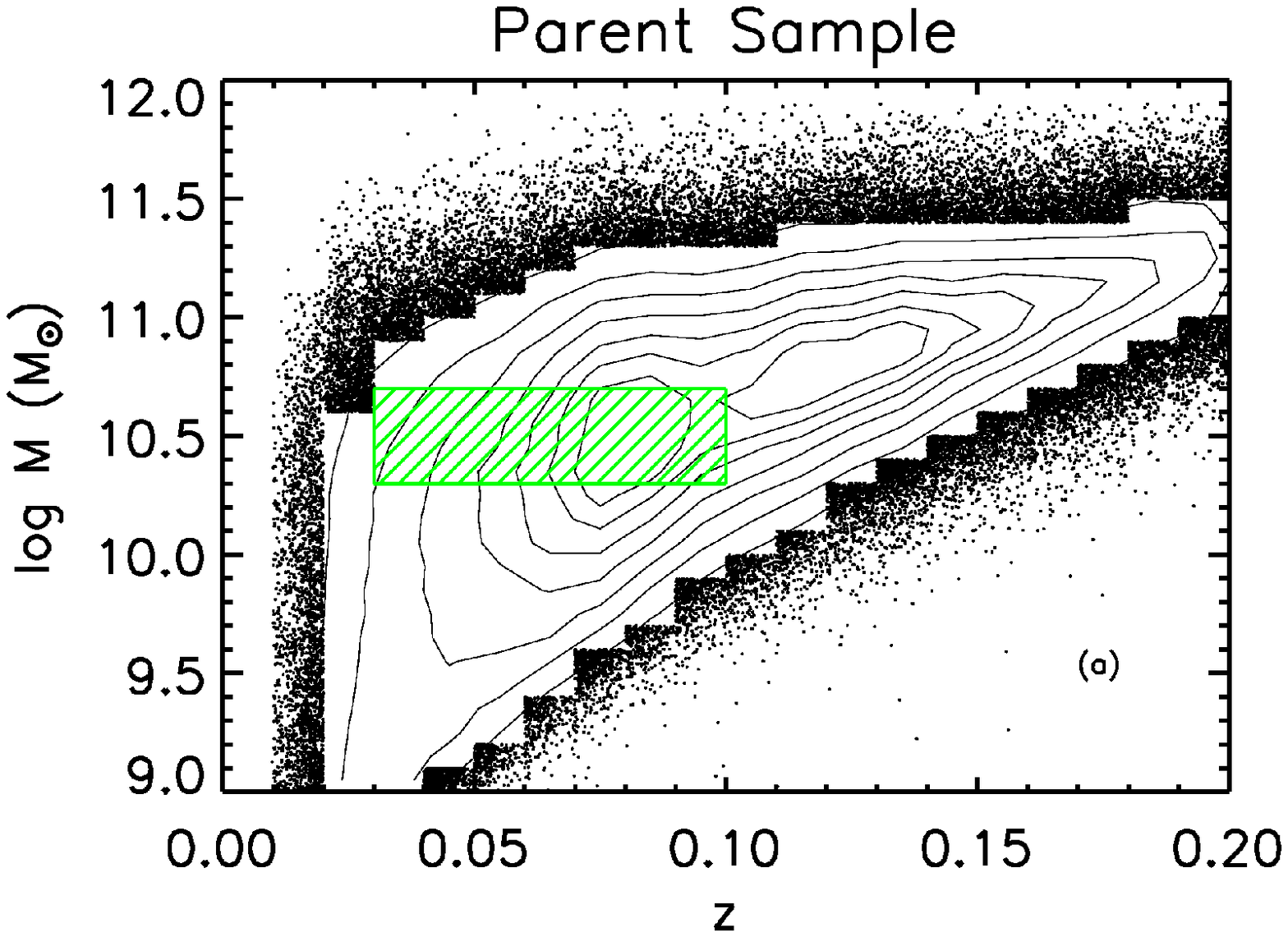}}
\subfigure{\includegraphics[width=3.5in]{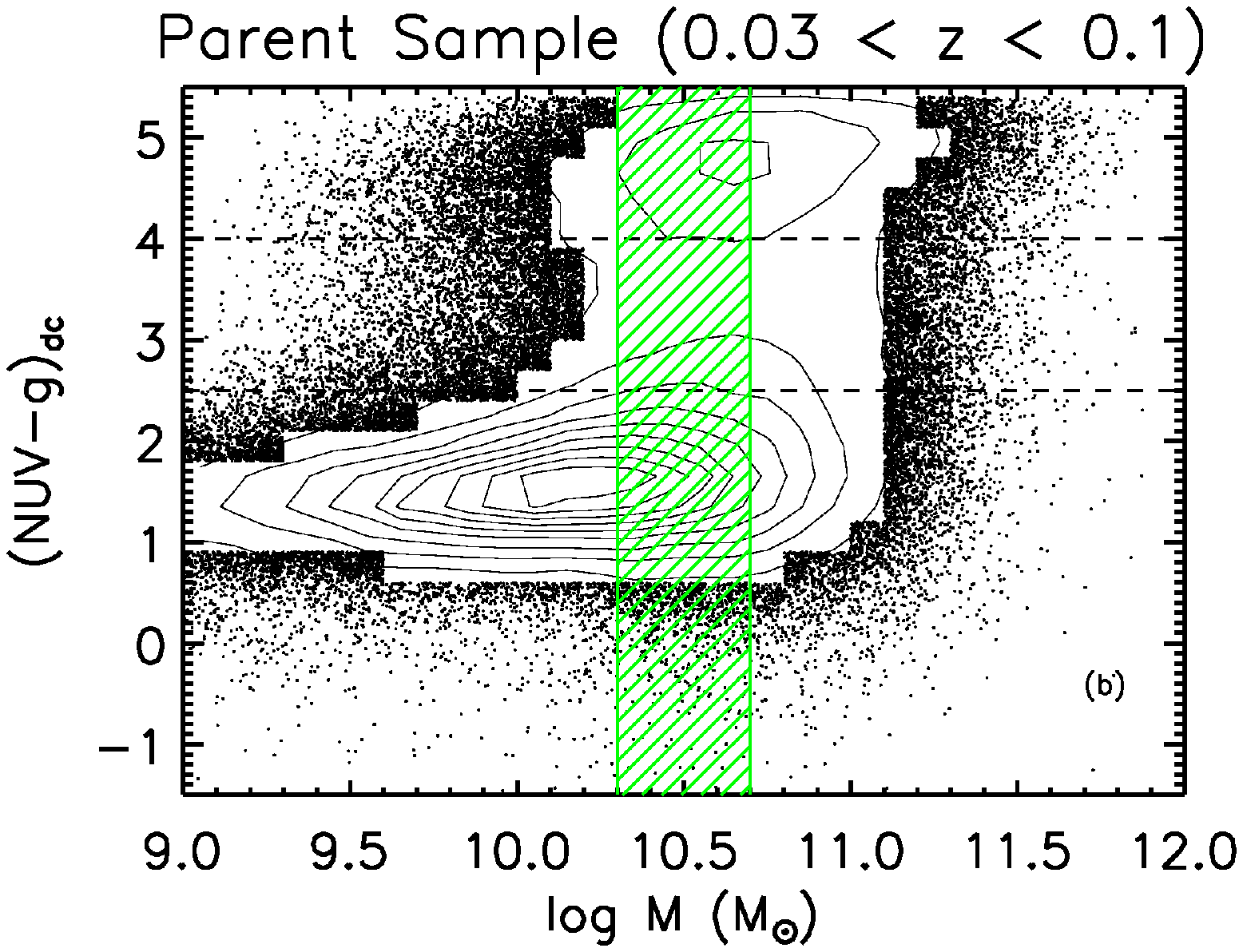}}}
\caption{Panel a): Redshift versus stellar mass. The black points are galaxies in the SDSS-GALEX-WISE-matched catalog. Panel b): Dust-corrected $NUV-g$ color versus stellar mass for galaxies in redshift range $0.03<z<0.1$. This study uses a volume-limited parent sample of galaxies in a narrow mass slice around the transition mass between the blue-cloud and the red-sequence. The hatched regions in both panels define the parent sample ($\log\,M (M_\odot) = 10.3-10.7$ and $0.03<z<0.1$). It is known that quenched post-starburst preferentially occupy the low-mass end of the green valley \citep{wong12}. Hence, we chose the lower mass end of the transition mass. The horizontal dashed lines approximately demarcate the blue cloud, the green valley and the red sequence.\label{fig:colormass}}
\end{figure*}

The basic sample selection is shown Figure~\ref{fig:colormass}. The sample consists of a SDSS/GALEX/WISE-matched volume-limited sample ($0.03<z<0.1$) in a narrow stellar mass range of $\log\,M (M_\odot) = 10.3-10.7$. We call this sample of $\sim 67,000$ galaxies the \emph{parent sample}. The chosen mass range roughly corresponds to the transition mass in the color-mass diagram (Figure~\ref{fig:colormass}b) from lower-mass star-forming blue galaxies to higher-mass quiescent red galaxies \citep{kauffmann03b}. We located the center of the mass bin on the lower end of the transition mass because post-starbursts are preferentially found in smaller-mass galaxies, unlike slowly transitioning galaxies which dominate at higher masses \citep[see also][]{wong12}. Moreover, restricting the redshift to be less than 0.1 ensures higher GALEX completeness of the parent sample to red-sequence galaxies. As discussed in \S1, the starburst-to-post-starburst evolution is followed in the narrow mass-slice because mass likely does not increase significantly more than a factor of 2 along the starburst sequence.

\begin{figure*}
\centering
\includegraphics[scale=0.5]{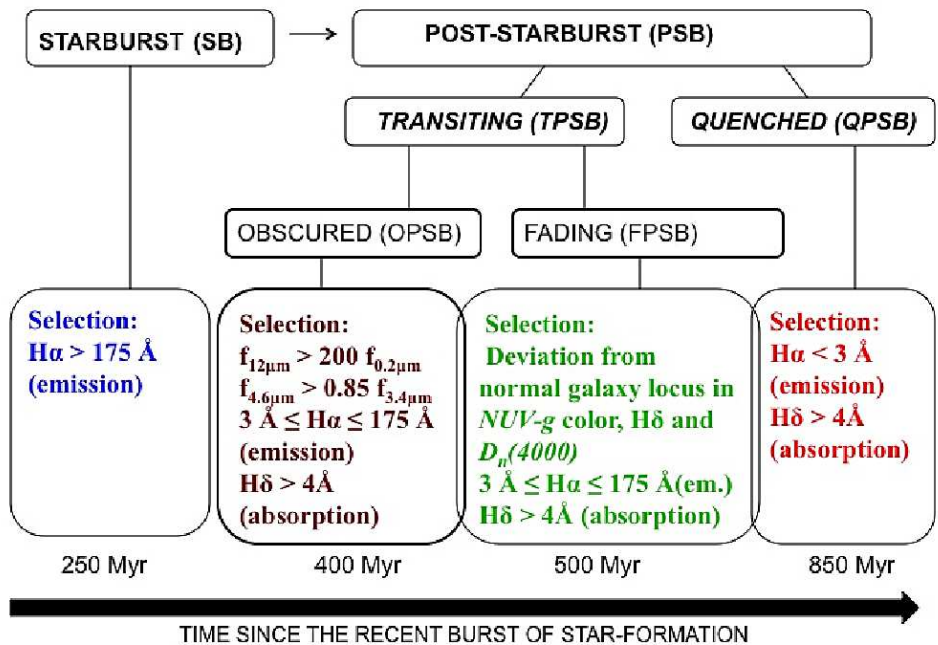}
\caption{Schematic outline of our starburst and post-starburst selection. Identifying a reasonably complete transiting post-starburst population between the starbursts and the quenched post-starbursts is the major new aspect of this paper. The TPSBs are identified by combining their GALEX and/or WISE photometery with with their optical photometery and spectral indices. \label{fig:flow}}
\end{figure*}

As schematically outlined in Figure~\ref{fig:flow}, the next three subsections describe in detail the selection of starbursts and post-starbursts from the parent sample. Starbursts are selected to have $\mathrm{H\alpha}$ emission equivalent width above 175\,\AA. The selection of post-starbursts generalizes the conventional definition to encompass both quenching and quenched objects. The conventional post-starbursts, which are characterized by weak or no emission lines but strong Balmer absorption lines, are termed as ``Quenched Post-starbursts (QPSBs)'' in this paper. Transiting post-starburst (TPSB) galaxies, which precede quenched post-starbursts but come after the starbursts, are selected in two ways. The first selection is based on the distinctive evolutionary path that starbursts and post-starbursts follow in the 3D parameter space defined by dust-corrected $NUV-g$ color, $\mathrm{H\delta}$ equivalent width and the 4000\,\AA\ break. Objects in this first class are called ``Fading Post-starbursts (FPSBs)''. They are clearly offset from normal galaxy locus in the parameter space that defines them.  The second selection of TPSBs uses GALEX and WISE photometery to identify dust-obscured transiting post-starbursts, which are simply referred as ``Obscured Post-starbursts (OPSBs)''. We will later show that both classes of transiting post-starbursts have similar properties (e.g., morphology) and that OPSBs generally precede the FPSBs. 

The above discussion has isolated four classes of SB and PSB galaxies, we now proceed to explain how the four classes are selected.

\subsection{Starbursts}

A starburst has been defined in at least three ways \citep{knapen09}. The definition we adopt considers a starburst as a galaxy with a temporarily higher current SFR than its past average by a factor of $2-3$ \citep[e.g.,][]{brinchmann04,kennicutt05}. This can be quantified by a threshold in the equivalent width of $\mathrm{H}\alpha$ ($\mathrm{W_{H\alpha}}$). Galaxies with ratios of current to past average SFR greater than two or three have (dust-extincted) $\mathrm{W_{H\alpha} \gtrsim 80-110}$ \AA\; \citep{lee09,mcquinn10}. 

We define a starburst as a galaxy with \textit{dust-corrected} $\mathrm{W_{H\alpha,dc}} >  175$\,\AA\;. This threshold corresponds to $\gtrsim 2\,\sigma$ deviation from the mean $\mathrm{W_{H\alpha,dc}}$ distribution of star-forming galaxies in the parent sample (star-forming galaxies are objects below the maximum starburst boundary of \cite{kewley01} in the BPT diagram). We note that a starburst with $\mathrm{W_{H\alpha} \gtrsim 80-110}$\,\AA\ and nebular extinction $\mathrm{A_V} = 1$ will have $\mathrm{W_{H\alpha,dc} \gtrsim 125-175}$\,\AA\ if the continuum is extincted less than the gas by a factor of two, as observed in starbursts \citep{calzetti00}. Our starbursts have a median $\mathrm{A_V}$ of 2.3 and (fiber) SFR of about $\mathrm{10\,M_\odot\,yr^{-1}}$ (specific SFR of about $\mathrm{10^{-9}\,yr^{-1}}$). For comparison, the typical SFR of a normal star-forming galaxy in the parent sample is about $\mathrm{1\,M_\odot\,yr^{-1}}$.

\subsection{Quenched Post-Starbursts (QPSB)}

The conventional post-starburst galaxies are characterized as having no detectable or weak current star formation, but with significant star formation in their recent past ($< 1$\,Gyr). These two underlying characteristics have been quantified using various spectral signatures. The lack of ongoing star formation is inferred from weak $\mathrm{H}\alpha$ and/or \OII\, emission lines. The episode of significant recent star formation is inferred from the presence of strong Balmer absorption lines ($H\delta \gtrsim 4$\,\AA), indicative of intermediate-age stars (A stars) or from the relative ratio of young to old stars or from the comparison of Balmer absorption lines to 4000\,\AA\ break strength \citep[e.g.,][]{zabludoff96,balogh99,blake04,poggianti04,quintero04,yang04,goto07,wild07,yan09}.

\begin{figure*}
\includegraphics[scale=0.7]{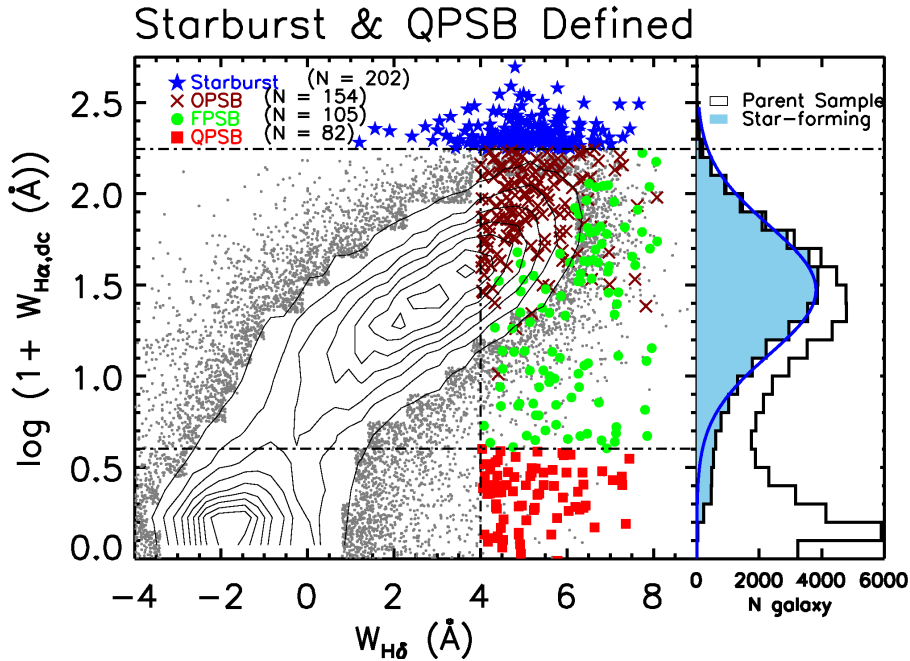}
\caption{The dust-corrected $\mathrm{H\alpha}$ emission equivalent width ($\mathrm{W_{H\alpha,dc}}$) against $\mathrm{H\delta_A}$ absorption strength($\mathrm{W_{H\delta}}$) for the galaxies in the parent sample (grey), starbursts (blue stars), fading post-starbursts (FPSBs; green circles), obscured post-starbursts (OPSBs; brown Xs), and quenched post-starbursts (QPSBs; red squares). This diagram defines starbursts and QPSBs only, as galaxies lying above the upper horizontal line and below the lower horizontal line; FPSBs and OPBS are defined by the next figures. The definition for each class is given in \S3 and the number of galaxies in each class is indicated on the plot. The histograms show distributions of the $\mathrm{H\alpha}$ EW for star-forming galaxies with well-measured emission lines (shaded sky-blue histogram) and the parent sample. As shown by the blue curve, the distribution of $\mathrm{H\alpha}$ EW for star-forming galaxies is well fit by a log-normal distribution with $\mu = 1.5$ and $\sigma = 0.35$. We define starbursts as objects with $\mathrm{W_{H\alpha,dc} > 175\,\AA}$, which is more than $2\,\sigma$ from the mean. \label{fig:hahd}}
\end{figure*}

Figure~\ref{fig:hahd} shows the dust-corrected equivalent width $\mathrm{W_{H\alpha,dc}}$ in emission versus $\mathrm{W_{H\delta}}$ in absorption, after the emission line infill correction. This diagram is used to define QPSBs, which will help us motivate and explore how such a conventional definition of post-starburst can be improved on to include emission line galaxies (AGN or star-forming post-starbursts). 

The grey points in the figure represent all galaxies in the parent sample. For the majority of galaxies, $\mathrm{W_{H\alpha,dc}}$ and $\mathrm{W_{H\delta}}$ are well-correlated with some scatter. Normal\footnote{The adjective ``normal'' is used throughout the paper to describe galaxies that have not undergone a large burst ($> 10\%$)  of star formation recently ($< 1$\,Gyr).} star-forming galaxies form an elongated concentration above $\mathrm{W_{H\alpha}} \gtrsim 10$\,\AA\; and $\mathrm{W_{H\delta}} \gtrsim 2$\,\AA\,while quiescent galaxies clump below $\mathrm{W_{H\alpha,dc}} \lesssim 3$\,\AA\,and $\mathrm{W_{H\delta}} \lesssim 1$\,\AA. Galaxies undergoing starburst or rapid quenching move vertically in this diagram \citep[e.g.,][]{shioya01,quintero04}. Galaxies undergoing strong starburst lie above the star formation sequence, while recently quenched post-starbursts lie below the sequence. 

We define quenched post-starbursts as galaxies with $\mathrm{W_{H\alpha,dc}} < 3$\,\AA\, and $\mathrm{W_{H\delta} > 4 }$\,\AA. We consider only QPSBs with well-measured $\mathrm{W_{H\delta}\, (\mathrm{SNR} > 3)}$ and $\mathrm{W_{H\alpha}}$ (contaminants with bad $\mathrm{H\alpha}$ equivalent width measurements due to spectral gaps around $\mathrm{H\alpha}$ are excluded). QPSBs are denoted by (red) squares and are found in the lower-right corner of Figure~\ref{fig:hahd}. The (blue) stars in the top-right corner represent the starburst galaxies selected in the previous subsection ($\mathrm{W_{H\alpha,dc}} >  175$\,\AA). A large gap exists between starbursts and QPSBs, which must contain many transiting objects if the basic picture of aging starbursts in this paper is correct. Identifying these transiting post-starbursts is the next goal of this paper. The (green) circles and the (brown) Xs represent the two types of transiting post-starbursts that are found in the next subsection.

\subsection{Transiting Post-starbursts (TPSB)}

This subsection will describe our two ways of identifying TPSBs. Because of its similarity to that of previous works, the selection of FPSBs is described first for convenience, but FPSBs actually come after the OPSBs in time.

\subsubsection{Fading Post-starbursts (FPSB)}

\begin{figure*}
\includegraphics{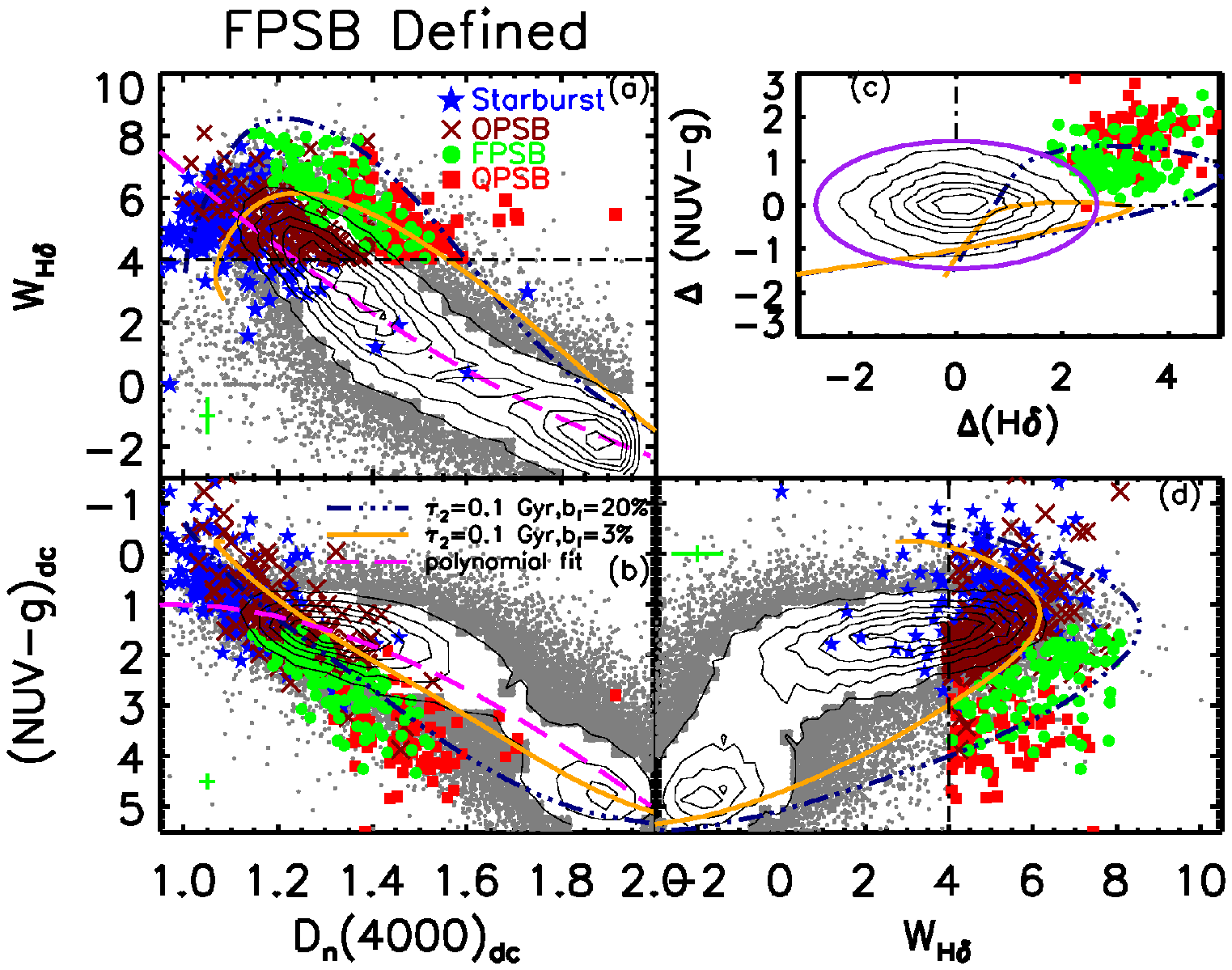}
\caption{The co-joined plots show the relation among dust-corrected $D_n(4000)$, $\mathrm{W_{H\delta}}$ and dust-corrected $NUV-g$ color for galaxies in the parent sample (grey), starbursts (blue stars), obscured post-starburst (OPSB; brown Xs), fading post-starbursts (FPSB; green circles), and quenched post-starbursts (QPSB; red squares). The (magenta) dashed curves in panel a and panel b are the polynomial fits to the main galaxy sequence (these are fits to the data, not burst models). The (dark blue) dash-dotted curve and the solid (orange) curve are BC03 burst tracks with star formation timescales, $\tau_2=0.1$\,Gyr and burst mass fraction ($b_f$) 3\% or 20\%. Panel c shows the difference in $\mathrm{W_{H\delta}}$ and $(NUV-g)_\mathrm{dc}$ from the polynomial fit at a given $D_n(4000)$. The FPSBs are selected if they are found in the upper right corner and outside the (purple) ellipse, which encloses normal galaxies at the $> 2\,\sigma$ level. The typical errors in each panel for transiting post-starbursts are shown as green crosses.\label{fig:hd_d4}}
\end{figure*}

Figure~\ref{fig:hahd} illustrates the point made in the introduction that it is difficult to identify the transiting post-starburst population without additional constraints because they mostly overlap in this figure with the normal (non-bursty) star-forming galaxy sequence. However, it is possible to find these transiting objects by using other combinations of colors and spectral indices. For this purpose, Figure~\ref{fig:hd_d4} shows 2D projections of the 3D parameter space defined by the 4000\,\AA\ break, $D_n(4000)_\mathrm{dc}$, $\mathrm{W_{H\delta}}$, and $(NUV-g)_\mathrm{dc}$. 

$D_n(4000)_\mathrm{dc}$ and $\mathrm{W_{H\delta}}$ are often used to distinguish recent star formation histories dominated by bursts from those that are more continuous \citep{kauffmann03a,martin07,wild07}. $D_n(4000)$, probes the mean temperature of the stars responsible for the continuum and is a good indicator of mean stellar population age \citep{bruzual83,kennicutt98,balogh99}. It is also much less sensitive, but not impervious, to dust effects \citep{macarthur05}. We correct for (possibly small) dust effects on $D_n(4000)$ using the average attenuation in the narrow wavelength range in which it is defined.

The $(NUV-g)_\mathrm{dc}$ color is sensitive to young massive stars and as a result it evolves rapidly in rapidly quenching galaxies. It provides an additional lever arm that can be used to cleanly separate galaxies that are rapidly quenching from the general star-forming population. The fact that $(NUV-g)_\mathrm{dc}$ color is an integrated galaxy-wide photometric quantity also makes it complementary to $D_n(4000)_\mathrm{dc}$ and $\mathrm{W_{H\delta}}$, which are spectroscopic quantities measured within a 3\arcsec\ aperture and therefore may not be representative values of the entire galaxy. We select the fading post-starbursts as galaxies that are outliers from normal galaxies in $(NUV-g)_\mathrm{dc}$ and $\mathrm{W_{H\delta}}$ at a given $D_n(4000)_\mathrm{dc}$.

As shown in Figure~\ref{fig:hd_d4}a, $D_n(4000)_\mathrm{dc}$ and $\mathrm{W_{H\delta}}$ are well-correlated for normal galaxies with smooth SFHs. Galaxies with bursty histories are found off the main relation, as shown by the (orange and navy) curved BC03 model tracks, which represent bursty SFHs. The (magenta) dashed curve across the main sequence denotes the fourth order polynomial fit to the normal data (see Appendix~\ref{sec:appC} for more information). 

Figure~\ref{fig:hd_d4}b plots dust-corrected $(NUV-g)_\mathrm{dc}$ color versus $D_n(4000)_\mathrm{dc}$. In this figure, two clouds of points are visible for normal galaxies, the blue-cloud of young star-forming galaxies to the upper left, and the quenched old and red galaxies to the lower right. The (magenta) dashed curve across the two clouds again denotes the fourth order polynomial fit to the normal data (see Appendix~\ref{sec:appC}). Galaxies with bursty star formation histories deviate off the main relation to the lower left.   

In both Figures~\ref{fig:hd_d4} a\,\&\,b, the starbursts and the quenched post-starbursts are located at the extrema of the burst tracks. SBs are found at the tip of the blue-cloud, with very blue $(NUV-g)_\mathrm{dc}$ color, low $D_n(4000)_\mathrm{dc}$ and relatively high $\mathrm{W_{H\delta}}$. Likewise, QPSBs are also located off the main relation for normal galaxies, with very red $(NUV-g)_\mathrm{dc}$ color, intermediate $D_n(4000)_\mathrm{dc}$ and relatively high $\mathrm{W_{H\delta}}$. The FPSBs are located in the intermediate region between starbursts and quenched post-starbursts in both figures. Hence, both $(NUV-g)_\mathrm{dc}$ and $\mathrm{W_{H\delta}}$ are useful to identify this population.

FPSBs are selected quantitatively as objects that are more than $2\,\sigma$ outliers from normal galaxies in  $(NUV-g)_\mathrm{dc}$ and $\mathrm{W_{H\delta}}$. This is illustrated in Figure ~\ref{fig:hd_d4}c, which depicts the difference in $(NUV-g)_\mathrm{dc}$ color, $\Delta(NUV-g)$, and the difference in $\mathrm{H\delta}$ equivalent width, $\Delta(\mathrm{H\delta})$, from the polynomial fit values at a given $D_n(4000)$. The FPSBs are indicated by the (green) circles. The (purple) ellipse encloses most normal galaxies at the $2\,\sigma$ level. Thus, FPSBs are selected to be well outside the normal galaxy locus (defined by the purple ellipse) with well-measured $\Delta$ quantities ($\Delta(\mathrm{H\delta})/\sigma (\mathrm{H\delta}) > 3$ and $\Delta(NUV-g)/\sigma (NUV-g) > 3,$ the $\sigma$s denoting the measurement errors of $\mathrm{H\delta}$ and $NUV-g$). This method of selecting post-starbursts recovers almost all of the quenched post-starbursts from Figure~\ref{fig:hahd} and identifies many FPSBs ($N \sim 105$). By using $(NUV-g)_\mathrm{dc}$ color as an additional selection criterion, specifically by requiring $\Delta(NUV-g) > 0$, a large number of contaminants ($N \sim 50$) are removed. A significant number of these contaminant galaxies show color gradient (have red centers but blue outer parts) and are (edge-on) disk galaxies.
 
\begin{figure*}
\includegraphics[scale=0.8]{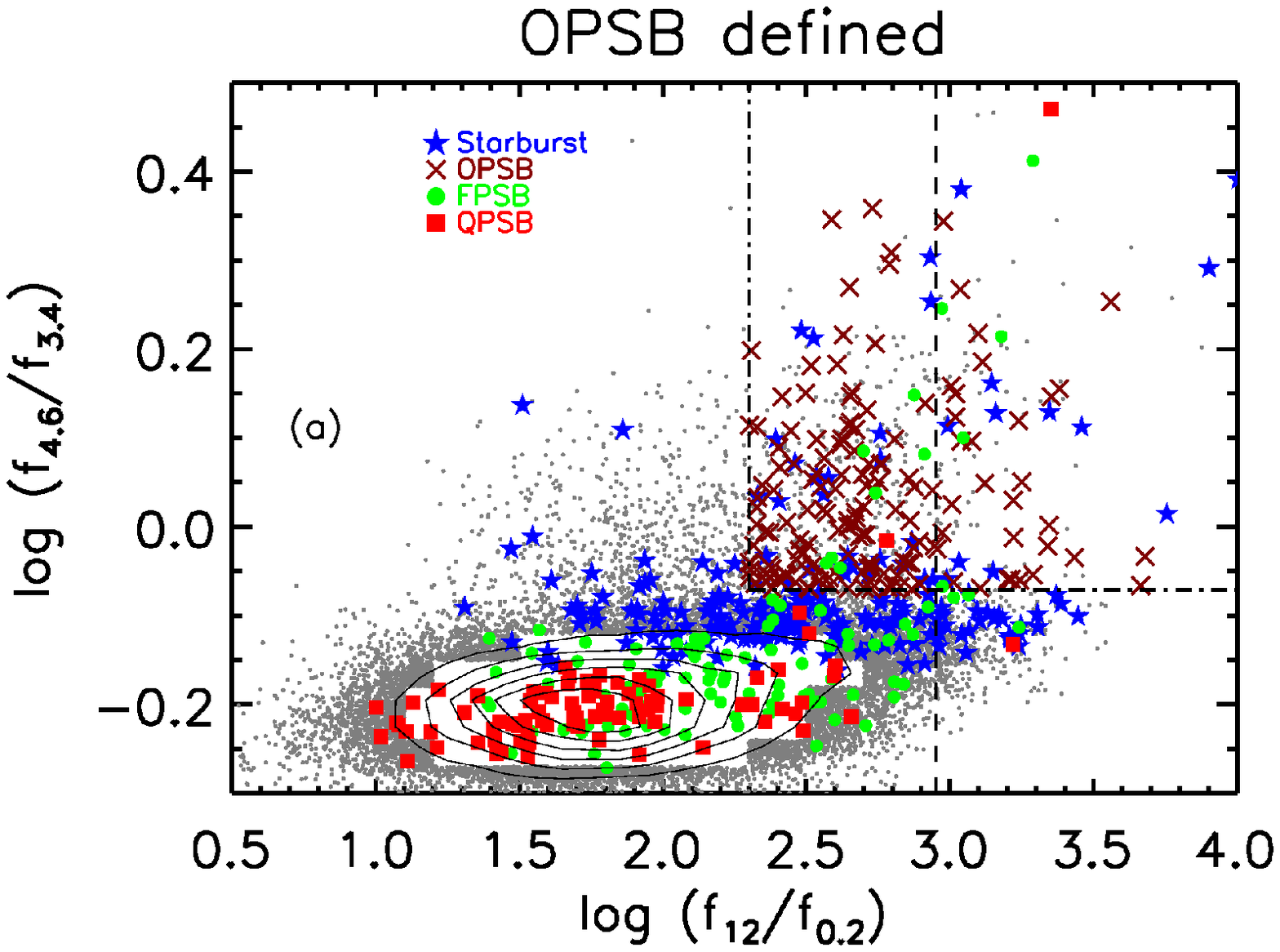}
\includegraphics[scale=0.8]{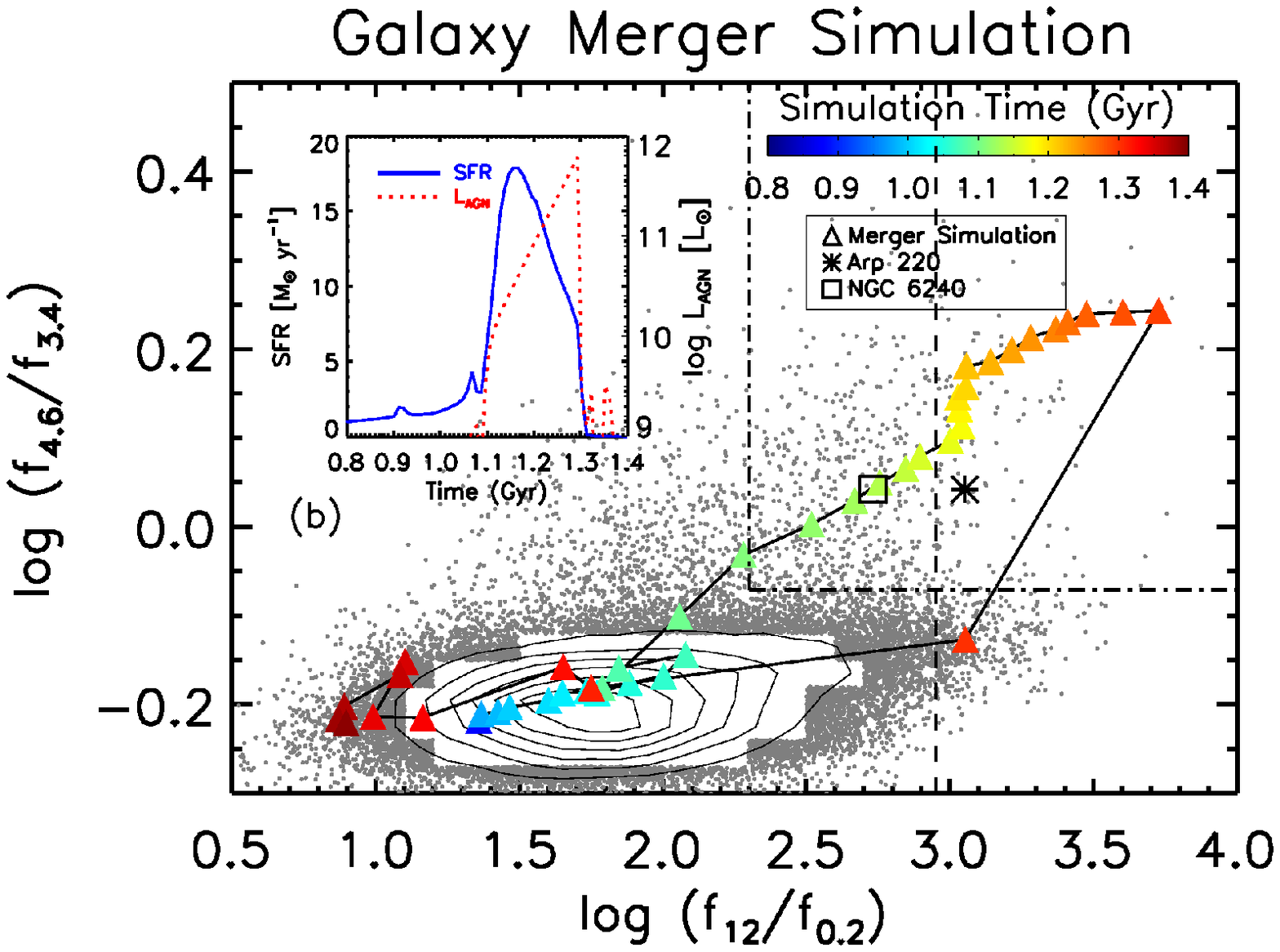}
\caption{Panel (a): The flux density ratio between WISE $12\,\mu \mathrm{m}$ and GALEX NUV, $\mathrm{f_{12\mu m}/f_{0.2\mu m}}$, versus the ratio of WISE $4.6\,\mu \mathrm{m}$ to WISE $3.4\,\mu \mathrm{m}$. The $\mathrm{f_{12\mu m}/f_{0.2\mu m}}$ ratio quantifies the amount of obscured star formation versus unobscured star formation while the $\mathrm{f_{4.6\mu m}/f_{3.4\mu m}}$ ratio quantifies hot dust emission from an AGN or a starburst. We define obscured post-starbursts (OPSBs) as galaxies in the upper right box and with $\mathrm{W_{H\delta} > 4}$\,\AA. Many of the previously-selected starbursts (69\%) and fading post-starbursts (45\%) are as obscured as OPSBs, even though this previous selection may be biased against obscured galaxies. We will later show these obscured post-starbursts galaxies have similar properties as the fading post-starbursts. The dashed vertical line denotes the boundary for local DOGs \citep{hwang13}. Panel (b): The evolutionary path of a simulated merger-induced starburst. The merger model (triangles colored-coded by time since the start of simulation) shows that starbursts pass through the DOG phase which coincides with AGN activity. Two classic dust-obscured AGN, Arp 220 and NGC 6240, are shown for a reference. The inset plot shows that star formation declines before the peak of AGN activity.\label{fig:wise2}}
\end{figure*}

In Figure~\ref{fig:hd_d4}d we plot $(NUV-g)_\mathrm{dc}$ color versus $\mathrm{W_{H\delta}}$, a variant of Figure~\ref{fig:hahd} in which $\mathrm{W_{H\alpha,dc}}$ is replaced by $(NUV-g)_\mathrm{dc}$ color. The overall trends of this figure and Figure~\ref{fig:hahd} are similar. We previously used $\Delta(NUV-g)$ and $\Delta(\mathrm{H\delta})$ in our selection because a starburst will cause these deviations in this diagram. By construction, the FPSBs do not overlap with normal galaxies in this diagram. There is also minimum overlap in Figure~\ref{fig:hd_d4}a, and the overlap in Figure~\ref{fig:hd_d4}b is a projection effect. The selection in 3D space cleanly separates FPSBs because it removes contaminants that are offset from normal galaxy in Figure~\ref{fig:hd_d4}a but not Figure~\ref{fig:hd_d4}d. Perhaps the offset of these contaminants in Figure~\ref{fig:hd_d4}a is due to the fiber effect of SDSS spectra.

The selection of the TPSBs employed so far only identifies objects that are significantly offset from the normal galaxy locus and therefore misses a subset that overlaps
with the normal galaxies (or those whose colors and spectral indices are not well-measured). This is evident from the small gaps between the SBs and the FPSBs in Figure~\ref{fig:hahd} \&~\ref{fig:hd_d4}. The next subsection will describe how some of these missing objects are identified.

\subsubsection{Obscured Post-starbursts (OPSB)}

As discussed in the introduction, we aim to test the merger-driven evolutionary framework for post-starbursts. Theoretically, it is thought that major mergers naturally result in highly dust-obscured galaxies \citep{hopkins06,jonsson06,Chakrabarti08,narayanan10,hayward12}. Since PSBs are believed to be the results of such mergers \citep[][]{hopkins06,hopkins08a,bekki01,bekki05,snyder11}, it is plausible that they exist in dust-obscured phase as they quench \citep{poggianti00,bekki01,shioya01}. Thus, we search for dust-obscured objects in our sample that likely bridge the gap between SBs and FPSBs. These objects have similar spectral indices and near UV colors as normal galaxies and therefore could not be identified in the previous section.

Figure~\ref{fig:wise2} plots the flux density ratio between WISE $12\,\mu \mathrm{m}$ and GALEX NUV, $\mathrm{f_{12\mu m}/f_{0.2\mu m}}$ versus the ratio of WISE $4.6\,\mu \mathrm{m}$ to WISE $3.4\,\mu \mathrm{m}$. The $\mathrm{f_{12\mu m}/f_{0.2\mu m}}$ ratio roughly quantifies the amount of obscured versus unobscured star formation \citep{narayanan10,hwang13}. We consider galaxies with $\mathrm{f_{12\mu m}/f_{0.2\mu m}} > 200$ as significantly dust-obscured \citep[cf.][]{narayanan10}. Local DOGs have $\mathrm{f_{12\mu m}/f_{0.2\mu m} > 892}$ \citep{hwang13}. According to our definition, 69\% of the starbursts, 45\% the FPSBs and the 20\% QPSBs are significantly dust-obscured. Likewise, 20\% of the starbursts and 8\% of the FPSBs are classified as DOGs. In comparison, only about 13\% of galaxies in the parent sample are significantly dust-obscured and only 0.8\% are DOGs. 

The fact that starbursts and post-starbursts selected thus far are significantly more dust-obscured than normal galaxies provides further motivation to select the second class of transiting post-starbursts using Figure~\ref{fig:wise2}. We define the obscured post-starbursts (OPSBs) as galaxies with $\mathrm{W_{H\delta}} > 4$\,\AA, $\mathrm{f_{12\mu m}/f_{0.2\mu m} > 200}$ and $\mathrm{f_{4.6\mu m}/f_{3.4\mu m}}> 0.85$ (the median value for SB is 0.8). Note that 20\% of the OPSBs are DOGs.

As further confirmation of the OPSB selection, Figure~\ref{fig:wise2}b shows how a simulated major merger evolves in the  $\mathrm{f_{4.6\mu m}/f_{3.4\mu m}}$ vs. $\mathrm{f_{12\mu m}/f_{0.2\mu m}}$ plot. The inset in this figure shows the time evolution of the star formation rate and AGN luminosity near the time of coalescence of the galaxies (at $\sim 1.13$ Gyr). As the galaxies coalesce, a strong starburst is induced. Simultaneously, the AGN luminosity increases rapidly as the black hole particles accrete gas. Because most of the gas in the galaxies is consumed or heated (by shocks and AGN feedback) during the starburst, the star formation rate rapidly decreases. The AGN continues to accrete for $\sim 100$ Myr after star formation is terminated because the gas inflow rate needed to sustain the black hole accretion is $\la 0.1 M_{\odot}$ yr$^{-1}$, which is orders of magnitude less than the star formation rate during the starburst. Dynamical effects can also cause a delay between the maxima in the star formation rate and black hole accretion rate \citep{hopkins12}. Note that gas consumption, not AGN feedback, is the dominant cause for the termination of the simulated starburst. The effect of the AGN feedback in the simulation is to further reduce the post-starburst star formation rate and expel the remaining gas and dust in the nuclear region \citep{hayward13,snyder11}.

\section{AGN and Their Connection to Post-Starbursts}
\label{sec:agn}

Having identified plausible candidate galaxies on the evolutionary pathway from starburst to quenched post-starbursts, we now explore the possible connection between AGN activity and quenching in these objects. The tight correlation between masses of galactic center super-massive black holes (SMBH) and properties of host galaxy bulges \citep[e.g.,][]{magorrian98,ferrarese00,tremaine02} imply that galaxy evolution and SMBH accretion occur in a long history of coupled growth and regulation (but see \citet{kormendy13} for a contrarian perspective on co-evolution). Many semi-analytical models and theoretical simulations require AGN feedback to quench star formation and correctly predict the observed color bi-modality of galaxies and the shape of the galaxy luminosity function \citep[e.g.,][]{kauffmann00,croton06,hopkins06,somerville08,gabor11}.

The rapid quenching of post-starburst galaxies makes them the ideal test-bed for AGN feedback models \citep[e.g.,][]{hopkins06,snyder11,cen12}. With our samples spanning the whole post-starburst evolutionary path, we quantify the fraction of AGN hosts among post-starbursts and their properties (stellar population age, AGN strength, dust properties, etc). These quantities may help us infer whether AGN are primarily responsible for quenching starbursts or not.  

\subsection{Optical AGN diagnostics}

\begin{figure*}
\centering
\mbox{\subfigure{\includegraphics[width=3.5in]{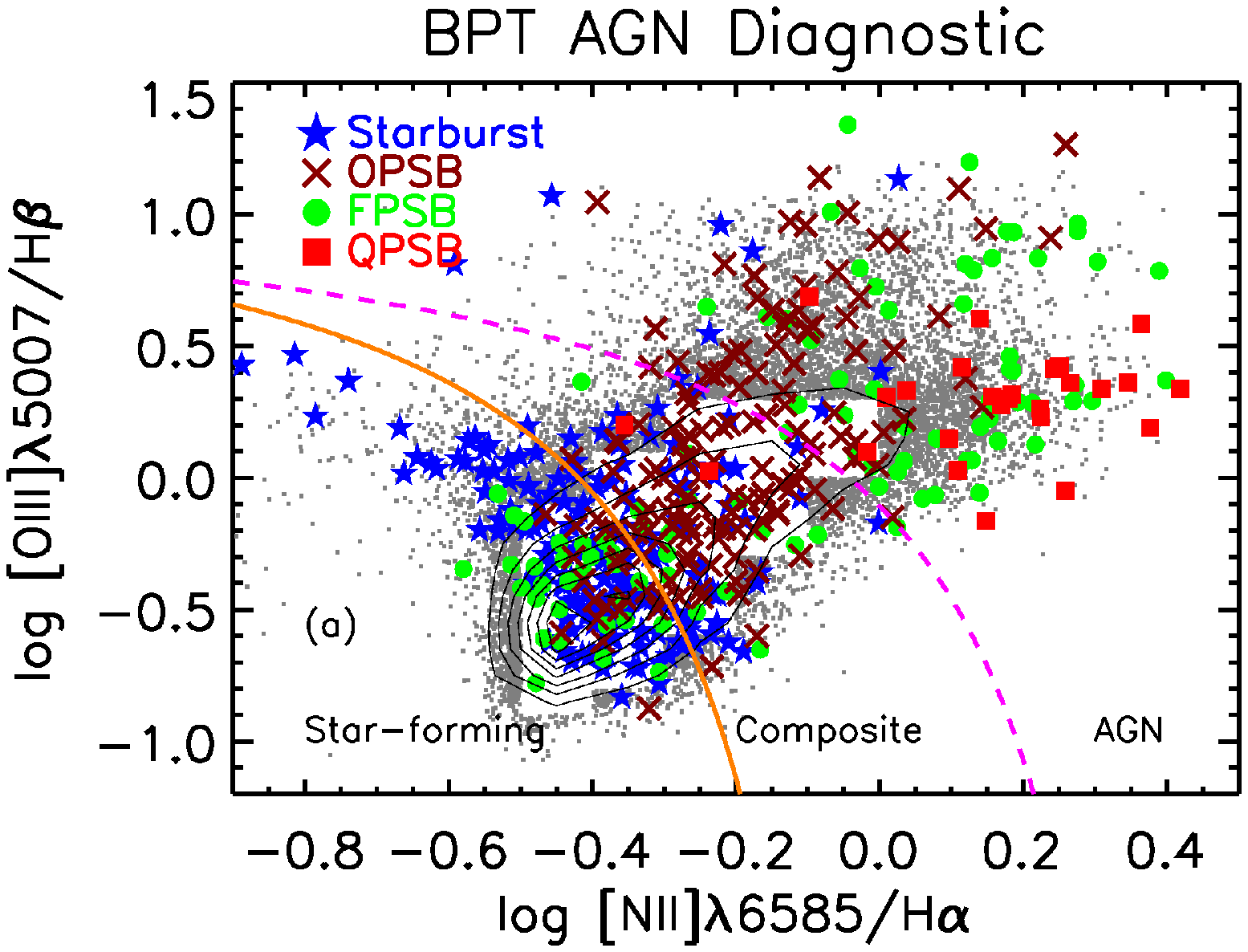}}
\subfigure{\includegraphics[width=3.5in]{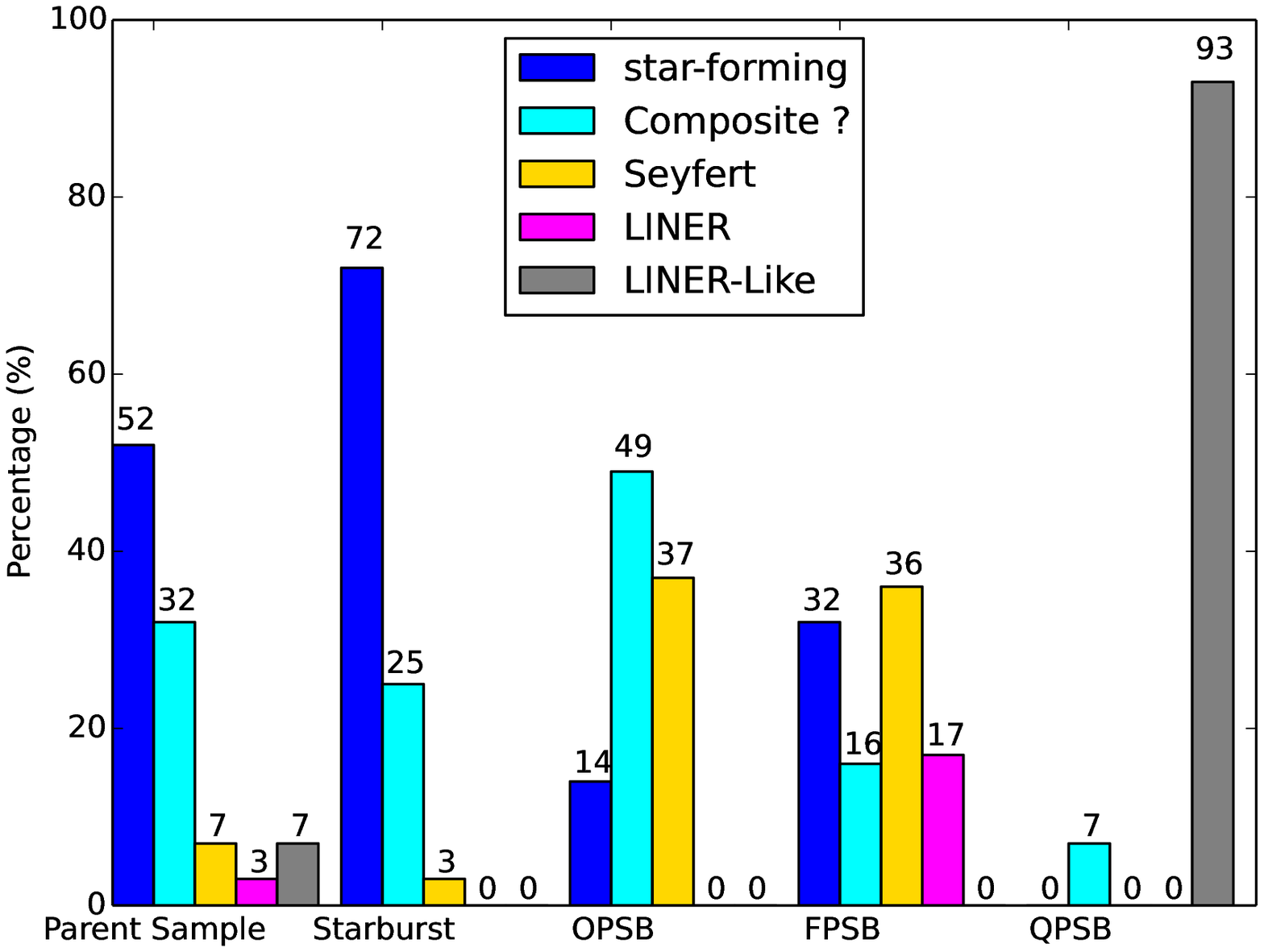}}}
\caption{Panel (a) shows the BPT emission-line ratio AGN diagnostic for the parent sample, starbursts and post-starbursts whose emission lines are detected with $\mathrm{SNR} >3$. The (magenta) dashed curve denotes the theoretical boundary for extreme starbursts \citep{kewley01} while the solid (orange) curve denotes the empirical boundary of pure star-forming galaxies \citep{kauffmann03c}. The diagram shows that the quenched post-starbursts have LINER-like emission while transiting post-starbursts have both star formation and AGN-dominated emission line ratios. The latter smoothly bridge the starbursts and the quenched post-starbursts. Panel (b) shows the percentage of AGN in starbursts, post-starbursts and galaxies in the parent sample. LINERs are objects in AGN region of the BPT diagram with $\mathrm{3\,\AA\; \le W_{H\alpha} \le 6\,\AA}$ while LINER-like objects are the corresponding objects with $\mathrm{W_{H\alpha} < 3\,\AA}$.  \label{fig:agn}}
\end{figure*}

In Figure~\ref{fig:agn}a, we show the BPT diagnostic using the \OIIIHB\;and \NIIHA\;line ratios \citep[][]{BPT,veilleux87}. The position of an object in this diagram depends on its nebular metallicity and the hardness of its radiation field. Thus, the BPT diagram distinguishes between emission lines from \HII\, regions and AGN. AGN-dominated galaxies have larger \OIIIHB\;and \NIIHA\;ratios and occupy the upper right of the diagram, while the softer ionization of \HII\;regions means star-forming galaxies occupy the lower left.

The dashed (magenta) curve demarcates the theoretical boundary for extreme starbursts, and galaxies above this curve probably host AGN \citep{kewley01}. The solid (orange) curve demarcates the empirical lower boundary for AGN \citep{kauffmann03c}. Objects below this curve are likely ``pure'' star-forming galaxies. Galaxies between the boundaries of extreme starbursts and ``pure'' star formation are thought to be mostly composites of star formation and AGN, although some have argued that unusual ionization in \HII\;regions can lead to starbursts without AGN lying in the composite region \citep[e.g.,][]{brinchmann08}. Similarly, galaxies in the AGN region may also have some star formation contribution, but their ionization state is dominated by the AGN. 

The starburst galaxies are distributed over the star-forming and composite regions (25\%) of the diagram and only 3\% are AGN. On the other hand, almost all (93\%) of the quenched post-starburst galaxies with well-measured emission lines lie in the AGN region of the BPT diagram \citep[cf.][]{yan06}. This might indicate weak AGN, although there is some evidence that photo-ionization in weak emission-line galaxies such as QPSBs can also be produced by shocks or post-asymptotic giant branch stars \citep{ho08,cid11,yan12,singh13}. For instance, \citet{cid11} have found that the ionization in galaxies with (dust-extincted) $\mathrm{W_{H\alpha}} < 3$\,\AA\ can be sufficiently accounted for by ionization from hot evolved stars without invoking AGN. The authors classified  AGN into Seyferts or Low Ionization Narrow Emission Region (LINER) galaxies if they have $\log(\NIIHA) > -0.4$ and $\mathrm{W_{H\alpha}\;> 6\,\AA}$ or $\log(\NIIHA) > -0.4$ and $\mathrm{3\,\AA\; \le W_{H\alpha} \le 6\,\AA}$ respectively. QPSBs are defined as objects with $\mathrm{W_{H\alpha,dc}}< 3\,\AA$ and accordingly they are not LINERs, but they are LINER-like (objects above the starburst boundary of \citep{kewley01} and with $\mathrm{W_{H\alpha,dc}}< 3\,\AA$).

The OPSBs and FPSBs bridge the starbursts and QPSBs. This is consistent with our evolutionary path from starburst to transiting to quenched post-starburst galaxies, with star formation decreasing along the sequence as AGN emerge. 53\% of the FPSBs and 37\% OPBSs are AGN while about 16\% FPSBs and 49\% of OPSBs are composite. Therefore, about 36\% and 35\% of transiting PSBs are AGN and composites respectively. In comparison, only 10\%  and 32\% of normal galaxies in the parent sample with $\mathrm{W_{H\alpha,dc}}> 3$\,\AA\ are AGN and composites respectively. 

Figure~\ref{fig:agn}b presents the AGN fraction in transiting post-starbursts using a bar chart. It subdivides the (BPT) AGN into Seyferts and LINERs if they have $\mathrm{W_{H\alpha,dc}\;> 6\,\AA}$ or $\mathrm{3\,\AA\; \le W_{H\alpha,dc} \le 6\,\AA}$ respectively. Seyferts are about 5 times more common in transiting post-starburst galaxies than in normal galaxies in our chosen mass range. LINER-like objects are shown in the figure for completeness, but our estimate of AGN fraction in PSBs does not include such objects.

\subsection{A delay between AGN and starbursts}

In this subsection, we quantify the time delay between the starburst and AGN phase.

\begin{figure*}
\includegraphics[width=3.5in,height=3.5in]{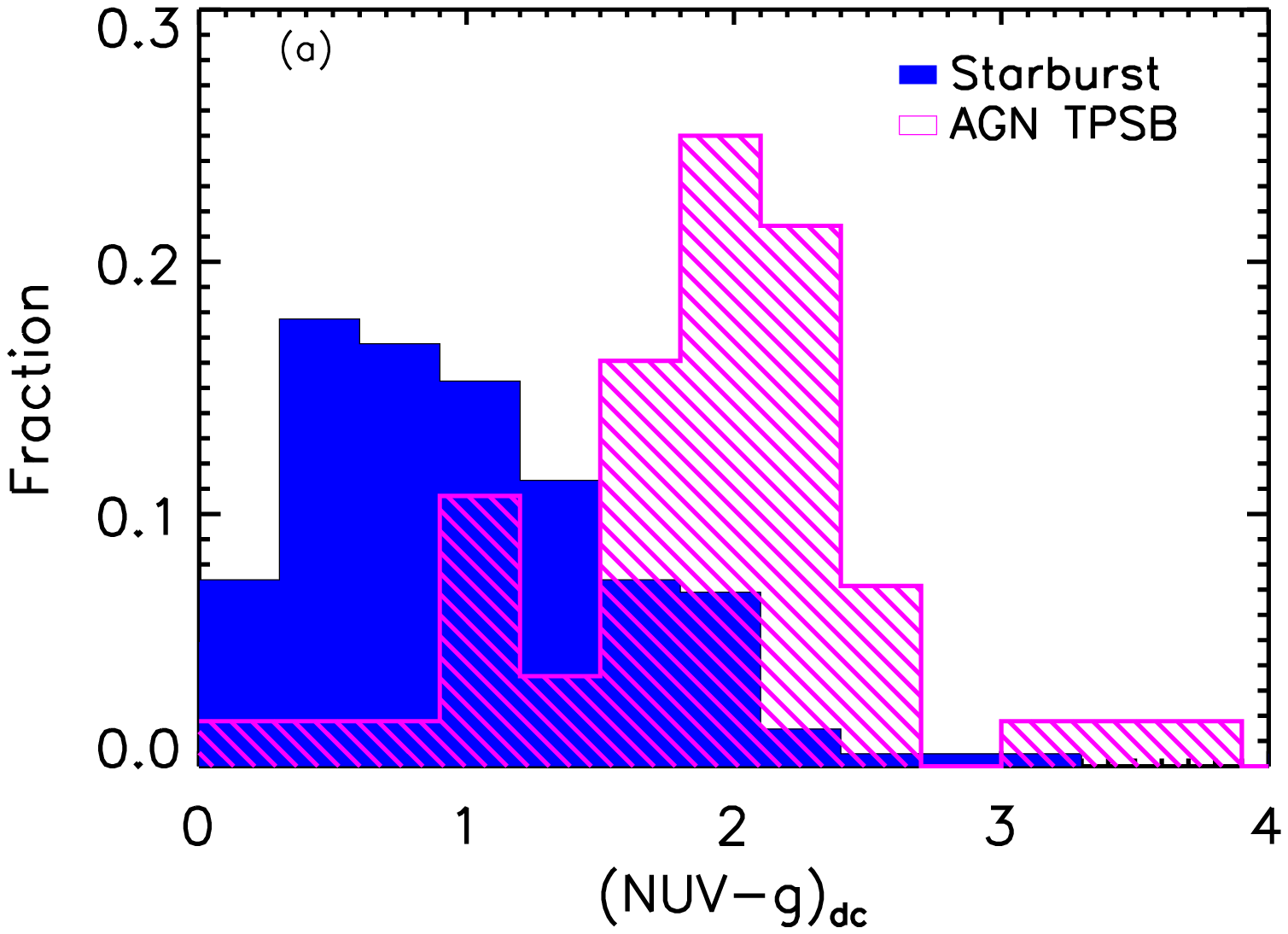}
\includegraphics[width=3.5in,height=3.5in]{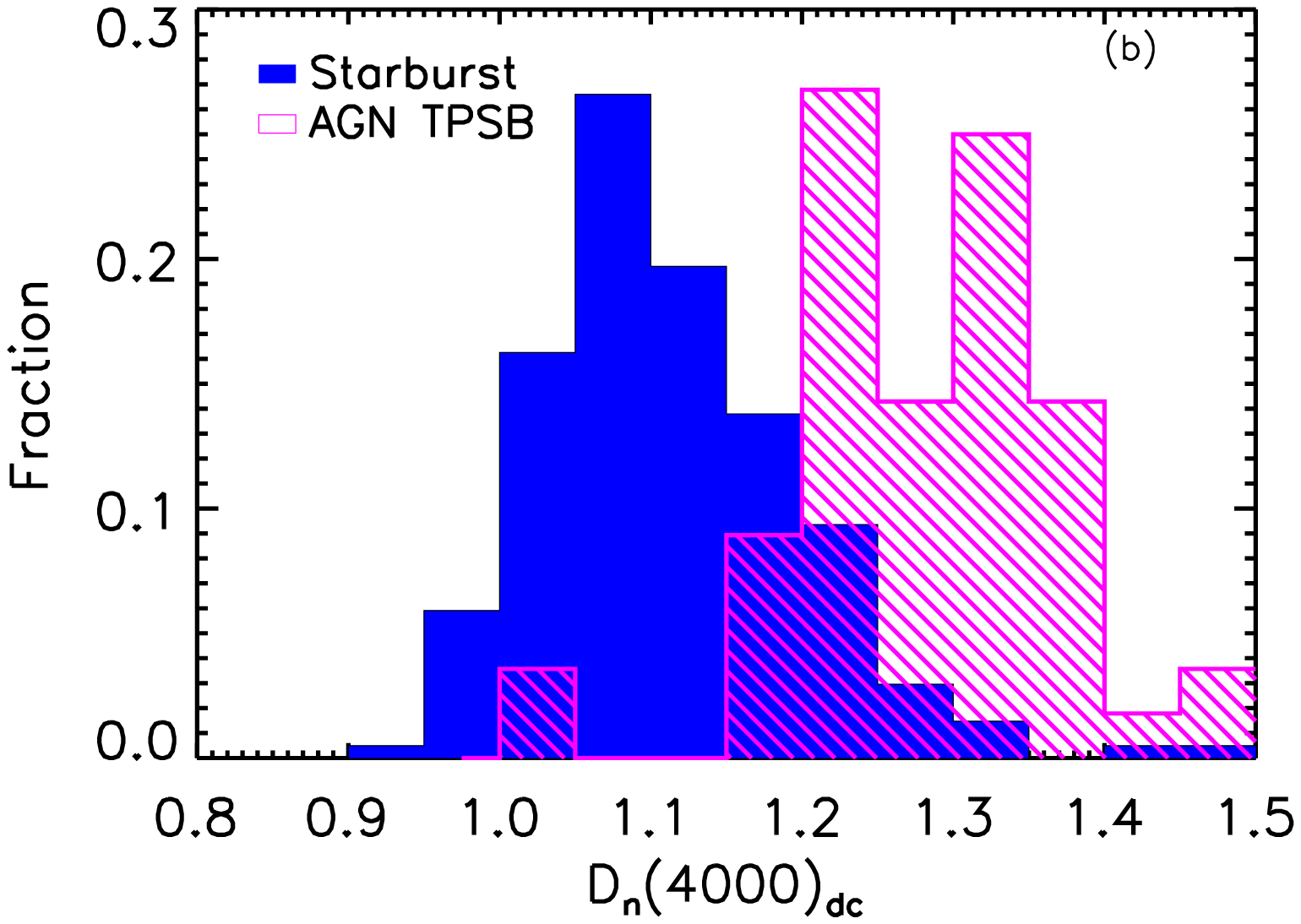}
\caption{The distributions of $(NUV-g)_\mathrm{dc}$ color and $D_n(4000)_\mathrm{dc}$ for starburst galaxies and transiting PSB with AGN. The offset between the peaks indicates that the two population are not coeval, with peak AGN activity appearing considerably after the peak star formation activity.\label{fig:age}}
\end{figure*}

Figure~\ref{fig:age} shows the distribution of $(NUV-g)_\mathrm{dc}$ color and $D_n(4000)_\mathrm{dc}$ (i.e, observable proxy for age) of starbursts and AGN in transiting post-starbursts.
The $(NUV-g)_\mathrm{dc}$ color and $D_n(4000)_\mathrm{dc}$ of TPSBs are significantly offset to higher values (older age) compare to values of starburst. The Kolmogorov-Smirnov test (K-S test) indicates that the null hypothesis that the $(NUV-g)_\mathrm{dc}$ color and $D_n(4000)_\mathrm{dc}$ of starbursts and TPSBs come from the same distribution (i.e, the two population are coeval) can be rejected at $\alpha < 0.001$ significance level.  

\begin{figure*}
\includegraphics{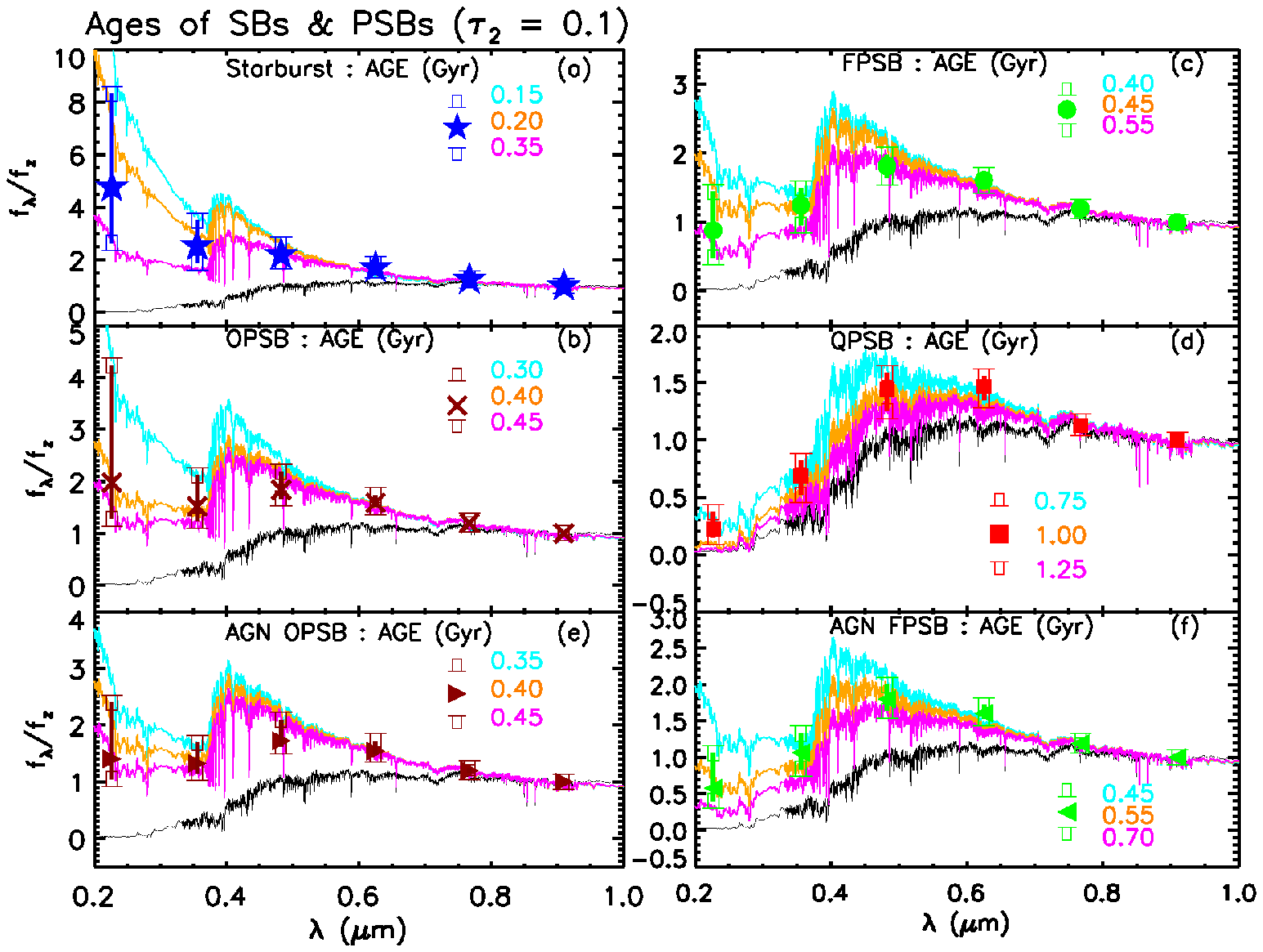}
\caption{The $z$ band-normalized median and quartile fluxes at the effective wavelengths of the $NUV,u,g,r,i,z$ bands (the flux ratios are dust-corrected). The (cyan, orange and magenta) overplotted spectra are \cite{bc03} burst models with SFR timescale $\tau_2=0.1$\,Gyr and burst fraction $b_f=20\%$ of different ages, as indicated on each panel. The lowest (black) spectrum in each panel is that of a 12.5 Gyr old galaxy (before the burst). The model spectra are not actual fits to the data but are chosen to be approximately consistent with data. Galaxies follow an age sequence from starbursts (panel a) to obscured PSBs (panel b) to fading PSBs (panel c) to quiescent PSBs (panel d). It is also notable that the SEDs of transiting PSBs hosting AGN (panels e and f) are significantly older than the starbursts, indicating a $\sim$200~Myr delay between a starburst and the appearance of an AGN.  This indicates that the AGN is not the primary source of quenching in starbursts.\label{fig:sed}}
\end{figure*}

Furthermore, Figure~\ref{fig:sed} shows the $z$ band-normalized median and quartile SEDs of galaxies evolving from the starburst to quenched post-starburst phase. We overplot BC03 models with SFR timescale, $\tau_2$, of 100\,Myr and burst mass-fraction $b_f$ of 20\%  at different ages in order to indicate the time after the second burst. This ballpark estimate shows that the median age of OPSBs is about 400-500\,Myr and there is $\gtrsim 200$ \,Myr gap between the median age of starbursts and and the AGN hosts among TPSB. Because of the burst mass-age degeneracy, the ages of the post-starbursts depends on the decay timescale ($\tau_2$) assumed. As shown in Appendix~\ref{sec:appB}, models tracks with $\tau_2 = 0.05-0.2$ can describe the starburst to post-starburst evolution while models with $\tau_2$ outside this range are excluded since they would not produce the observed population of post-starburst galaxies \citep[cf.][]{wild10}. Therefore, in agreement with the findings of several recent observational works \citep[e.g,][]{davis07,bennert08,schawinski09,wild10}, the time delay might range between $100-400$\,Myr depending on the assumed $\tau_2$.

The significance of this time delay is that it strongly suggests that AGN do not directly quench starbursts. Recent theoretical works are converging to a view that, in merger-fueled post-starburst evolution, AGN may play a secondary or limited role in quenching \citep{croton06,wild09,snyder11,cen12,hayward13}. In other words, a post-starburst results from exhaustion of a bulk of its gas supply in a starburst and/or from its expulsion by stellar feedback; AGN feedback mainly reheats or ejects the remaining gas that would otherwise fuel low-level star formation over the next few billion years.

In particular, \cite{cen12} proposed a new evolutionary model of galaxies and their SMBH. In this model, starbursts and AGN are not coeval and AGN do not quench starbursts. They argued that the main SMBH growth occurs in the post-starburst phase, fueled by recycled gas \citep[cf.][]{scoville88,ciotti07,wild10,hopkins12} from aging stars in a self-regulated fashion on a timescale that is substantially longer than 100 Myr. Our analysis supports the \cite{cen12} model in that AGN are more frequent in post-starbursts and they appear significantly delayed from the starbursts phase. But as we will show later, we do not find observational support for the model's prediction that a substantial ($\times 10$) black hole growth occurs in the post-starburst phase compared to the starburst phase.

\subsection{Dust properties of AGN host}

In Figure~\ref{fig:wise2}, we showed that more than two-thirds of starbursts and more than a third of FPSBs are significantly more dust-obscured compared to normal star-forming galaxies. We also identified heavily dust-obscured PSBs that precede the FPSBs. Therefore, our finding that quenched post-starbursts were once heavily dust-obscured, and that some dust-obscured AGN are likely post-starbursts, is consistent with the later removal of obscuring gas and dust by AGN feedback. However, beyond this consistency, there is no clear observational evidence yet that AGN clear away the remaining gas and dust in post-starburst galaxies \citep[e.g.,][]{tremonti07,coil11}. Therefore, future study of post-starburst with strong AGN identified in this work, may provide further clues on the (secondary) role of AGN and its relationship with its host galaxy.

\begin{figure}
\includegraphics[width=3.5in,height=3.5in]{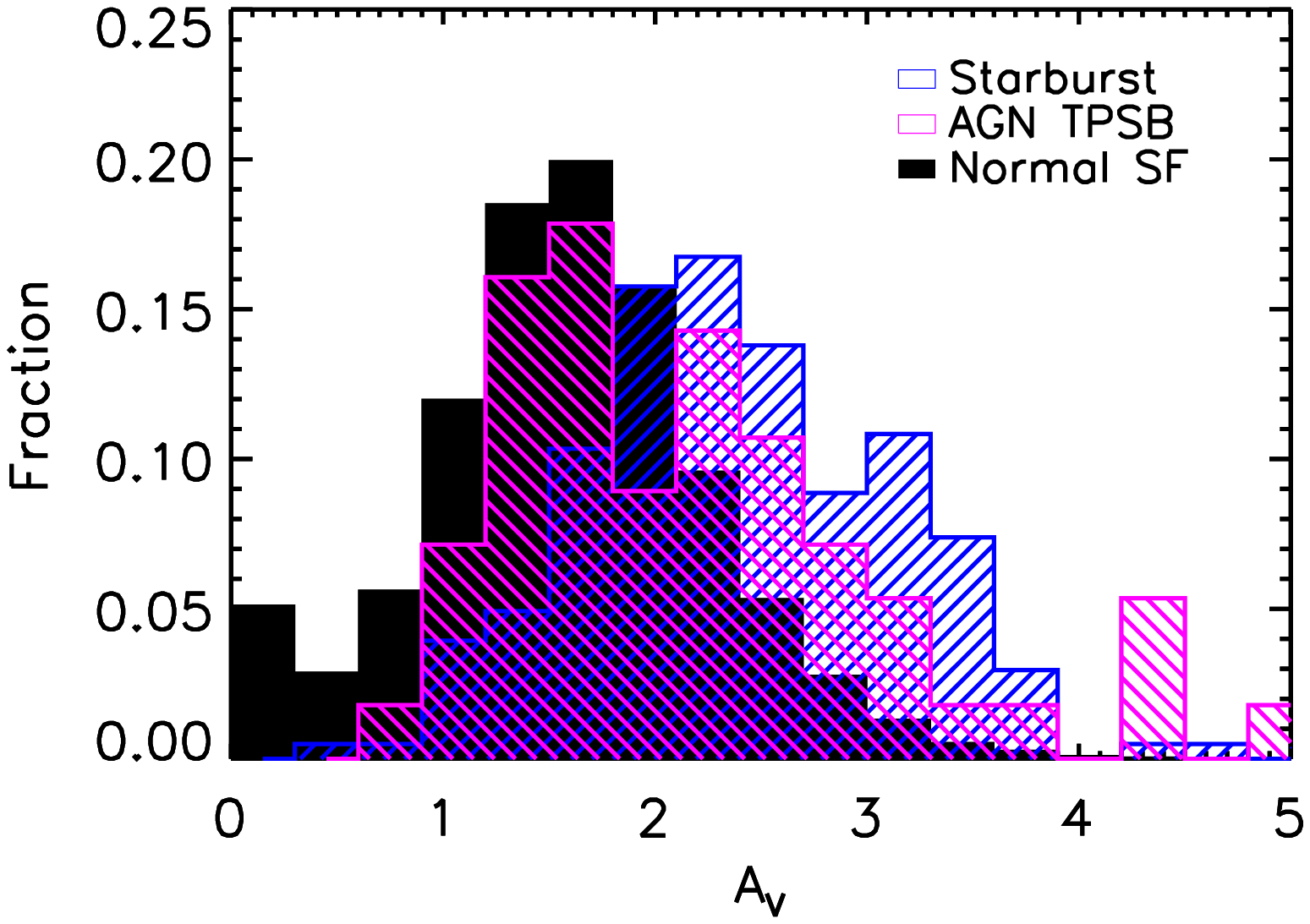}
\caption{The distribution of V-band nebular attenuation $A_V$ for normal star-forming galaxies, starbursts and AGN in transiting post-starbursts.\label{fig:av_dist}. The continuum attenuation are approximately half the nebular attenuation. The AGN hosted by TPSBs are significantly less dusty than the starbursts, consistent with the removal of dust by AGN feedback.}
\end{figure}

Figure~\ref{fig:av_dist} shows the distribution of V-band nebular attenuation $\mathrm{A_V}$ for normal star-forming galaxies, starbursts and transiting post-starbursts. SBs and AGN in TPSB have higher dust attenuation ($\mathrm{A_V}= 2.7 \pm 3.4$ and $\mathrm{A_V}=2.2 \pm 0.9$ respectively) than normal star-forming galaxies ($\mathrm{A_V}=1.6\pm 0.7)$. K-S test indicates that the null hypothesis that the $\mathrm{A_V}$ distribution of SBs or AGN TPSBs come from that of normal star-forming galaxies can be rejected at $\alpha < 0.001$ significance while $\mathrm{A_V}$ distribution of SBs and AGN TPSBs are similar only at $\alpha \lesssim 0.05$. This observation is consistent with a removal of dust by AGN feedback.

So far we have shown: 1) Starbursts and post-starbursts are likely more dust-obscured than normal star-forming galaxies. The starburst to quenched post-starbursts evolutionary sequence is a decreasing dust sequence. 2) AGN are about three times more common in transiting post-starbursts than in normal galaxies. However, we found, similar to previous works, that there a significant time delay between starburst and the peak of AGN activity in both obscured and fading post-starbursts.

\subsection{Broad-Line AGN (BLAGN)}

Special techniques are often required to disentangle AGN and galaxy emission in BLAGN host galaxies. \citet{trump13} have recently used SDSS aperture photometry and $z$ band concentration index to disentangle the light of broad-line AGN and their host galaxies. By doing so, they have assembled a large sample of BLAGN with host galaxy colors and stellar mass measurements. 

The selection criteria of post-starbursts discussed in previous subsections will not identify post-starbursts galaxies hosting BLAGN because their NUV fluxes and spectral indices are rendered immeasurable by the bright AGN. Nevertheless, to constrain how BLAGN fit in our starburst sequence, we select a subset of broad-line AGN from \citet{trump13} that have similar stellar mass and redshift range as the parent sample. The properties of these objects are discussed in Appendix~\ref{sec:appD} .

\section{The Bulge properties of post-starbursts and its necessity for quenching}

The overall aim of this section is to provide a complimentary check on our sample selection by showing that the starbursts and post-starbursts are both bulge-dominated, unlike most normal star-forming galaxies. We show that their morphology is consistent with that of galaxies transitioning between blue and red galaxies. In a future paper we will present other structural parameters that better discriminate between post-starbursts and the slowly quenching normal galaxies.
 
\begin{figure*}
\centering
\mbox{\subfigure{\includegraphics[width=3.5in]{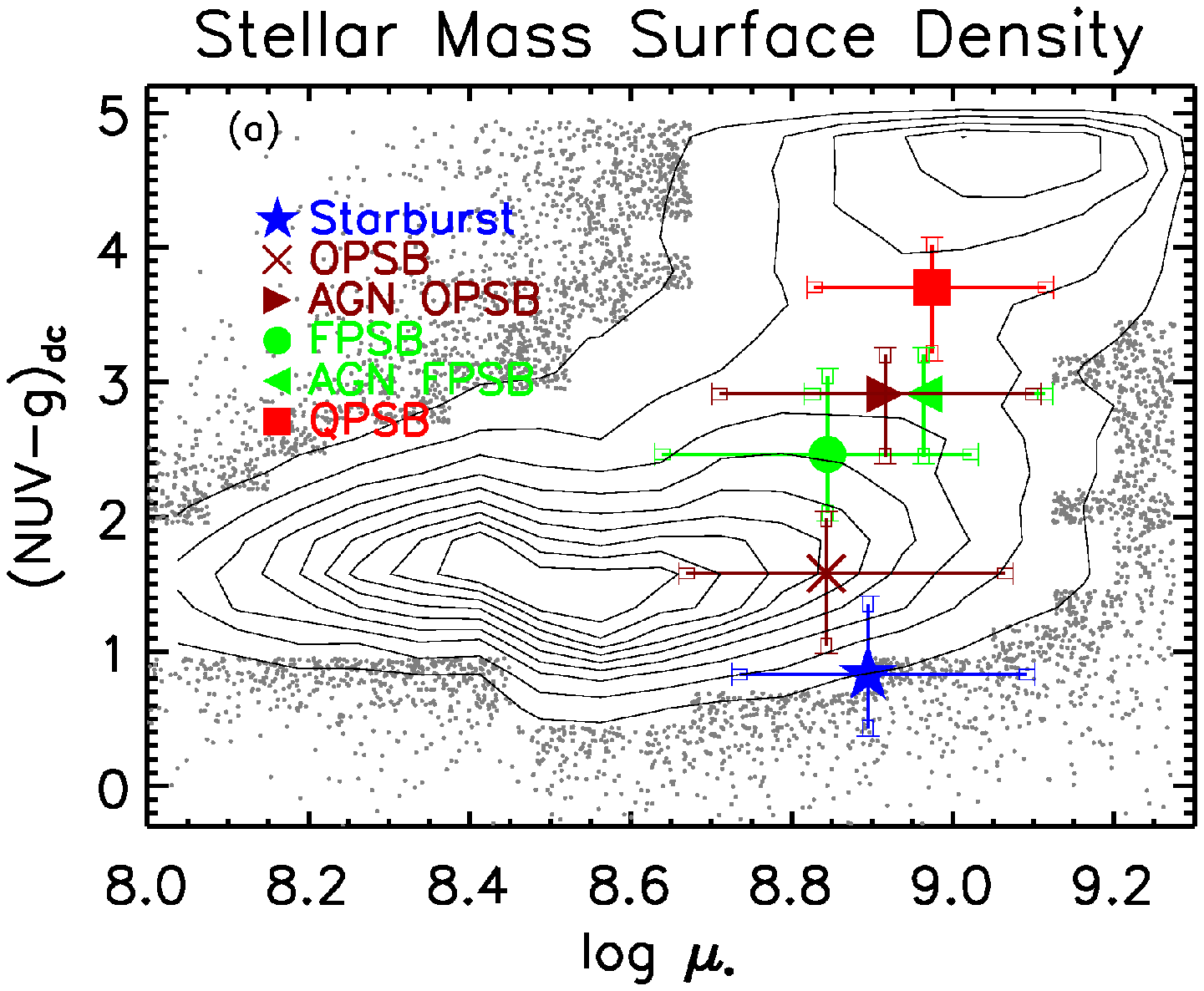}}
\subfigure{\includegraphics[width=3.5in]{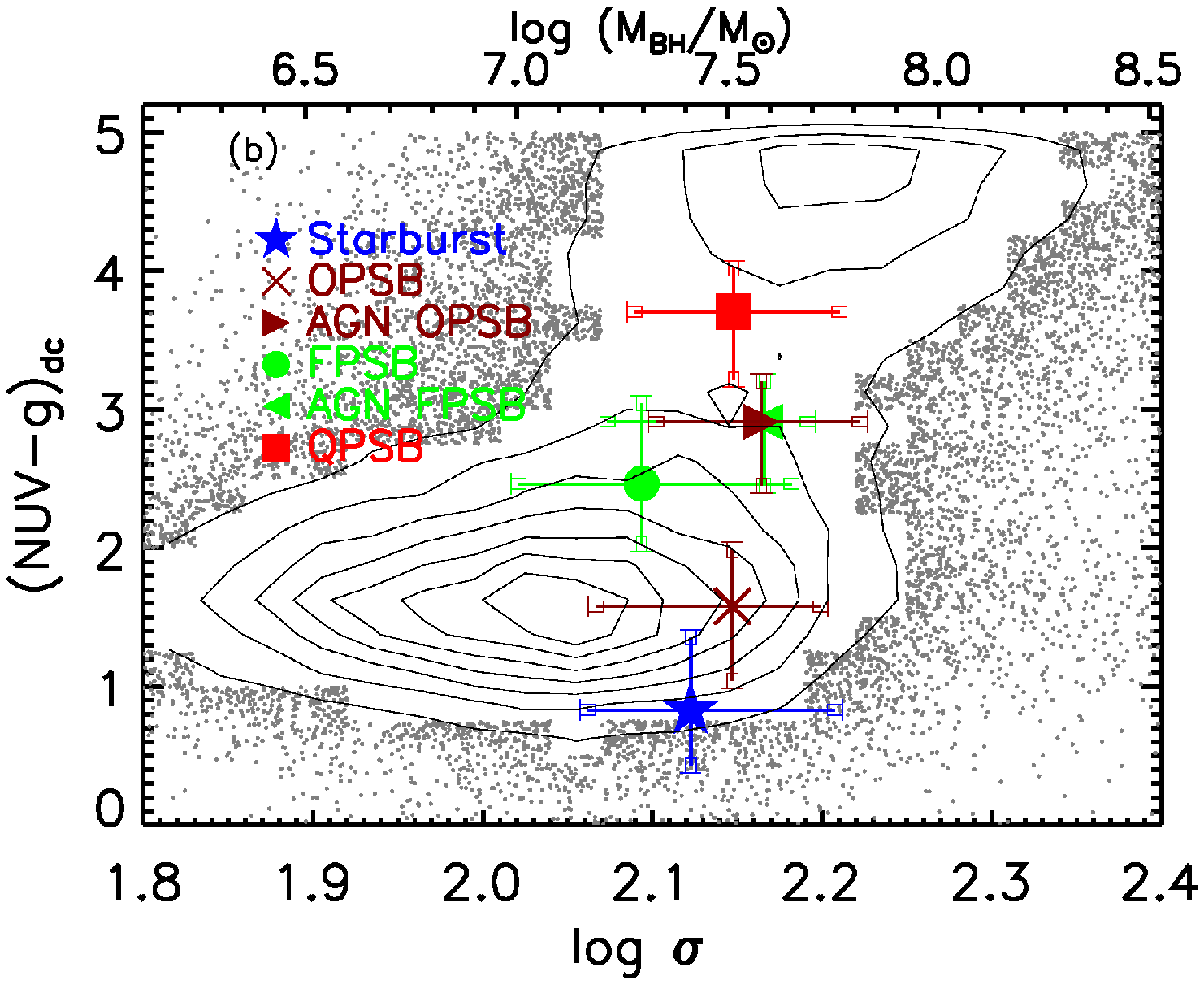} }}

\caption{The median and upper/lower quartile values of $(NUV-g)_\mathrm{dc}$ color versus stellar mass surface density in Panel (a) and $(NUV-g)_\mathrm{dc}$ color versus velocity dispersion in panel (b) are plotted. Starbursts and post-starbursts have similar morphology and they are offset from normal star-forming galaxies in mass surface density and velocity dispersion (i.e, have more prominent bulges).  If we assume that velocity dispersion correlates with black hole mass following the $M_{BH}-\sigma$ relation from \citet{tremaine02}, then there is little black hole growth from SBs to QPSBs (in contrast with the prediction of \citet{cen12}).\label{fig:sprm}}
\end{figure*}

Figure~\ref{fig:sprm}a shows the relationship between the stellar mass surface density, $\mu_\ast$, and the dust-corrected $(NUV-g)_\mathrm{dc}$ color. NUV-optical color and $\mu_\ast$ are known to trace gas consumption time and the change in SFH that takes place as galaxies transition from disc-dominated to bulge-dominated systems \citep{kauffmann06,catinella10}. Comparisons between stellar surface mass density and bulge-to-total ratio by Wild et al.(in preparation) shows that galaxies with $\mu_\ast >  3.0 \times 10^8$\,\mkpc\ are classical bulge-dominated galaxies while ones with $1.0 \times 10^8 $\,\mkpc\ $< \mu_\ast < 3.0 \times 10^8$\,\mkpc\ are pseudo-bulges. About 67\% (95\%) of starbursts and 65\% (91\%) of post-starburst have $\mu_\ast > 6 \times 10^8\,(3 \times 10^8$)\,\mkpc. In comparison, only about 30\% (68\%) of normal star-forming galaxies have $\mu_\ast > 6 \times 10^8\, (3 \times 10^8$)\mkpc. K-S test also indicates that the distribution $\mu_\ast$ for starbursts and post-starbursts are significantly different from normal star-forming galaxies (they are drawn from same distribution at $\alpha \lesssim 0.001$). The compactness of large majority of starbursts and post-starbursts is consistent with the necessity of bulge build-up for quenching \citep{wuyts11,bell12,cheung12,fang13,mendel13}.
   
Similarly, Figure~\ref{fig:sprm}b shows $(NUV-g)_\mathrm{dc}$ color as a function of the velocity dispersion, $\sigma$. The velocity dispersion is the best correlated parameter with galaxy color and star formation history \citep{wake12,fang13}. The $M_\mathrm{BH}-\sigma$ relation \citep{magorrian98} also means that velocity dispersion is a tracer of black hole mass: the upper $x$-axis in Figure~\ref{fig:sprm}b shows the inferred black hole mass using the \citet{tremaine02} relation.

The general galaxy population forms the blue cloud at lower $\sigma$ (median $\mathrm{\sigma=108\,km\,s^{-1}}$) and the red-sequence at higher $\sigma$ (median $\mathrm{\sigma=160\,km\,s^{-1}}$). As expected for quenching/recently quenched galaxies, the starbursts and the three post-starbursts classes are located in the transition region between the blue cloud and the red sequence, at intermediate velocity dispersion ($\sigma \sim 125-140\,\mathrm{km\,s^{-1}})$. 

The SB to QPSB sequence is offset as a whole from the normal SFR galaxies by about a factor of two in black hole mass. However, from starburst to transiting to quenched post-starbursts, there is little or no black hole growth along the evolutionary sequence. This observation does not not support the prediction of substantial ($\times 10$) black hole growth in the post-starburst phase compared to the starburst phase \citep{cen12}.

In summary, we have shown that the three post-starburst classes are bulge-dominated unlike most normal star-forming galaxies. The fact both SBs and PSBs have similar morphology is independent evidence that these two populations are linked. Similarly, the fact FPSBs and OPSBs have structural properties that are fully consistent with each other supports that they are objects in the same category despite their different selection criteria.

\section{Discussion and Conclusion}

\subsection{Building the red-sequence through post-starbursts}

The quenching process happens in both slow and fast-mode \citep[e.g.,][]{cheung12,barro13,dekel13,fang13}. We attempt to constrain the transit time and the fraction of galaxies evolving through the two modes of quenching using simple crude estimates. Assuming that the starbursts are triggered by mergers or by some other phenomenon that has a redshift dependence and using our thorough and fairly complete post-starbursts sample today, one can constrain how many of each kind of product evolved to the red sequence through the two quenching modes in the past 10\,Gyr.

The number of galaxies in the parent sample is $\sim 67,000$ of which $\sim 40,400$ galaxies are located in the blue cloud, $\sim 14,700$ galaxies are in the red-sequence and $\sim 11,900$ galaxies are in the green valley (see Figure~\ref{fig:colormass}b). If half of galaxies currently on the red sequence had under gone a dry major merger since $z \sim 2$ \citep[][]{bell06,hopkins10b}, accounting for galaxies that might have evolved out of the parent sample, a total of $\lesssim 22,000$ red-sequence galaxies must have been in the parent sample since $z \sim 2$. 

On the other hand, from Figure~\ref{fig:hahd} the total number of post-starburst galaxies in the parent sample is 341. If we take the difference between the median age of SB and QPSB to be the quenching time or the transit time to the red-sequence, this transit time is $\sim 600$\,Myr for star formation timescale of $\tau_2=0.1$\,Gyr (as shown Appendix~\ref{sec:appB}, the transit time may range between 400-900\,Myr). This suggests that about 570 galaxies per Gyr are currently moving to the red-sequence through the post-starbursts path at the constant mass.

Theoretical models argue that post-starbursts are the end-products of galaxy mergers \citep[][]{hopkins06,hopkins08a,bekki01,bekki05,snyder11}. Assuming a uniform merger (production) rate since $z \sim 2$ (last 10 Gyr), then the total production of post-starbursts in our adopted mass range would be about 5700 galaxies. This is about 40\% of the galaxies on the red-sequence in the parent sample today. The major-merger rate however is thought to increase with redshift roughly as $\propto (1 + z)^{2-3}$ \citep{Kartaltepe07,hopkins10b,lotz11}. In this case, the transit rate through post-starbursts integrated to $z\sim 2$ gives $3-6$ times more post-starbursts than the estimate that assumes a uniform merger rate. Therefore, integrated over time post-starbursts are an important pathway to the red-sequence. They can account for at least a quarter, and up to essentially all of the red-sequence galaxies that are (were) in the parent sample. 

At high redshift, disk instability-induced starbursts may be more common than merger-induced starbursts \citep[][]{bournaud08,dekel09}. Our estimate of post-starburst fraction above does not include post-starbursts that might have resulted from this mechanism. In addition, we also have not accounted for post-starbursts that host broad-line AGN (which we do not have a way of identifying). For these and other reasons, the total contribution of the post-starburst path over time to the build-up of the red sequence is certainly above 25\%. Similarly, \citet{wild09} found that about 40\% of the mass growth of the red sequence at $z \sim 1$ is likely due to galaxies passing through the post-starbursts phase while \citet{barro13} found that almost all quiescent galaxies at $z\gtrsim 2$ are descendants of rapidly quenching compact star-forming galaxies. 

If we conservatively assume that $\gtrsim 25\% $ of the red-sequence galaxies in the parent sample (over the past 10 Gyr) descended from post-starbursts, we can constrain the transit time across green valley for slowly quenching galaxies. Excluding the $\sim 5700$ galaxies that might have descended from PSBs, about 16,300 out of the total of $\lesssim 22,000$ red-sequence galaxies must have gone through the slow mode of quenching over the past 10 billion years. Assuming a constant transit time across the green valley \citep{faber07}, the fact that we currently observe $\sim 11,900$ slowly fading normal galaxies in green valley implies that the transit time through GV for the slow track is $\gtrsim 7$\,Gyr. This lower limit is a factor of two higher than the transit time found by \citet{martin07}. They estimated that slow fading blue galaxies take $\sim 3$\,Gyr to arrive in green valley, plus additional $\sim 3$\,Gyr to reach the red sequence. However, the  $\sim 3$\,Gyr estimate of \citet{martin07} is strictly speaking a lower limit because it includes bursting and dust-extincted galaxies among green valley galaxies.

Moreover, even though PSBs may account for essentially all of the red-sequence, the evidence for evolution via both the slow and fast track is indisputable. Previous studies suggest that nearly half of the red sequence galaxies have disk-like morphologies \citep[e.g.,][]{bundy10,cheng11,vanderwel11}, indicating that the two modes of quenching are about equally important. Similarly, \citet{fang12} find that a non-negligible fraction of green valley galaxies have disk-like morphologies \citep{salim12} and can remain in the GV for several Gyr, which both point to the slow mode.

Likewise, disks in quenched galaxies are not rare at high redshift despite the expected dominance of mergers then. \citet{bundy10} studied quiescent galaxies at $1 < z < 2$ and found that passive disks with typically Sa-Sb morphological types represent nearly one-half of all red sequence galaxies. Similarly, \citet{vanderwel11} investigated morphology of massive, quiescent galaxies at $z \sim 2$. They estimate that more than $65\% $ of these galaxies are disk dominated. At a similar redshift, \citet{kocevski12} found that moderate luminosity, X-ray-selected AGN do not exhibit a significant excess of distorted morphologies relative to a mass-matched control sample. About half of the AGN reside in galaxies with discernible disks. The observed high disk fraction in AGN hosts is hard to reconcile with the merger picture of AGN fueling discussed in \S1.

Despite its theoretical appeal, compelling observational evidence linking mergers with AGN activity has been elusive, with results in favor of \citep{silverman11,ellison11,liu12}
and against \citep{jahnke11,cisternas11,kocevski12,schawinski12,villforth14} this picture. There are several effects that make it difficult to identify the connection between AGN activity
and mergers. One is the extreme dust obscuration that can be expected in such systems \citep{hopkins06}, making AGN detection challenging. The second is the significant time delay between the onset of the merger and the peak of the AGN activity. Because of this delay, the most obvious merger signatures may have faded by the time the merger remnant is identified as a bright AGN. 

We have looked at mergers fraction in starbursts and post-starbursts in our sample. We \emph{tentatively} find that starbursts are more disturbed than normal star-forming galaxies (the disturbance could be due to major or minor mergers). We visually classified about 30\% the starbursts as merging or disturbed galaxies (they show either tidal tails or strong asymmetries or have close companions). In contrast,  only about 3 \% of 200 randomly selected normal star-forming galaxies show merger signatures. Likewise, according to the Galaxy Zoo classification \citep{lintott11}, which rather tend to be conservative in calling something a merger, about 10\% of the starbursts have a weighted merger fraction $f_\mathrm{m} > 0.4$ while only 1\% of normal star-forming galaxies have a weighted merger fraction above this value. The $f_\mathrm{m}$ is calculated by taking the ratio of the number of merger classifications for a given galaxy to the total number of classifications for that galaxy multiplied by a weighting factor that measures the quality of the classifiers that have classified the galaxy. \citet{darg10} have shown that almost all galaxies with $f_\mathrm{m} > 0.4$ are robust major mergers. However, we also find that merger signatures disappear after the starburst phase, and the transiting and quenched post-starburst galaxies in general are much smoother than the starbursts. We classify about 15\% of TPSBs as as merging or disturbed galaxies and 6\% have $f_\mathrm{m} > 0.4$. Perhaps the merger signature are washed out because of the substantial time lag between the starburst and the PSB (AGN) phases. Galaxy merger simulations estimate that major merger signatures have a timescale of 200-400 Myr \citep{Lotz10}. Our estimated age of the transiting post-starburst phase ($\gtrsim 300$\,Myr) or the time delay between starbursts and AGN ($\gtrsim 200 \pm 100 $\,Myr) is in accord with the timescale for the disappearance of merger signatures. The color gradient and metallicity of starbursts and PSB are also consistent with the merger origin of these galaxies (see Appendix~\ref{sec:appE} \& \,\ref{sec:appF}).

The above tentative result supports that the fast-track, in local universe, is triggered by merger starbursts, whose signatures are washed out in the post-starbursts phase. We have also shown that velocity dispersion and global mass surface density are high, presumably from mergers, leaving post-starburst remnants which are smaller, more compact, and with high stellar surface mass density than non-bursty star-forming galaxies. However, despite their high velocity dispersion and global mass surface density, the post-starbursts still overlap in morphology with many slowly quenching galaxies. Future work will  explore better morphological discriminants between the fast and slow mode (Yesuf et al., in preparation). 

Deep high resolution studies of handful of K+A galaxies and post-starburst quasars however find significant morphological disturbances in these objects \citep[e.g.,][]{canalizo01,bennert08,yang08,cales11}. Galaxies we classified as undisturbed using the SDSS images may have faint tidal features visible in deeper images. Therefore, deep high resolution studies with more robust measurements of merger signatures in transiting post-starburst galaxies will be useful to test merger origin of post-starbursts and to understand the AGN triggering mechanism in post-starbursts.

\subsection{Conclusion and summary}

The unique spectral properties of quenched post-starburst galaxies hint that these objects are recently quenched starbursts. We investigated this inferred relationship in detail by directly tracing them back to the starbursts through a newly identified population of ``transiting'' post-starbursts in the midst of quenching. We showed that dust-obscured post-starbursts comprise the majority of the transiting post-starburst population. 

With our new sample of post-starbursts, we studied the connection between quenching and AGN in post-starbursts. We found that AGN are more commonly hosted by post-starbursts than by normal galaxies. Post-starburst AGN hosts make up  $\gtrsim 36 \pm 8\%$ of transiting post-starbursts. Despite the high frequency of AGN in post-starbursts, we found that the clear presence of AGN is significantly delayed from the peak of starbursts by $\gtrsim 200 \pm 100$\,Myr. 

As long as the AGN appearance is delayed, our results are generally consistent with ``merger hypothesis'' of post-starbursts \citep{hopkins06,snyder11,cen12}, where mergers between gas-rich galaxies drive nuclear inflows of gas thereby leading to nuclear starbursts, bulge formation, AGN activity, and eventually to quenched post-starbursts. In support of the merger hypothesis, we \emph{tentatively} find that the starbursts are relatively metal-poor at earlier stages, exhibit clear merger signatures, and have shallower color gradients and prominent young bulges. On the other hand, consistent with the time delay, merger signatures disappear after the starburst phase and that our three post-starburst classes also have shallower color gradients and prominent young bulges.  

We also showed that starbursts and transiting post-starbursts are significantly more dust-obscured than normal galaxies and quenched post-starbursts. The fact that starbursts and post-starbursts evolve through a heavily dust-obscured phase which also seems to coincide with AGN activity, is consistent with later removal of dust by AGN feedback. We therefore conclude that AGN may not \emph{primarily} quench starbursts but may play an important role in quenching or preventing low-level star formation in post-starbursts. We acknowledge that the large extent of the SDSS fiber beyond the nuclear region of a galaxy could be a major concern since the line ratios of an AGN may be diluted by on-going star formation inside the fiber. Future works with spatially resolved line ratios or with other AGN selection criteria unbiased by the host properties will hopefully provide a more definitive test on the time delay between the AGN phase and the starburst phase, and they will also help estimate the AGN fraction in post-starbursts more accurately than we have attempted in this work. Similarly, a more direct evidence on the role of AGN in removing a leftover gas and dust during the post-starburst phase may come to light from observations of TPSBs using new facilities such as ALMA.

\acknowledgments
We are grateful for the referee for his/her detailed comments which significantly improved the clarity and content of the paper. We thank Renbin Yan, Edmond Cheung, Guillermo Barro, Dale Kocevski and Charlie Conroy for helpful discussion and assistance on the paper. We acknowledge financial support from NSF grant AST-0808133. CCH is grateful to the Klaus Tschira Foundation for financial support.

Funding for the SDSS and SDSS-II has been provided by the Alfred P. Sloan Foundation, the Participating Institutions, the National Science Foundation, the U.S. Department of Energy, the National Aeronautics and Space Administration, the Japanese Monbukagakusho, the Max Planck Society, and the Higher Education Funding Council for England. The SDSS Web Site is http://www.sdss.org/.

The SDSS is managed by the Astrophysical Research Consortium for the Participating Institutions. The Participating Institutions are the American Museum of Natural History, Astrophysical Institute Potsdam, University of Basel, University of Cambridge, Case Western Reserve University, University of Chicago, Drexel University, Fermilab, the Institute for Advanced Study, the Japan Participation Group, Johns Hopkins University, the Joint Institute for Nuclear Astrophysics, the Kavli Institute for Particle Astrophysics and Cosmology, the Korean Scientist Group, the Chinese Academy of Sciences (LAMOST), Los Alamos National Laboratory, the Max-Planck-Institute for Astronomy (MPIA), the Max-Planck-Institute for Astrophysics (MPA), New Mexico State University, Ohio State University, University of Pittsburgh, University of Portsmouth, Princeton University, the United States Naval Observatory, and the University of Washington.

\clearpage
\appendix

\section{A: Details of Dust Correction}
\label{sec:appA}

\subsubsection{Methods of dust correction}
We correct for dust effects on emission-line luminosities (\OII, $\mathrm{H\alpha}$, and etc), GALEX and SDSS colors, the $\mathrm{H\alpha}$ equivalent width, $W_\mathrm{H\alpha}$, and $D_n(4000)$. For emission line extinction curve, we use eqn.\,A1 and for the continuum extinction curve, we use eqn.\,A2 \citep{charlot00,wild11b,wild11a}. In this section, continuum quantities will be denoted by `$\ast$' superscript.

\begin{equation}
Q_\lambda= (1-\mu)\left(\lambda/5500\right)^{-1.3}+\mu\left(\lambda/5500\right)^{-0.7} \;\mathrm{where\;} \mu=0.4
\label{eqn:ql}
\end{equation}

\begin{equation}
Q_\lambda^\ast= 1/N\left[\left(\lambda/\lambda_{b_1}\right)^{n\,s_1}+\left(\lambda/\lambda_{b_1}\right)^{n\,s_2}+\left(\lambda/\lambda_{b_2}\right)^{n\,s_3}+\left(\lambda/\lambda_{b_3}\right)^{n\,s_4}\right]^{-1/n}
\label{eqn:qc}
\end{equation}

$Q_\lambda^\ast$ is composed of four power-law functions with exponents $s_{[1-4]}$ smoothly joined with a smoothness parameter $n = 20$. The power-law exponents vary with both axis ratio, $b/a$, and fiber specific star formation rate, $\psi_s$, according to linear functions given in \citet{wild11b} eqn. 22-25.  The $\lambda_{b[1-3]}$ are related to the position of the three break points at 0.2175\,$\mu \mathrm{m}$, 0.3\,$\mu \mathrm{m}$ and 0.8\,$\mu \mathrm{m}$ and the power-law exponents according to eqn.\,19-21 in \citet{wild11b}. $N$ is the normalization, defined such that $Q_\lambda^\ast$ is unity at $5500\,\AA$.

The line optical depth is given by eqn.\,A3 and uses the expression of $\tau_V$ in eqn.\,A4. 

\begin{equation}
\tau_\lambda = \tau_VQ_\lambda
\end{equation}

\begin{equation}
\tau_V = 0.921\times 2.5/\left(Q_{4861}-Q_{6563}\right)\times \log\left(\mathrm{H}\alpha/\mathrm{H}\beta \times \left(\mathrm{H}\alpha/\mathrm{H}\beta\right)_\mathrm{int}^{-1}\right)
\end{equation}

We require SNR $> 1\sigma$ on $\mathrm{H\alpha}$ and $\mathrm{H\beta}$ lines. We assume dust-free $(\mathrm{H\alpha/H\beta)_{df} = 2.86}$  for star-forming galaxies \citep{osterbrock89} and $(\mathrm{H\alpha/H\beta)_{df} = 3.1}$ for AGN \citep{veilleux87}. For example, using the above equations, the dust corrected \OII\;flux is given by:

\begin{equation}
f_\mathrm{OII,dc} = f_\mathrm{OII} \times 10^{0.4\times 1.086 \times \tau_{3727}}
\end{equation}

To correct for galaxy fluxes (colors) we use the ratios of $\tau_V/\tau_V^\ast$ in eqn.\;13-16 \citet{wild11b}, which are found to vary strongly with galaxy properties such as axis-ratio and specific $\mathrm{SFR}$. We use the ratios that give maximal stellar extinction. In other words, min\{$\tau_V/\tau_V^\ast (\psi_s),\tau_V/\tau_V^\ast (b/a)\}$.  We prefer maximal stellar extinction because the optical depth ratios in \citet{wild11b} are generally smaller but asymptote to 2.08, the measured values in starbursts \citep{calzetti00}. In estimating the optical depth ratios, we use star formation rates calculated from dust-corrected $\mathrm{H\alpha}$ using the conversion factor of \citet{kennicutt98}. The SFRs will be overestimated if there is a significant contribution from AGN to the $\mathrm{H\alpha}$ emission line. We used the optical depth ratios estimated from axis-ratio (i.e, inclination) only as a check, and the possible over-estimation of SFR due to AGN does not significantly affect our results. It should be noted that we do not purposely use the star formation rate measurements provided in SDSS DR8 which are derived from photometry for AGN, because they may be systematically underestimated for dusty galaxies \citep[including AGN,][]{wild11b}.

\begin{equation}
\tau_\lambda ^\ast = Q_\lambda^\ast \times \tau_V^\ast = Q_\lambda^\ast \times (\tau_V^\ast /\tau_V) \times \tau_V
\end{equation}

\begin{equation}
A_\lambda ^\ast = 1.086\tau_\lambda ^\ast
\end{equation}

Because we are correcting for the global galaxy colors, while our estimate of $\tau$ is based on fiber quantities, we will approximately correct for gradient (aperture bias) in $\tau_V$  by dividing $A_\lambda ^\ast$ with a correction factor $f_\nabla= 1.0-1.25$ according to Figure 6 of \citet{wild11b}: for bulge-dominated galaxies we 
use eqn.\,A8 while for disk-dominated galaxies we use eqn.\,A9.
\begin{equation}
f_\nabla = \left\{
  \begin{array}{l l}
    1.0  & \quad \text{if $R_{fib}/R_{90} \geq 1$ else}\\
    1.05 &  \quad \text{if $R_{fib}/R_{90} < 1$\; \& \;$\log \psi \leq -9.9$}\\
    1.1 & \quad \text{if $R_{fib}/R_{50} \geq 1$ \;\& \;$-9.9 < \log \psi < -9.6$}\\
    1.15 & \quad \text{if $R_{fib}/R_{50} < 1$ \;\& \;$-9.9 < \log \psi < -9.6$}\\
    1.2  & \quad \text{if $R_{fib}/R_{50} \geq 1$ \;\& \;$\log \psi > -9.6$}\\
    1.25 & \quad \text{if $R_{fib}/R_{50} < 1$ \;\& \;$\log \psi > -9.6$}
\end{array} \right.
\end{equation}

\begin{equation}
f_\nabla = \left\{
  \begin{array}{l l}
    1.0  & \quad \text{if $R_{fib}/R_{90} \geq 1 $ else}\\
    1.05 &  \quad \text{if $R_{fib}/R_{90} < 1$\; \& \;$\log \psi \leq -9.9$}\\
    1.05 &  \quad \text{if $R_{fib}/R_{50} \geq 1$\; \& \;$-9.9 < \log \psi < -9.6$}\\
    1.1 & \quad \text{if $R_{fib}/R_{50} \geq 1 \;\& \;\log \psi \geq -9.6$}\\
    1.15 & \quad \text{if $R_{fib}/R_{50} < 1$ \;\& \;$-9.9 < \log \psi < -9.6$}\\
    1.2  & \quad \text{if $R_{fib}/R_{50} < 1$ \;\& \;$\log \psi \geq -9.6$}
\end{array} \right.
\end{equation}

For instance, $(NUV-g)_\mathrm{dc}$ is given by eqn.\,A10 below. We use the effective wavelengths of SDSS bands given in \citet{fukugita96}. 

\begin{equation}
(NUV-g)_\mathrm{dc} = (NUV-g)-(A_{2267}^\ast - A_{4825}^\ast)/f_\nabla
\end{equation}
 
We correct the $W_\mathrm{H\alpha}$ using the following equation:

\begin{equation}
W_\mathrm{H\alpha,dc} = W_\mathrm{H\alpha} \times \frac{10^{0.4\times 1.086 \times \tau_{6563}}}{10^{0.4\times 1.086 \times \tau_{6563}^\ast}} = W_\mathrm{H\alpha} \times 10^{0.4\times 1.086 \times (\tau_{6563} -\tau_{6563}^\ast)}
\end{equation}

$D_n(4000)$ is defined as a flux ratio of a narrow continuum range red-ward of $4000$\,\AA\, break ($4000-4100$) to a narrow continuum range blue-ward of the break \citep[$3850-3950$,][]{balogh05}. 

Therefore, $D_n(4000)_\mathrm{dc} \approx D_n(4000) \times 10^{0.4 \times (A_\mathrm{red}^\ast - A_\mathrm{blue}^\ast)}$. where $A_\mathrm{red}^\ast = (A_{4000}^\ast+A_{4050}^\ast+A_{4100}^\ast)/3.0$ and $A_\mathrm{blue}^\ast = (A_{3850}^\ast+A_{3900}^\ast+A_{3950}^\ast)/3.0$

\begin{figure*}
\includegraphics[scale=0.8]{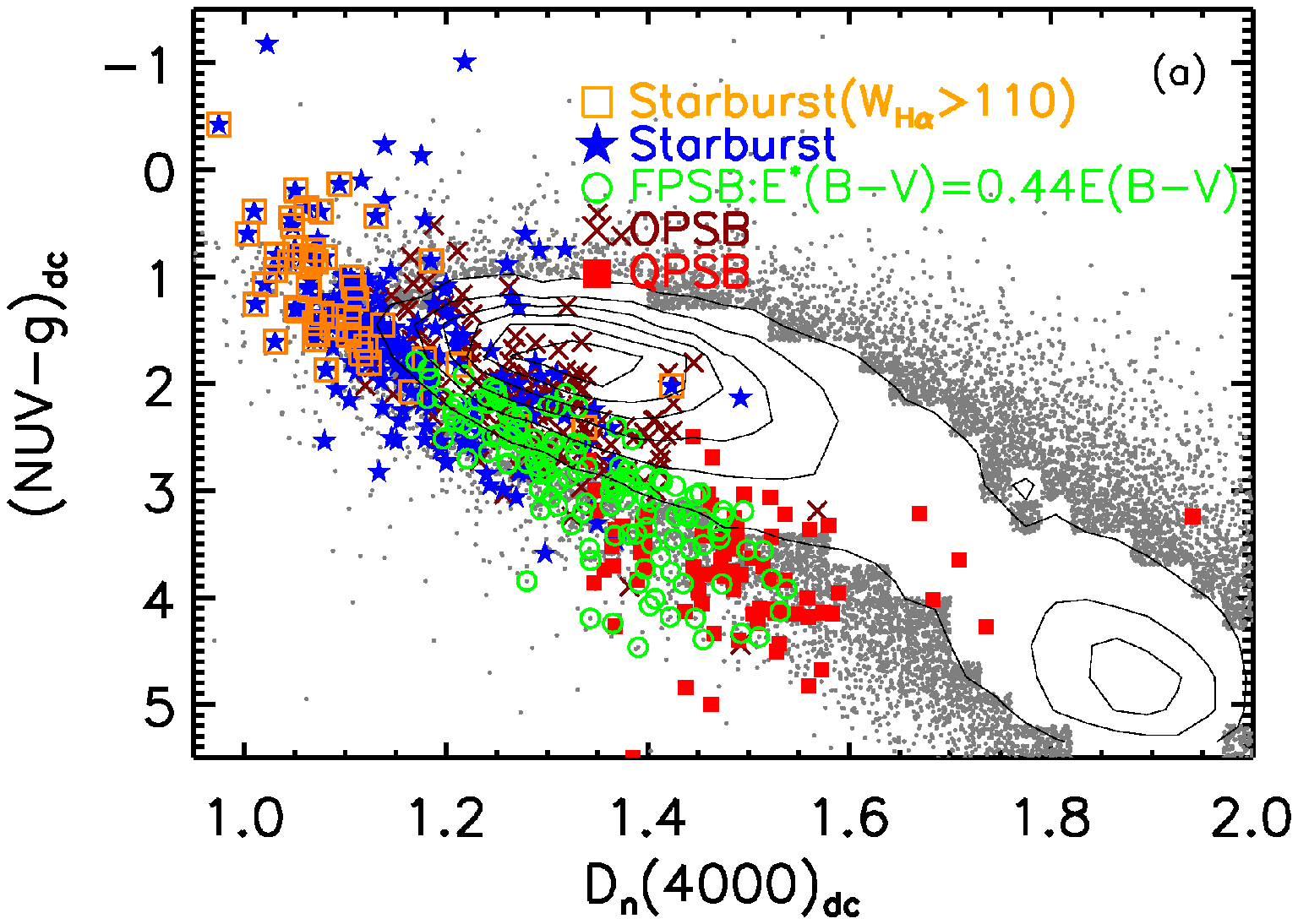}
\includegraphics[scale=0.8]{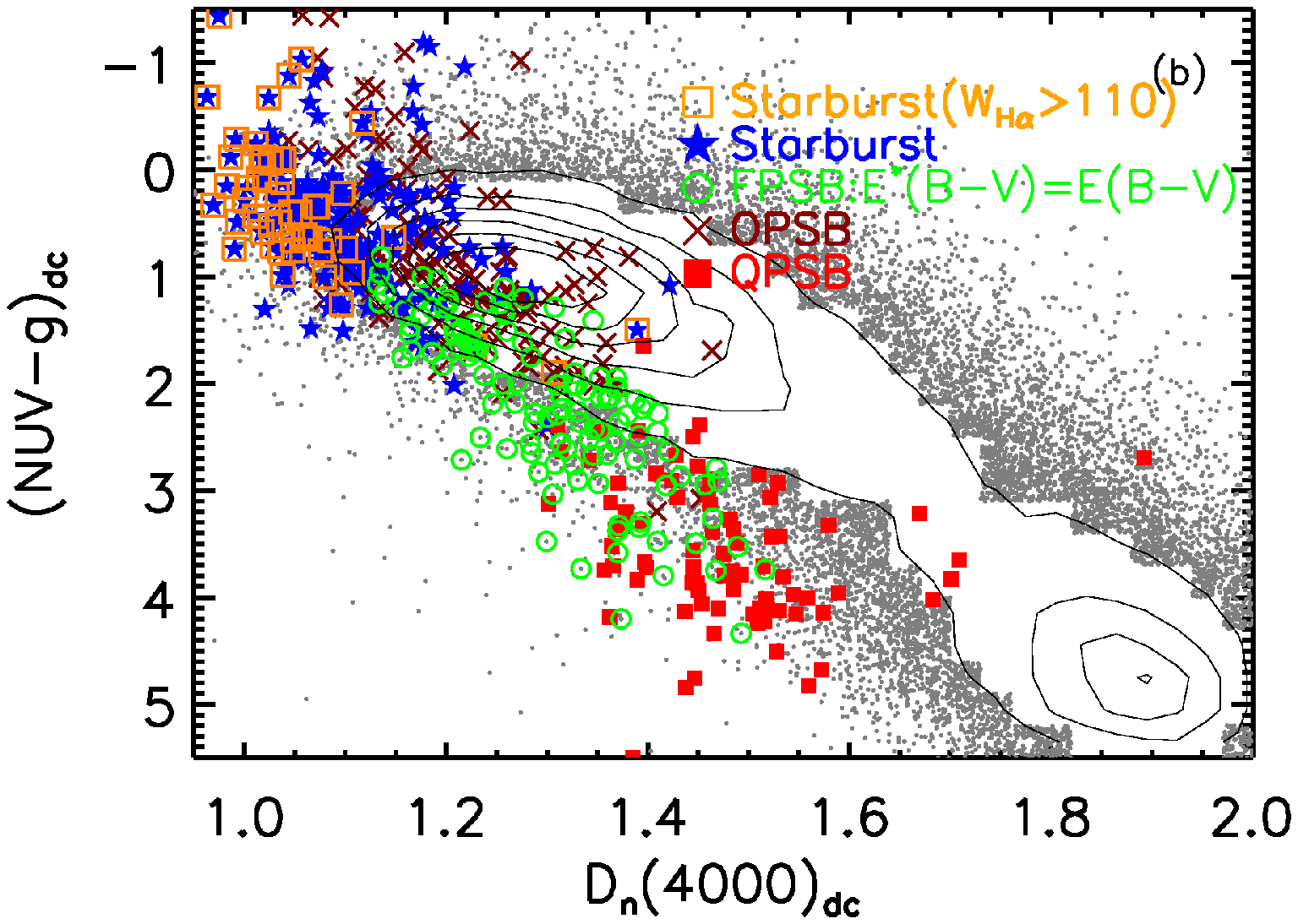}
\caption{ Panel (a): Dust-corrected $NUV-g$ color versus $D(4000)$. The FPSBs are selected assuming the Calzetti extinction curve with $E^\ast(B-V)/E(B-V) = 0.44$. In panel (b) the FPSB selection assumes the Calzetti extinction curve with $E^\ast(B-V)/E(B-V) = 1.0$ instead. The main point of the figure is that the details of the dust-correction are not important for the post-starburst selection as long as dust in all galaxies are similar. On the other hand, most starbursts would not have been identified if it were not for the dust correction.\label{fig:nuvd4ap}}
\end{figure*}

In the rest of this section we will show that our post-starburst selection does not significantly depend on the detail assumptions of the dust correction described above. To that end, Figure~\ref{fig:nuvd4ap} replots the $NUV-g$ color versus $\mathrm{D_n(4000)}$ diagram for different dust-correction assumptions. In Panel a and b, we show the version of the figure in which $NUV-g$ color versus $\mathrm{D_n(4000)}$ are corrected using the \citet{calzetti00} extinction curve with the ratio of excess $B-V$ colors of gas to stars is, $E^\ast(B-V)/E(B-V) = 0.44$ and 1.0. Note that, as described in \S2.4, we adopted in the main text the empirical attenuation curve of \citep{wild11b} and their prescription to estimate $E^\ast(B-V)/E(B-V)$ ratios. Since the selection of FPSBs explicitly depends on the dust correction, we show in panel a to b, the alternative selection of this class for the the given dust-correction prescription adopted in each panel. In panel a 136 FPSBs and in panel b 126 FPSBs are identified. About 15-25\% of the FPSBs are previously (\S3) unidentified but about 85\% of the FPSBs identified in section three are also identified in panel a to b. The \citet{calzetti00} curve lacks the 2175\,\AA\, bump and \citet{wild11b} find typically $0.3-1.0$ magnitudes more attenuation in the NUV compared to the Calzetti extinction curve. Thus, the previously unidentified FPSBs may be dusty contaminants. Overall, the fact that each panel identified comparable number of FPSBs and recovered 85\% of FPSBs defined in \S3, suggests that the details of dust-correction are not important for the selection of these objects. Furthermore, the figure also shows (in orange square) the subset starbursts with (dust-extincted) $\mathrm{W_{H\alpha}>110}$, that is to say, those that satisfying the \citet{lee09} definition of starbursts. This subset only account for 25\% of all starbursts we have identified. Therefore, the dust correction of $\mathrm{H\alpha}$ is important to identify majority of dust-extincted starbursts.

The AGN fraction for FPSBs selected in panel a and pane b is 45\% and 48\% respectively. In comparison, the AGN fraction for FPSBs selected in the main text (\S 3.4) is 53\%. Therefore, the error on AGN fraction of transiting post-starbursts may be as high as 8\% (or even higher if BPT composite galaxies indeed host AGN). Even with 8\% error, the AGN fraction in post-starbursts is still more than two times higher than that of normal galaxies.

\subsubsection{The color-color diagram: the intrinsic colors of obscured post-starbursts}

\begin{figure*}
\centering
\includegraphics[scale=0.8]{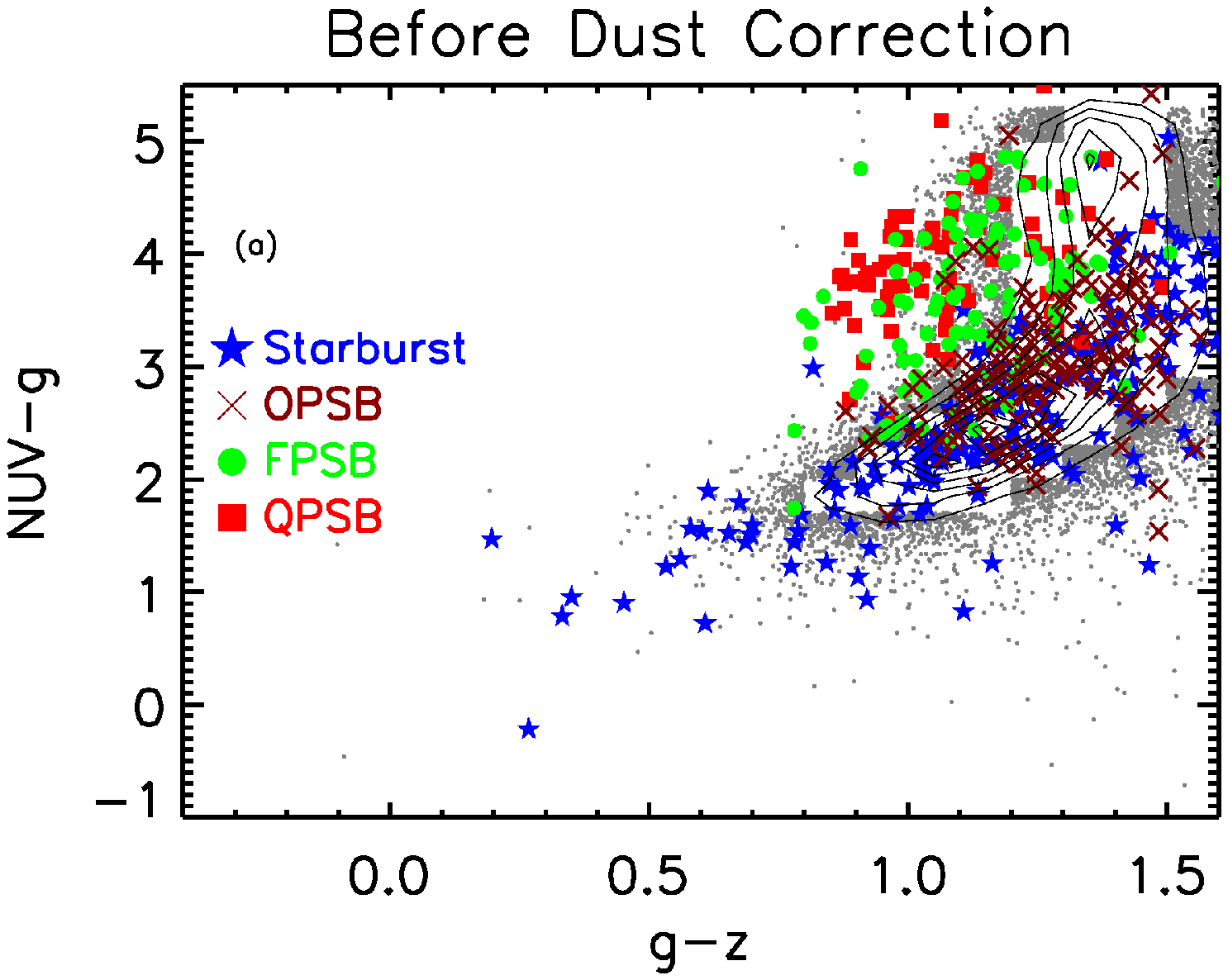}
\includegraphics[scale=0.8]{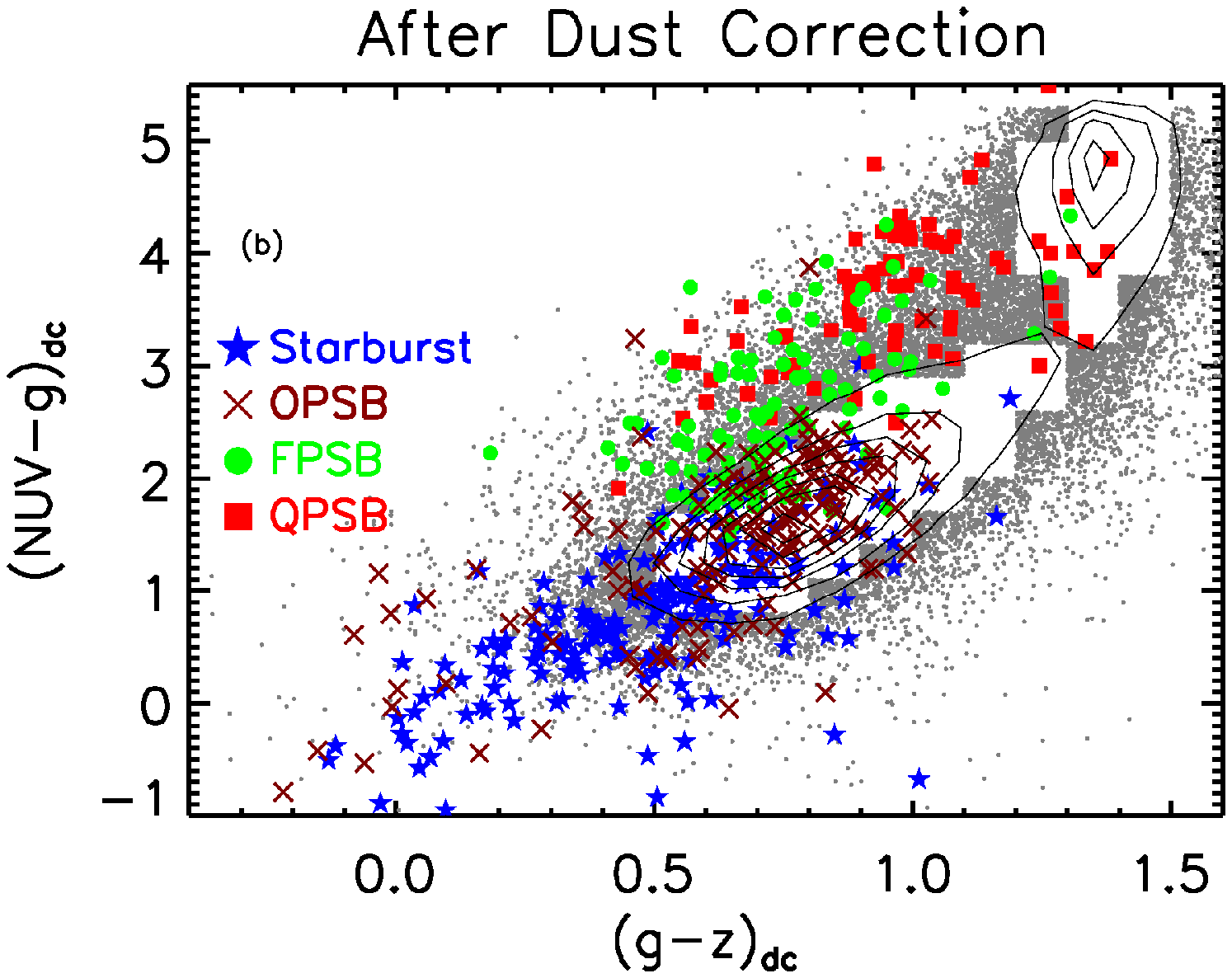}
\caption{Panel (a) : Rest-frame $NUV-g$ vs. $g-z$ color-color diagram, using observed (not dust-corrected) magnitudes. Panel (b) : Dust-corrected rest-frame $NUV-g$ vs. $g-z$ color-color diagram. The rapid quenching/strong burst model tracks nicely describe the sequence of starburst to transiting post-starburst to classical quenched post-starburst. The dust-obscured galaxies are also consistent with the transiting post-starbursts along this track.  We use this with our other evidence (in Figures~\ref{fig:hd_d4} and~\ref{fig:sed}) to infer that, like the fading post-starburst population, the dust-obscured post-starbursts represent an intermediate phase from starbursts to post-starbursts. } \label{fig:uvgz}
\end{figure*}

Moreover, in this subsection we aim to show that our dust-correction works and our starburst evolutionary path is plausible. Figure~\ref{fig:uvgz} shows the $UVgz$ diagram ($NUV-g$ vs. $g-z$), a variant of the widely used $UVJ$ diagram in galaxy evolution studies \citep[e.g.,][]{wuyts07,williams09,whitaker12}. In these diagrams dusty star-forming, non-dusty star-forming and quiescent galaxies are well separated. Star-forming galaxies form a diagonal track which extends from blue to red colors. The red end of this track is populated by dusty galaxies. The quiescent galaxies form a separate clump above the dusty star-forming galaxies. We show the $UVgz$ diagram before and after the dust correction.

After dust-correction, blue star-forming and red quiescent galaxies are cleanly separated in the $UVgz$ diagram. The starbursts are significantly bluer after the dust correction and they lie well off the blue cloud to the lower left. The dust correction is difficult for the quenched post-starbursts because of their weak emission lines. However their location in the upper right corner of the $UVgz$ diagram is consistent with little or no dust extinction\citep[cf.][]{balogh05,kaviraj07,brown09,chilingarian12,whitaker12}. We also showed in Figure~\ref{fig:wise2} that about 80\% of quenched post-starbursts do not show significantly dust-obscured star formation ( have $\mathrm{f_{12\mu m}/f_{0.2\mu m} < 200}$). 

Transiting post-starbursts show more significant dust-reddening than do quenched post-starbursts: their observed colors are significantly redder but they are indeed intrinsically bluer. 
Despite the large overlap with normal galaxies, the OPSBs generally have intermediate intrinsic colors between that of QPSBs and FPSBs. Their overlap with normal galaxies is not a problem because the the overplotted model tracks also pass through normal galaxies. In contrast to the dust-unreddened colors, the observed $g-z$ colors of PSBs get redder from OPSBs to FPSBs to QPSB, suggesting a decreasing dust sequence we have seen in previous diagrams. The plausible arrangement of SBs and PSBs in color-color space is also further evidence that the dust corrections work.

\section{B: Details of Stellar Population Modeling}
\label{sec:appB}
We modeled SFHs of a post-starburst as a superposition of an old stellar population initially formed at time $t=0$ following a delayed exponential SFH of a form $\psi \propto t\exp(-t/\tau_1)$ with e-folding time, $\tau_1=1$\,Gyr \citep[cf.][]{kriek11} and a young stellar population formed in a recent burst at $t=12.5$\,Gyr ($z \sim 0.1$) with exponentially declining SFH, $\psi \propto \exp(-t/\tau_2)$, of $\tau_2 = 0.1$\,Gyr \citep[cf.][]{kaviraj07,falkenberg09a}. Because of the burst mass-age degeneracy, the ages of the post-starbursts depends on the decay timescale $\tau_2$ we assumed. In this section, we quantify the effect of using different decay timescales  ($\tau_2 = 0.05$  or $\tau_2 = 0.2$) instead of our fiducial value of $\tau_2 = 0.1$\,Gyr used in the main text .

\begin{figure*}
\centering
\includegraphics[scale=0.8]{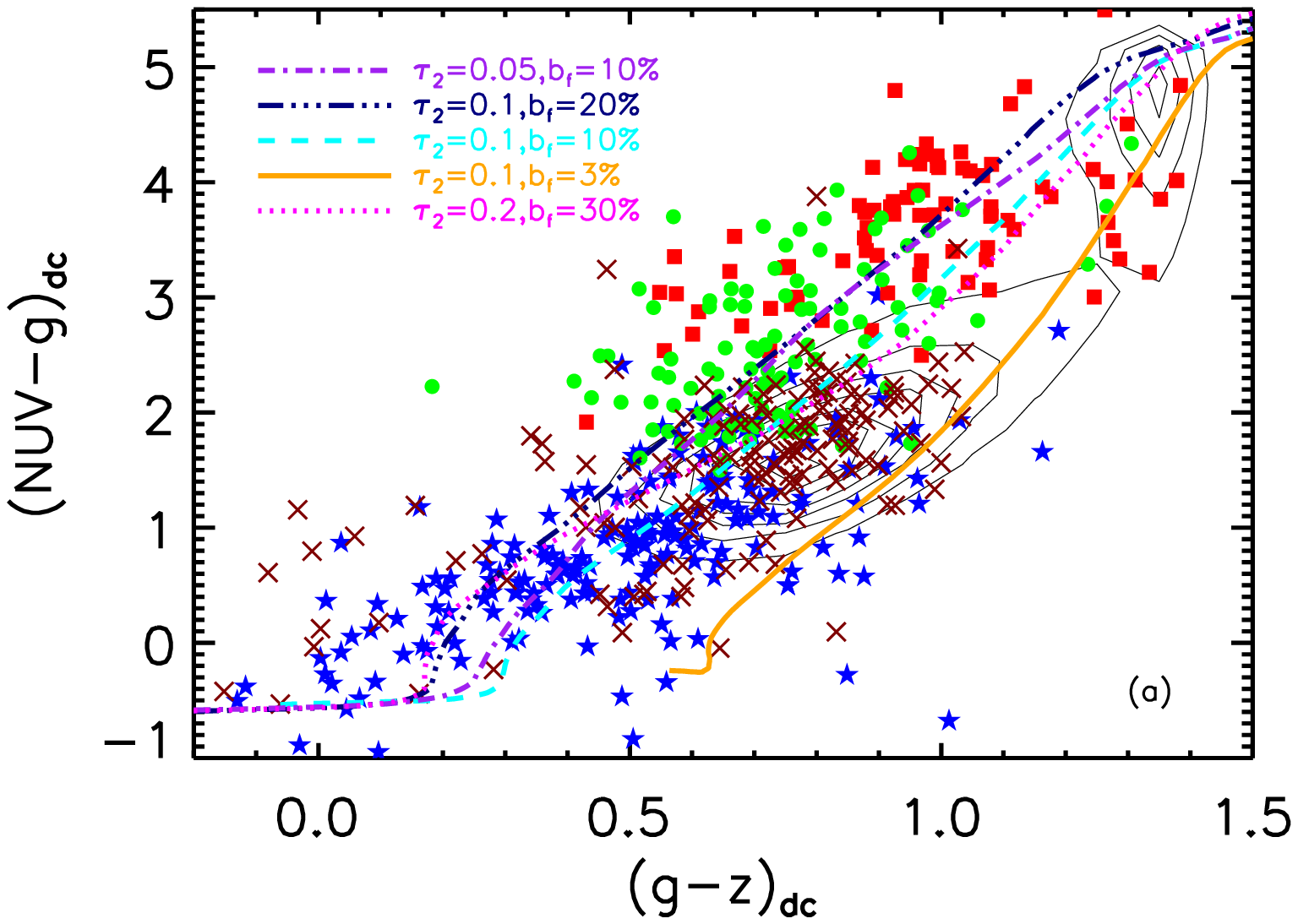}
\includegraphics[scale=0.8]{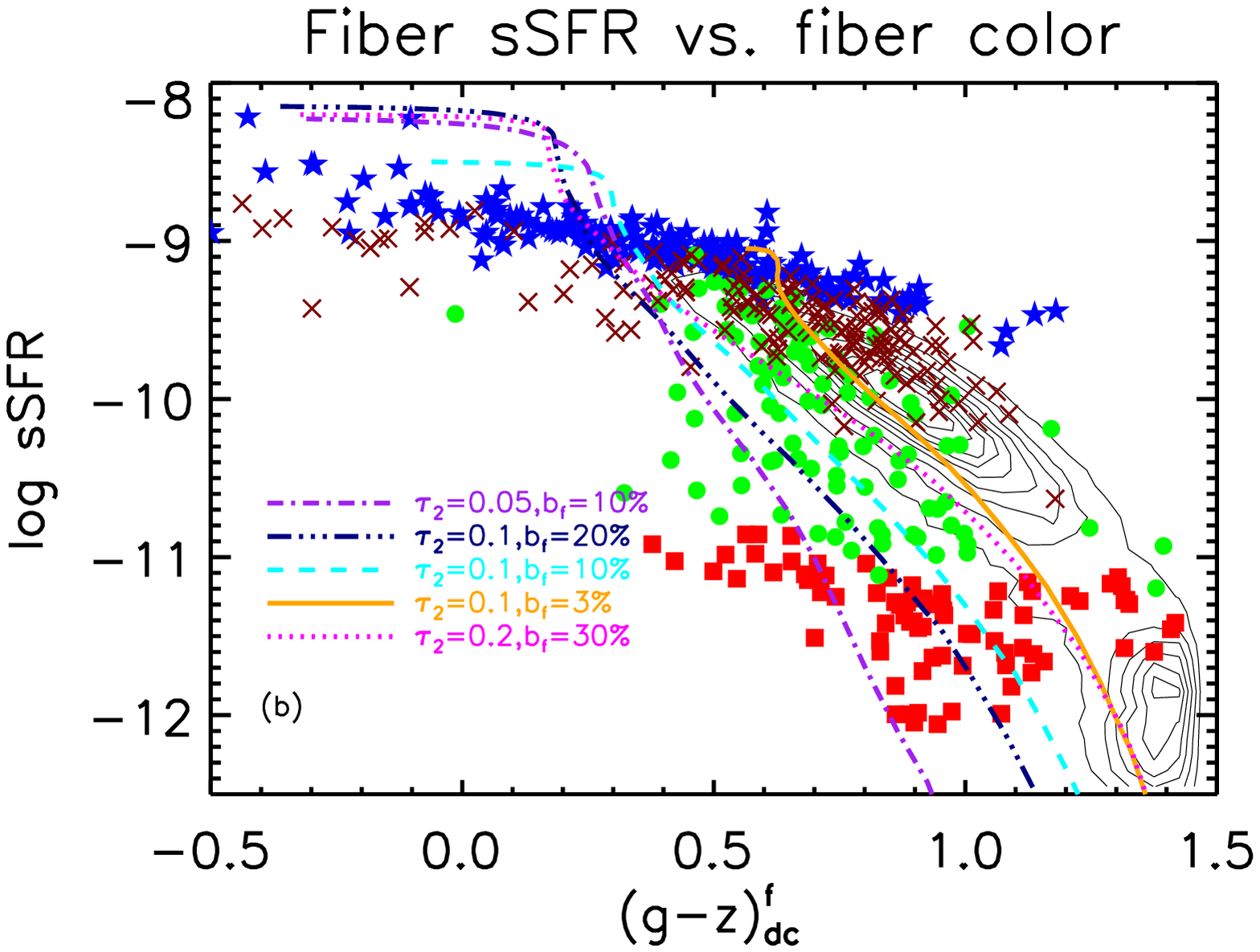}

\caption{Panel (a): dust-corrected rest-frame $NUV-g$ and $g-z$ color-color diagram. Overplotted are burst modeled tracks of $\tau_2 = 0.05,0.1$ and $0.2$\,Gyr \citep{bc03}. Panel (b): Specific fiber star formation rate versus dust-corrected fiber $g-z$ color \label{fig:ugz2}. These diagrams exclude burst models outside $\tau_2 = 0.05-0.2$ range.
}
\end{figure*}

\begin{figure*}
\centering
\includegraphics[scale=0.93]{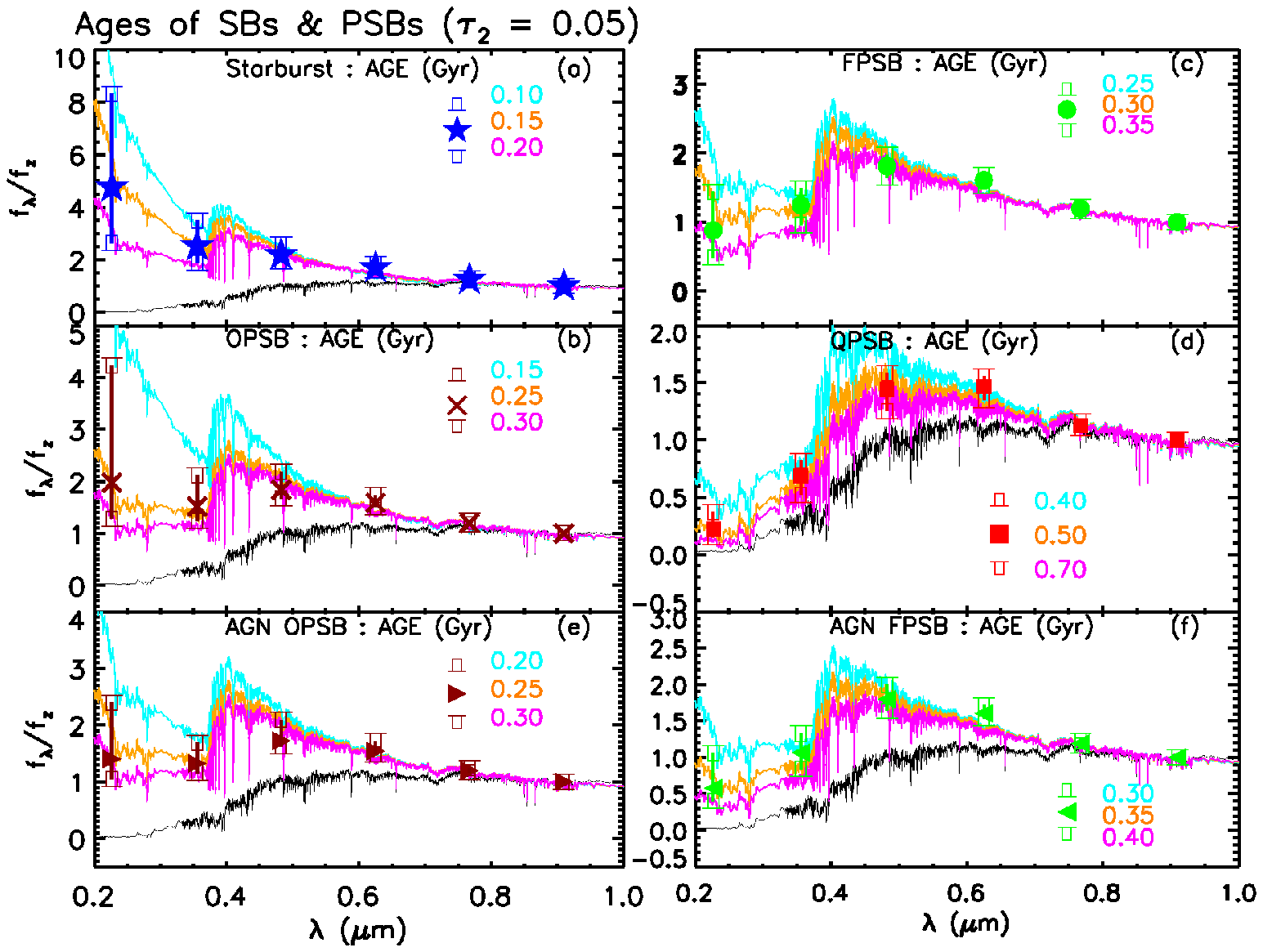}
\includegraphics[scale=0.93]{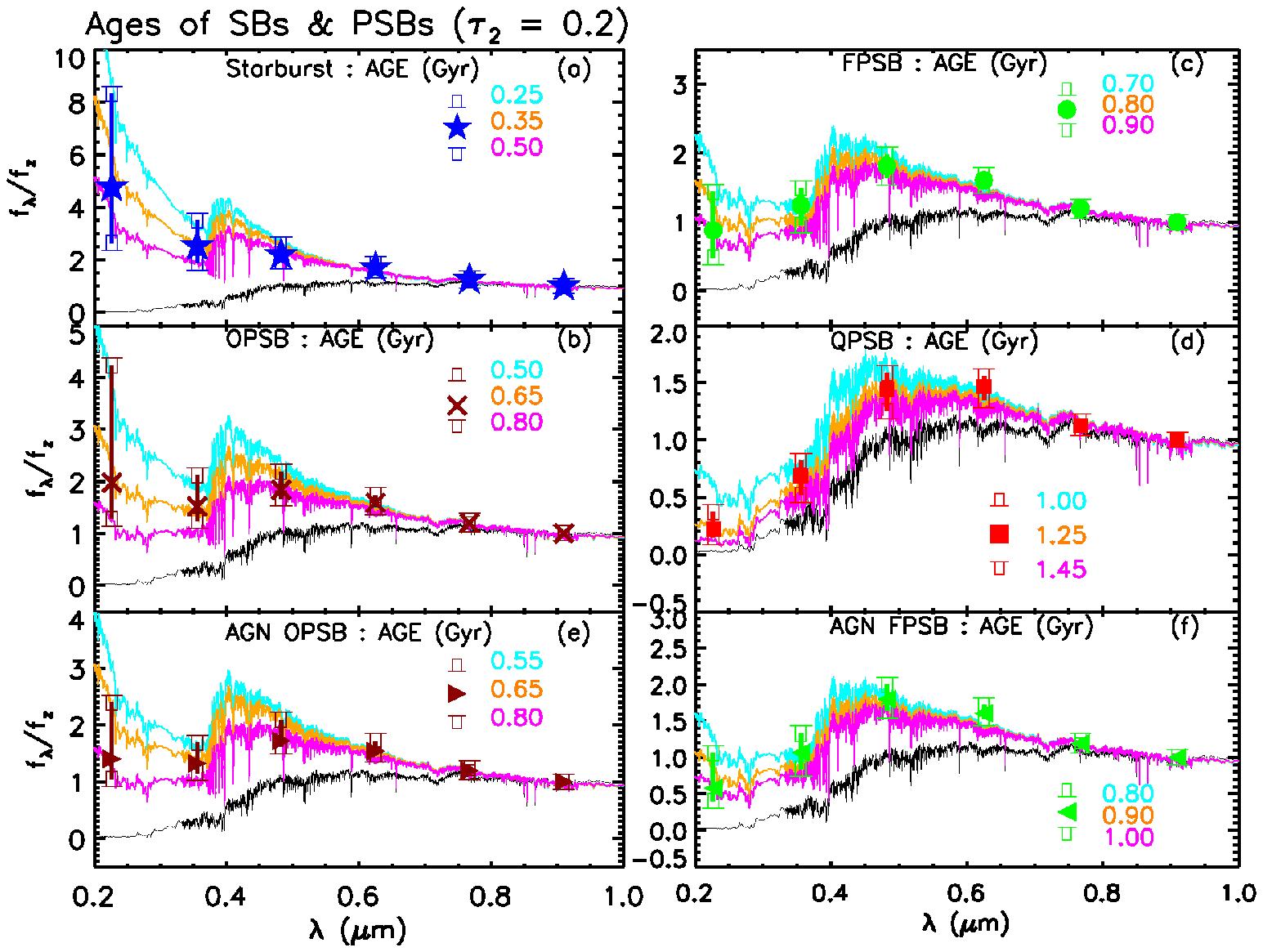}

\caption{The $z$ band normalized median and quartile fluxes at the effective wavelengths of the $NUV,u,g,r,i,z$ bands. This figure is the similar to Figure~\ref{fig:sed} but overplots \cite{bc03} burst models with different SFR timescale $\tau_2$ and burst fraction $b_f$. Top figure: the overplotted spectra are $\tau_2=0.05$\,Gyr and $b_f=10\%$ of different ages, as indicated on each panel. Bottom figure uses $\tau_2=0.2$ and $b_f=30\%$ instead. The main point of the figure is that starbursts and AGN are not coeval, they are separated at least by about 100-400 Myr. \label{fig:sed2}
}
\end{figure*}

Figure~\ref{fig:ugz2}a replots the dust-corrected rest-frame $NUV-g$ and $g-z$ color-color diagram to show that the starbursts to quenched post-starbursts evolution can alternatively be modeled with $\tau_2 = 0.05$\,Gyr and burst fraction $b_f=10\%$ or $\tau_2 = 0.2$\,Gyr and burst fraction $b_f=30\%$. Likewise, Figure~\ref{fig:ugz2}b plots fiber specific SFR against dust-corrected fiber $g-z$ color to make a similar point. Thus, models with $\tau_2$ outside the range 0.05-0.2  are excluded since they do not produce the observed population of post-starburst galaxies. The fiber specific star formation rates are estimated from dust-corrected $\mathrm{H\alpha}$ (\S~\ref{sec:dustcorr}) using the conversion factor of \citet{kennicutt98} and the fiber stellar mass. Figure~\ref{fig:sed2} estimates the age of starbursts and post-starbursts for these alternative burst models using the $z$ band normalized median and quartile SEDs of these galaxies. Accordingly, the time lag between the starburst and AGN phase may be between 100 and 400\,Myr.

\section{C: Details of Post-starburst Selection}
\label{sec:appC}
The following equations specify the fourth order polynomial fits to the data of main sequence galaxies in Figure~\ref{fig:hd_d4}. 
 
\begin{equation}
\mathrm{W_{H\delta}} = 23.112-19.700\times x+3.355\times x^2+0.0817\times x^3-0.00871\times x^4
\end{equation}

\begin{equation}
(NUV-g)_\mathrm{dc} = 7.010-13.547\times x+8.895\times x^2-1.397\times x^3+0.0462\times x^4
\end{equation}
where $x= D_n(4000)_\mathrm{dc}$

\section{D: More AGN Properties of Post-starbursts}
\label{sec:appD}
\subsection{WISE AGN diagnostic}

\begin{figure*}
\subfigure{\includegraphics[scale=0.8]{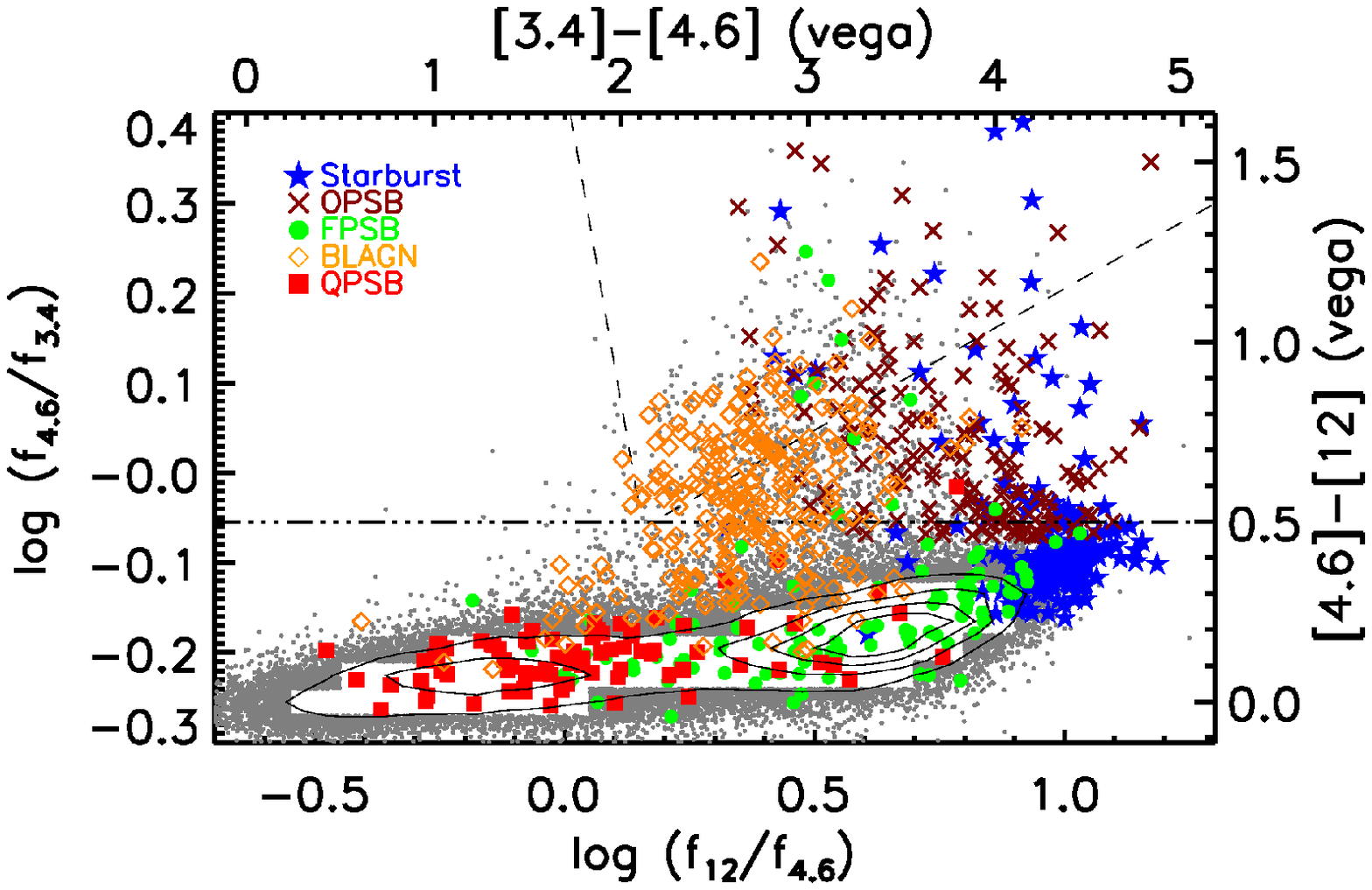}}
\caption{The WISE color-color plot which can reliably identify luminous (obscured and unobscured) AGN.  Starbursts (blue star), fading post-starbursts (green circles), obscured post-starbursts (brown Xs), quenched post-starbursts (red squares) and broad-line AGN (orange diamonds) are overplotted on the figure. The $\mathrm{f_{12\,\mu m}/f_{4.6\,\mu m}}$ flux ratio is sensitive to PAH emission and is a first order star formation indicator while $\mathrm{f_{4.6\,\mu m}/f_{3.4\,\mu m}}$ is sensitive to hot dust emission from AGN. The dashed wedge denotes the \citet{mateos12} AGN selection criteria while the horizontal dash-dotted line demarcates that of \citet{ashby09}. The latter is more complete but less pure. For comparison, we overplot broad line AGN of comparable mass and redshift studied by \citet{trump13}. The figure confirms that almost all starbursts and strongly star-forming transiting PSBs do not host strong or obscured AGN. Some transiting post-starbursts show hot dust emission from AGN. The BLAGN have lower $\mathrm{f_{12\mu m}/f_{4.6\mu m}} $ ratios (less star-forming) than obscured post-starbursts but higher flux ratios than quenched post-starbursts, suggesting that they may come after the obscured AGN phase \citep{hopkins06}, if they are related to post-starbursts.\label{fig:wise}}
\end{figure*}

An AGN has a spectral energy distribution (SED) that rises from $\sim 3-5\,\mu \mathrm{m}$ due to hot dust emission from its dusty torus \citep[]{nenkova08}, while a starburst has a composite stellar spectrum which peaks at $1.6\,\mu\mathrm{m}$ and declines over the range from $\sim 3-5\,\mu \mathrm{m}$. Mid-IR color-color diagnostic diagrams use this idea to distinguish AGN dominated galaxies from starburst dominated ones \citep[e.g.,][]{lacy04,stern05,donley12}. The IR color-color diagrams select only luminous AGN, and do not detect weak AGN \citep[][]{donley12}.

Figure~\ref{fig:wise} plots the WISE color-color diagram:  $\mathrm{log (f_{12\mu m}/f_{4.6\mu m})}$ versus $\mathrm{log (f_{4.6\mu m}/f_{3.4\mu m})}$ \citep{wright10,izotov11,assef12,lake12,stern12}. 
This figure is similar to Figure~\ref{fig:wise2} and it is presented here for completeness. Stellar populations younger than 0.6 Gyr dominate $12\,\mu \mathrm{m}$ emission and, as a result, $[4.6\,\mu \mathrm{m}]-[12\,\mu \mathrm{m}]$ color is known to correlate well with SFR \citep{donoso12}. Normal galaxies in the parent sample form a tight and elongated bi-modal sequence with some vertical scatter at $\mathrm{f_{12\mu m}/f_{4.6\mu m} \gtrsim 1}$. Generally, the starbursts are located at the right-most high-SFR end of the sequence, while quenched post-starbursts typically have lower $\mathrm{f_{12\mu m}/f_{4.6\mu m}} $ ratios like quiescent galaxies. The transiting post-starbursts mostly have intermediate $\mathrm{f_{12\mu m}/f_{4.6\mu m}} $ ratios between SBs and QPSBs. The arrangement of SBs, FPSBs and QPSBs in decreasing order of redness due to dust is another independent confirmation for the consistency of our evolutionary sequence. As expected from their selection, the obscured post-starbursts are found in between the FPSBs and SBs.

The $\mathrm{f_{4.6\mu m}/f_{3.4\mu m}} $ ratio may indicate emission from hot dust, ionized gas or stars. The simple $\mathrm{f_{4.6\mu m}/f_{3.4\mu m}} >0.88$ criterion can identify hot dust emission from AGN but only with $\sim$50\% reliability \citep{ashby09,stern12}. The dashed wedge, which is calibrated by X-ray-selected AGN, identifies a highly complete and reliable sample of luminous (hard X-ray luminosity, $L_{2-10\,\mathrm{kev}}> 10^{44}$\,erg\,s$^{-1}$) AGN \citep{mateos12}. For a reference, we also overplot broad-line AGN in the similar mass and redshift range.

According to the \citet{mateos12} classification, only $7\%$ of the FPSBs and 21\% of the OPSBs show clear AGN signatures in WISE, and the majority of these galaxies are already classified as Seyferts by the optical emission line diagnostics. This indicates that most of AGN found in the transiting post-starbursts, including the ones in composite regions of the BPT diagram, must be weak ($L_{2-10\,\mathrm{kev}} < 10^{44}$\,erg\,s$^{-1}$) if their presence is hidden by dilution from stellar emission.

\subsection{The star formation rates of broad-line AGN}

\begin{figure*}
\includegraphics[scale=0.8]{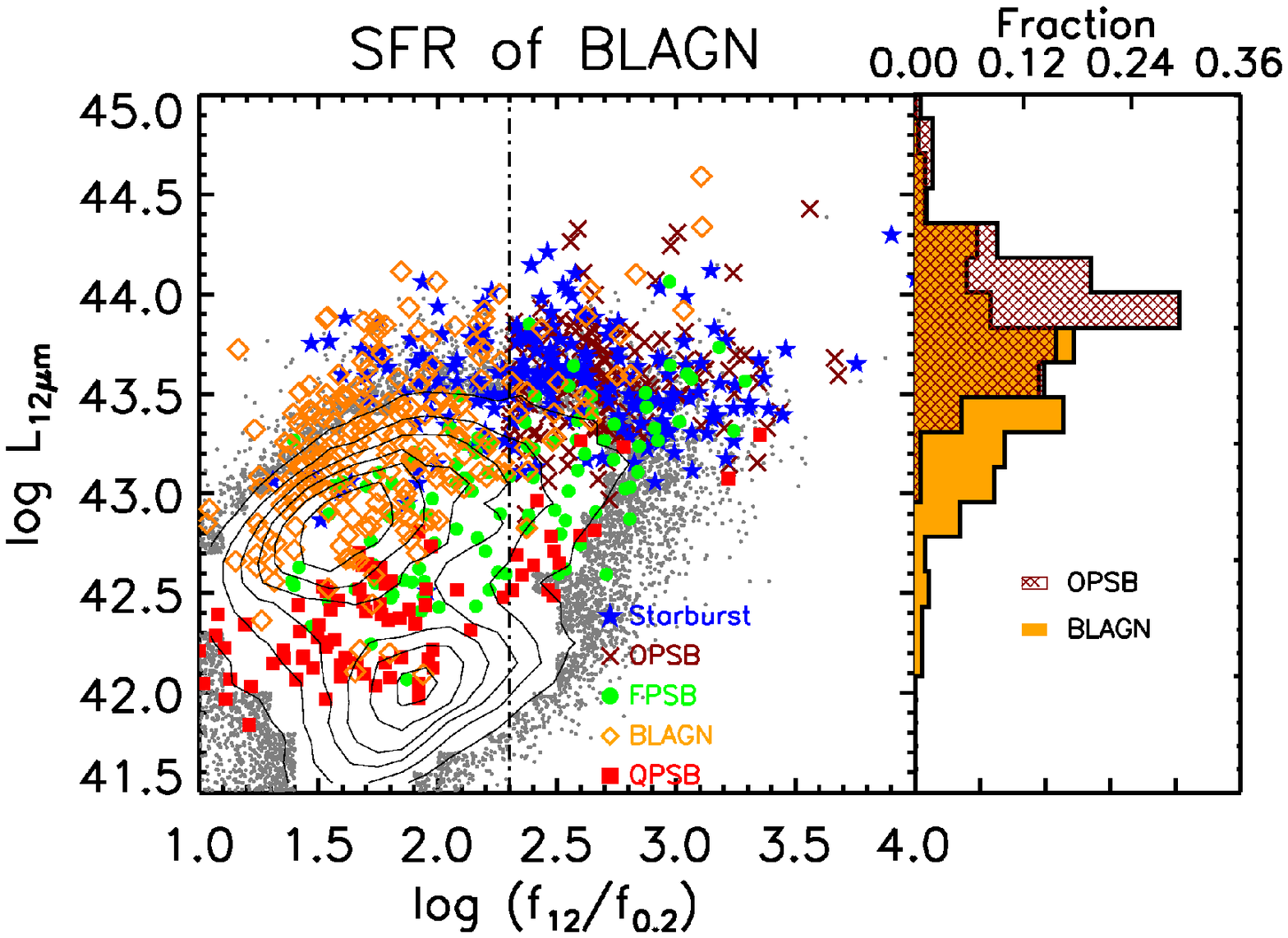}
\caption{The flux density ratio between WISE $12\,\mu \mathrm{m}$ and GALEX NUV, $\mathrm{f_{12\mu m}/f_{0.2\mu m}}$, versus the WISE $12\,\mu \mathrm{m}$ luminosity. The $12\, \mu \mathrm{m}$ luminosity is used as a proxy for the (obscured) star formation rate \citep[ an upper limit in AGN, see][]{donoso12}. $\mathrm{f_{12\mu m}/f_{0.2\mu m}}$ ratio may indicate the amount of dust-obscuration. The histograms on the right show the distribution of $12\,\mu \mathrm{m}$ luminosities for BLAGN and OPSBs respectively. Even with some contribution of AGN to the $12\,\mu \mathrm{m}$ luminosity, BLAGN are generally less star-forming than OPSBs (and majority of BLAGN are also likely less obscured). Therefore, BLAGN do not play a primary role quenching starbursts.  Their properties in this diagram are consistent with the idea that BLAGN come after the obscured AGN phase \citep{hopkins06}.}\label{fig:lum12}
\end{figure*}

Note that our post-starburst selection does not apply to broad-line AGN (BLAGN) hosts because many of the indicators that we have used to characterize the main evolutionary PSB sequence are diluted or polluted by the strong AGN signature in BLAGN (e.g., optical-UV colors and optical spectral signatures). Therefore, we cannot directly constrain the role of broad-line AGN in our post-starburst evolutionary sequence. However, in the following analysis we use $12\, \mu \mathrm{m}$ luminosities of BLAGN hosts to infer upper limits on their star formation rates \citep{chary01} and argue that their exclusion from the parent sample is not a problem. Their inferred star formation rates suggest that they either come after the obscured post-starburst AGN \citep[e.g.,][]{hopkins06} or they are not part of our evolutionary sequence at all. 

The fact that BLAGN seem older than SBs and OPSBs was already suggested by their intermediate $[4.6\,\mu \mathrm{m}]-[12\,\mu \mathrm{m}]$ color in Figure~\ref{fig:wise}. However, some AGN are known to exhibit suppressed aromatic features short-ward of $11.3\,\mu \mathrm{m}$ \citep[][]{smith07,diamond-stanic10}, suggesting that the $[4.6\,\mu \mathrm{m}]-[12\,\mu \mathrm{m}]$ color might underestimate the SFR.

Figure~\ref{fig:lum12} shows $12\,\mu \mathrm{m}$ luminosity against the flux density ratio of WISE $12\,\mu \mathrm{m}$ to GALEX NUV, $\mathrm{f_{12\mu m}/f_{0.2\mu m}}$. The $12\, \mu \mathrm{m}$ luminosity is dominated by stellar populations younger than 0.6\,Gyr in star-forming galaxies and in type $\Rmnum{2}$ AGN \citep{donoso12}. The $\mathrm{f_{12\mu m}/f_{0.2\mu m}}$ ratio roughly quantifies the ratio of obscured to unobscured SFR in star-forming galaxies and in type \Rmnum{2} AGN. It is not clear what $\mathrm{f_{12\mu m}/f_{0.2\mu m}}$ ratio exactly means for BLAGN because we do not know how much of their NUV and IR flux comes from stars and how much from the AGN. For this reason, we place more emphasis on the 12 $\mu$m luminosity as an upper limit on star formation.

The general galaxy population shows a bi-modality in $12\, \mu \mathrm{m}$ luminosity, reflecting the global bi-modality in star formation rates. As expected, the starbursts have higher $12\, \mu \mathrm{m}$ luminosity than normal star-forming galaxies while quenched post-starbursts have intermediate $12\, \mu \mathrm{m}$ luminosity between quiescent and star-forming galaxies. Most obscured post-starbursts have comparable $12\, \mu \mathrm{m}$ luminosity to that of starbursts. This, at face value, is inconsistent with the fact their SFRs as indicated by their $\mathrm{H\alpha}$ and $NUV-g$ colors are lower than those of starbursts (Figure~\ref{fig:hahd} \& ~\ref{fig:hd_d4}). However the excess mid-IR emission in OPSBs may be due to additional dust heating from their intermediate age ($\sim 0.4$\,Gyr) stellar populations \citep{salim09,kelson10}.

The FPSBs have high to intermediate $12\, \mu \mathrm{m}$ luminosity but they are clearly offset to the right from normal star-forming galaxies, that is, they are more dust-obscured. On the other hand, the BLAGN have similar $12\, \mu \mathrm{m}$ luminosities to those of FPSBs but most of them coincide with normal blue star-forming galaxies, i.e, they are less obscured.     

As the histograms of $12 \mu \mathrm{m}$ luminosities appended to the right of the plot shows, obscured post-starbursts are on average more luminous than BLAGN in $12\,\mu \mathrm{m}$. K-S test indicates that distribution of $12\, \mu \mathrm{m}$ luminosities of OPSBs and BLAGN are significantly different ($D=0.45,p_\mathrm{ks}= 5.0\times 10^{-18}$, i.e, $\alpha < 0.001$). Therefore, BLAGN are likely less star-forming (older) than obscured post-starbursts. This indicates that our conclusion that AGN and starbursts are not coeval is not likely to be affected by the exclusion of BLAGN from our post-starburst sample. More work is needed to directly constrain the age (after the starburst phase) of the very luminous BLAGN. Other studies have shown that BLAGN hosts have comparable age ($0.7 - 2$\,Gyr) to that of quenched post-starbursts   but they are substantially older than starbursts \citep{jahnke04,canalizo13}. At high redshift, a recent far-infrared Herschel/PACS study by \citet{rosario13} found that the mean SFRs of BLAGN hosts are consistent with those of normal massive star-forming galaxies and do not show strong enhancement in their SFRs to suggest that they are starbursting systems.

The fact that BLAGN have intermediate $[4.6\,\mu \mathrm{m}]-[12\,\mu \mathrm{m}]$ color and $12\, \mu \mathrm{m}$ luminosity and are likely less dust-obscured is consistent with the expectation that AGN might quench or prevent low-level star formation in post-starburst galaxies by removing leftover gas and dust after the starburst events \citep[][]{hopkins06}. Similarly, \citet{zakamska08} have shown that type \Rmnum{2} quasar hosts have increased star formation than type \Rmnum{1} quasar hosts, thereby supporting the suggestion that obscured quasars come before naked quasars. 

We conclude that broad-line AGN are unlikely to play \emph{a primary} role in the initial quenching of starbursts and their exclusion from our post-starburst sample does not affect our main conclusions. Future work to robustly constrain the instantaneous star formation rate of local BLAGN hosts would be very useful to understand whether BLAGN are associated with quenching of starbursts or low star-forming post-starburst galaxies.

\section{E: Flat color gradients of starbursts \& post-starbursts} 
\label{sec:appE}
Normal star-forming galaxies have red centers. A major merger likely alters or erases a color gradient of a pre-merger normal galaxy by inducing star formation at the center or throughout the galaxy. Since the truncation of the starburst is abrupt, post-starburst galaxies should still carry the imprint of their merger origin by having flatter or more positive color gradients than normal star-forming galaxies.

\begin{figure*}
\centering
\mbox{\subfigure{\includegraphics[width=3.5in]{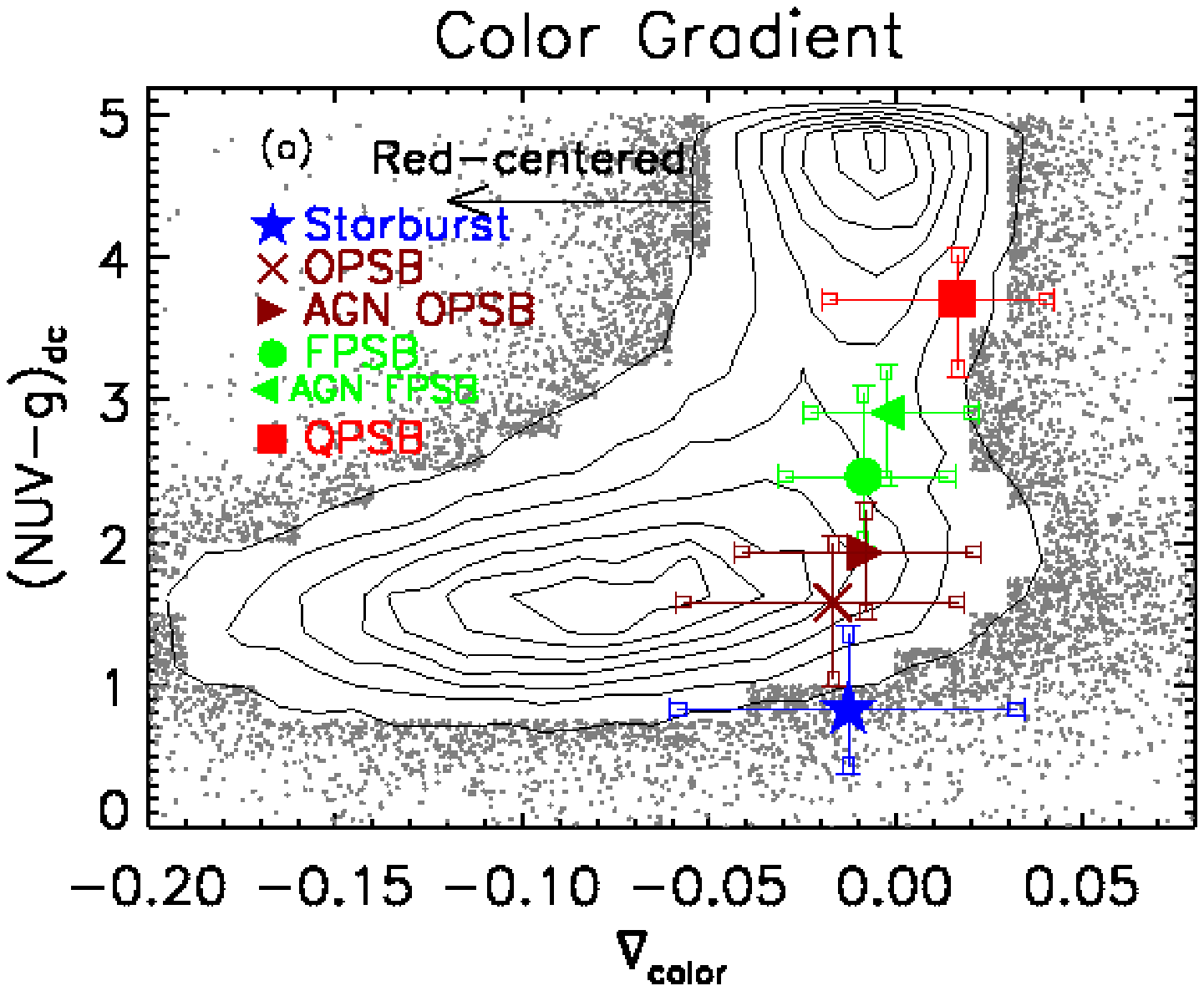}}
\subfigure{\includegraphics[width=3.5in]{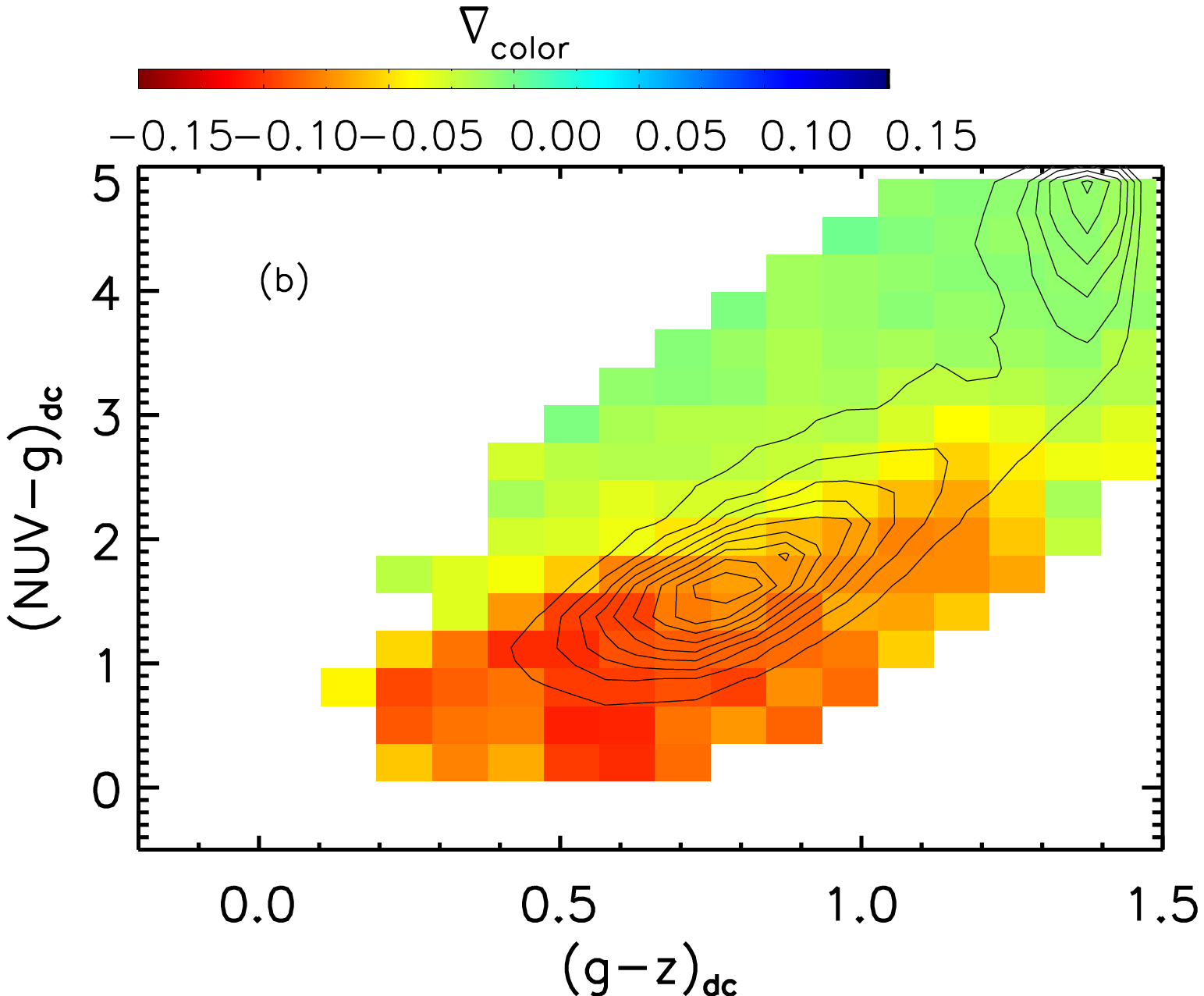} }}

\caption{Panel a: the median and upper/lower quartile values of $(NUV-g)_\mathrm{dc}$ color versus color gradient, $\nabla_\mathrm{color}$. Panel b: NUVgz diagram color-coded by color gradient. The contours represent the number density of all galaxies in the parent sample. Where the number of galaxies within a bin is more than 15, we color code by the median values of the color gradient or the concentration index. Otherwise, the individual values for the galaxy is used.\label{fig:cgrad}}
\end{figure*}

Figure~\ref{fig:cgrad}a shows the color gradient, $\nabla_\mathrm{color}$, versus $(NUV-g)_\mathrm{dc}$ color while Figure~\ref{fig:cgrad}b plots dust-corrected $NUV-g$ and $g-z$ diagram color-coded by the color gradient. Starbursts and their descendants (transiting and quenched post-starbursts) have much shallower color gradients than the bulk of normal star-forming galaxies.

As expected, red quiescent galaxies have flat color gradients and are red throughout but blue galaxies show an interesting regularity in their color gradients: blue galaxies with  bluer in $(NUV-g)_\mathrm{dc}$ color (or lower $\mathrm{H}\alpha$) have negative color gradients (i.e, show large reddening in their centers) while blue galaxies with redder $(NUV-g)_\mathrm{dc}$ colors (or higher $\mathrm{H}\alpha$) have flat color gradients. In other words, galaxies with red centers have most of their star formation in their outer blue disks and have small quiescent bulges. On the other hand, galaxies which are blue throughout are either experiencing nuclear bursts (have star formation rates above the average) or their nuclear bursts are abruptly terminated (have redder $NUV-g$ colors). They have young bulges.

The fact that both starbursts and post-starbursts have flat color gradients (young blue bulges) suggests that star formation is uniformly distributed throughout these galaxies, remaining so even as the burst quenches throughout the galaxy. It also suggests that the centers of these galaxies must have been unreddened from the typical red centers of disk galaxies, perhaps by gas inflow during a merger. Likewise, the flat color gradients of obscured post-starbursts suggests that the dust-obscuration in these objects is likely a galaxy wide phenomenon. Previous studies have also shown that A stars in K+A galaxies are widespread and are not confined to their nuclear regions \citep{kauffmann03c,swinbank05,goto08,pracy09,swinbank12}.

\section{F: Metalicity of post-starbursts}
\label{sec:appF}

\begin{figure*}

\centering

\includegraphics[scale=0.8]{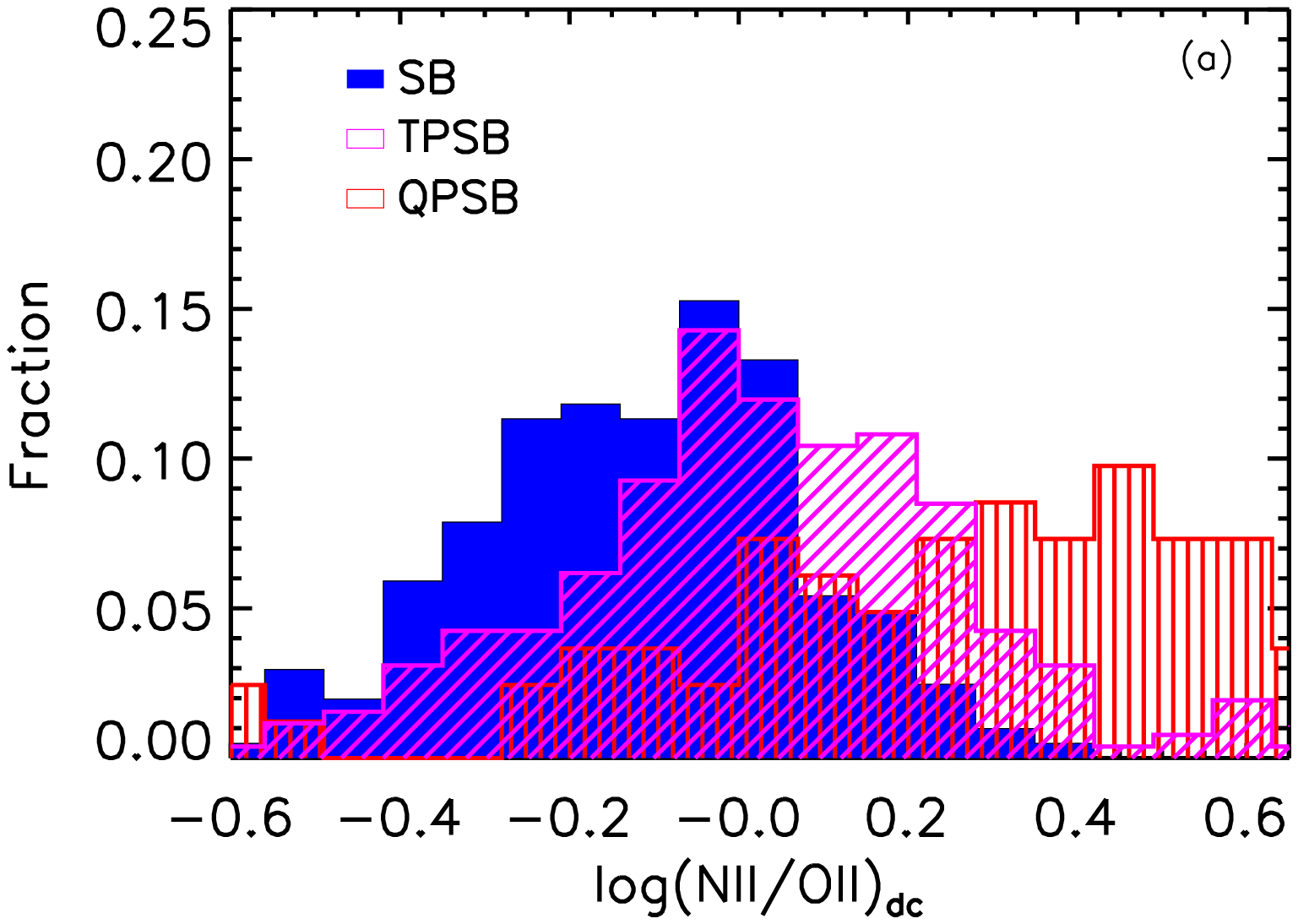}
\includegraphics[scale=0.8]{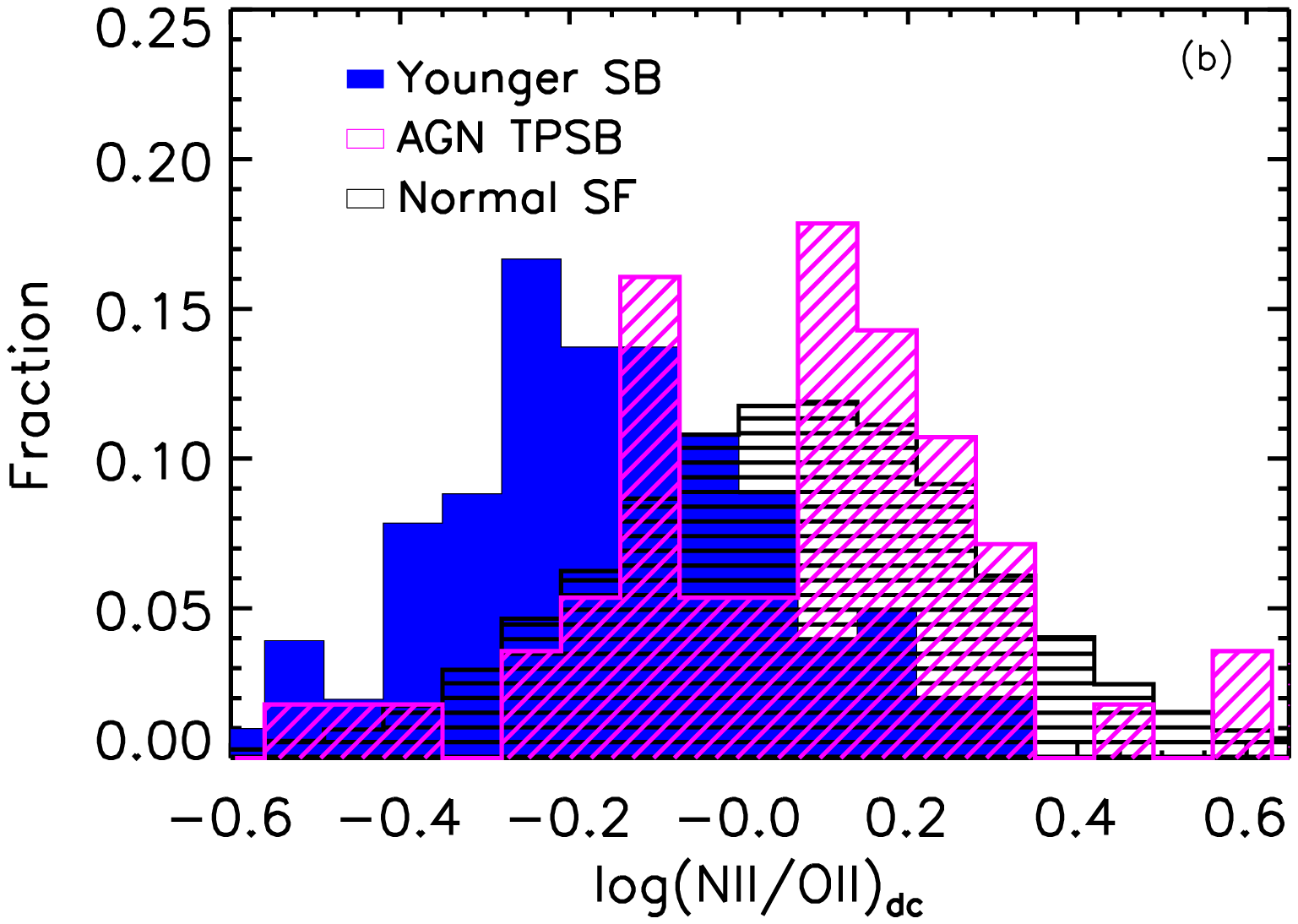}

\caption{Panel (a): The distribution dust-corrected \NIIOII\;ratios for the starbursts, transiting post-starburst and quenched post-starbursts. The metallicity increases from the starbursts
to the quenched post-starbursts. Panel (b):The distribution dust-corrected \NIIOII\;ratios for the young starbursts ($\mathrm{D_n(4000)} < 1.1$), normal star-forming galaxies and AGN in TPSBs. Starbursts are significantly metal poor especially at younger ages, consistent with gas inflow during merger-induced starbursts. The AGN is TPSBs have significantly higher metallicity than starbursts, suggesting a time delay between the starburst and AGN phase.\label{fig:o3han2}}
\end{figure*}

Figure~\ref{fig:o3han2} depicts the distribution of dust-corrected (\NIIOII)$_\mathrm{dc}$ ratios for normal star-forming galaxies, starbursts and post-starbursts. This ratio is a very reliable metallicity diagnostic and it is not very sensitive to the ionization level \citep[][]{kewley02}. Panel a compares the (\NIIOII)$_\mathrm{dc}$ ratios of transiting PSBs with those of quenched PSBs and starbursts. The (\NIIOII)$_\mathrm{dc}$ ratio increases as the starbursts progressively evolve to the transiting and quenched post-starbursts. The starbursts have  $(\NIIOII)_\mathrm{dc}$ ratios that correspond to $1-1.5\,Z_\odot$ solar metallicity range \citep[][]{kewley02} while the QPSBs have ratios corresponding to $2-3\,Z_\odot$ solar metallicity, although their (\NIIOII)$_\mathrm{dc}$ ratio might not be well measured because their emission lines are weak. The TPSBs have intermediate metallicity between starbursts and QPSBs. K-S test indicates that the distributions of metallicity of starbursts and TPSBs are significantly different ($D=0.28,p_\mathrm{KS}= 2.2 \times 10^{-8}$, i.e, $\alpha < 0.001$) and so are the distributions of TPSBs and QPSBs ($D=0.46,p_\mathrm{KS}= 2.3 \times 10^{-12}$, i.e, $\alpha < 0.001$).

Furthermore, panel b shows that starbursts which are younger than the median age ($D_n(4000)_\mathrm{dc} < 1.1$) have even lower metallicity than the AGN hosts in TPSBs \citep[cf.][]{groves06} and normal star-forming galaxies. K-S test also indicates that the distributions of metallicity of younger starbursts (or all SBs) are significantly different ( at $\alpha < 0.001$ level) from AGN in TPSBs or normal star-forming galaxies. The transiently lower metallicity of younger starbursts is consistent with metal poor gas inflows during merger-induced starbursts \citep{barnes91,barnes96} while the higher metallicity in post-starburst AGN is consistent with time delay between AGN and starbursts \citep[e.g.,][]{wild10,hopkins12,cen12}.

\clearpage

\bibliographystyle{apj}

\end{document}